\documentclass[journal]{IEEEtran}
\usepackage{times}
\usepackage{epsfig}
\usepackage{graphicx}
\usepackage{amsmath}
\usepackage{amsthm}
\usepackage{amssymb}
\usepackage{booktabs}
\usepackage{bm}
\usepackage{algorithm}
\usepackage{algorithmic}
\usepackage{mathtools}
\usepackage{enumerate}
\usepackage{paralist}
\usepackage{subcaption}
\usepackage{color}
\usepackage{cite}
\usepackage{colortbl}

\usepackage{caption}
\usepackage{tikz}
\usepackage{pgfplots}
\usetikzlibrary{spy,calc}
\usepackage{adjustbox}
\usepackage{makecell}
\usepackage{multirow}
\usepackage{siunitx}
\usepackage{etoolbox,siunitx}
\renewcommand{\bfseries}{\fontseries{b}\selectfont}
\robustify\bfseries
\newrobustcmd{\BF}{\bfseries}

\newif\ifblackandwhitecycle
\gdef\patternnumber{0}

\pgfkeys{/tikz/.cd,
zoombox paths/.style={
    draw=orange,
    very thick
},
black and white/.is choice,
black and white/.default=static,
black and white/static/.style={
    draw=white,
    zoombox paths/.append style={
        draw=white,
        postaction={
            draw=black,
            loosely dashed
        }
    }
},
black and white/static/.code={
    \gdef\patternnumber{1}
},
black and white/cycle/.code={
    \blackandwhitecycletrue
    \gdef\patternnumber{1}
},
black and white pattern/.is choice,
black and white pattern/0/.style={},
black and white pattern/1/.style={
    draw=white,
    postaction={
        draw=black,
        dash pattern=on 2pt off 2pt
    }
},
black and white pattern/2/.style={
    draw=white,
    postaction={
        draw=black,
        dash pattern=on 4pt off 4pt
    }
},
black and white pattern/3/.style={
    draw=white,
    postaction={
        draw=black,
        dash pattern=on 4pt off 4pt on 1pt off 4pt
    }
},
black and white pattern/4/.style={
    draw=white,
    postaction={
        draw=black,
        dash pattern=on 4pt off 2pt on 2 pt off 2pt on 2 pt off 2pt
    }
},
zoomboxarray inner gap/.initial=5pt,
zoomboxarray columns/.initial=2,
zoomboxarray rows/.initial=2,
subfigurename/.initial={},
figurename/.initial={zoombox},
zoomboxarray/.style={
    execute at begin picture={
        \begin{scope}[
            spy using outlines={%
                zoombox paths,
                width=\imagewidth / \pgfkeysvalueof{/tikz/zoomboxarray columns} - (\pgfkeysvalueof{/tikz/zoomboxarray columns} - 1) / \pgfkeysvalueof{/tikz/zoomboxarray columns} * \pgfkeysvalueof{/tikz/zoomboxarray inner gap} -\pgflinewidth,
                height=\imageheight/2 / \pgfkeysvalueof{/tikz/zoomboxarray rows} - (\pgfkeysvalueof{/tikz/zoomboxarray rows} - 1) / \pgfkeysvalueof{/tikz/zoomboxarray rows} * \pgfkeysvalueof{/tikz/zoomboxarray inner gap}-\pgflinewidth,
                magnification=3,
                every spy on node/.style={
                    zoombox paths
                },
                every spy in node/.style={
                    zoombox paths
                }
            }
            ]
        },
        execute at end picture={
        \end{scope}
        \gdef\patternnumber{0}
    },
    spymargin/.initial=0.5em,
    zoomboxes xshift/.initial=1,
    zoomboxes right/.code=\pgfkeys{/tikz/zoomboxes xshift=1},
    zoomboxes left/.code=\pgfkeys{/tikz/zoomboxes xshift=-1},
    zoomboxes yshift/.initial=0,
    zoomboxes above/.code={
        \pgfkeys{/tikz/zoomboxes yshift=1},
        \pgfkeys{/tikz/zoomboxes xshift=0}
    },
    zoomboxes below/.code={
        \pgfkeys{/tikz/zoomboxes yshift=-1},
        \pgfkeys{/tikz/zoomboxes xshift=0}
    },
    caption margin/.initial=1pt,
},
adjust caption spacing/.code={},
image container/.style={
    inner sep=0pt,
    at=(image.north),
    anchor=north,
    adjust caption spacing
},
zoomboxes container/.style={
    inner sep=0pt,
    at=(image.north),
    anchor=north,
    name=zoomboxes container,
    xshift=\pgfkeysvalueof{/tikz/zoomboxes xshift}*(\imagewidth+\pgfkeysvalueof{/tikz/spymargin}),
    yshift=\pgfkeysvalueof{/tikz/zoomboxes yshift}*(\imageheight+\pgfkeysvalueof{/tikz/spymargin}+\pgfkeysvalueof{/tikz/caption margin}),
    adjust caption spacing
},
calculate dimensions/.code={
    \pgfpointdiff{\pgfpointanchor{image}{south west} }{\pgfpointanchor{image}{north east} }
    \pgfgetlastxy{\imagewidth}{\imageheight}
    \global\let\imagewidth=\imagewidth
    \global\let\imageheight=\imageheight
    \gdef\columncount{1}
    \gdef\rowcount{1}
    
},
image node/.style={
    inner sep=0pt,
    name=image,
    anchor=south west,
    append after command={
        [calculate dimensions]
        node [image container,subfigurename=\pgfkeysvalueof{/tikz/figurename}-image] {\phantomimage}
        node [zoomboxes container,subfigurename=\pgfkeysvalueof{/tikz/figurename}-zoom] {\phantomimage}
    }
},
color code/.style={
    zoombox paths/.append style={draw=#1}
},
connect zoomboxes/.style={
    spy connection path={\draw[draw=none,zoombox paths] (tikzspyonnode) -- (tikzspyinnode);}
},
help grid code/.code={
    \begin{scope}[
        x={(image.south east)},
        y={(image.north west)},
        font=\footnotesize,
        help lines,
        overlay
        ]
        \foreach \x in {0,1,...,9} {
            \draw(\x/10,0) -- (\x/10,1);
            \node [anchor=north] at (\x/10,0) {0.\x};
        }
        \foreach \y in {0,1,...,9} {
            \draw(0,\y/10) -- (1,\y/10);                        \node [anchor=east] at (0,\y/10) {0.\y};
        }
    \end{scope}
},
help grid/.style={
    append after command={
        [help grid code]
    }
},
}

\newcommand\phantomimage{%
\phantom{%
    \rule{\imagewidth}{\imageheight/2}%
}%
}
\newcommand\zoombox[2][]{
\begin{scope}[zoombox paths]
    \pgfmathsetmacro\xpos{
        (\columncount-1)*(\imagewidth / \pgfkeysvalueof{/tikz/zoomboxarray columns} + \pgfkeysvalueof{/tikz/zoomboxarray inner gap} / \pgfkeysvalueof{/tikz/zoomboxarray columns} ) + \pgflinewidth
    }
    \pgfmathsetmacro\ypos{
        (\rowcount-1)*( \imageheight / \pgfkeysvalueof{/tikz/zoomboxarray rows} + \pgfkeysvalueof{/tikz/zoomboxarray inner gap} / \pgfkeysvalueof{/tikz/zoomboxarray rows} ) + 0.5*\pgflinewidth
    }
    \edef\dospy{\noexpand\spy [
        #1,
        zoombox paths/.append style={
            black and white pattern=\patternnumber
        },
        every spy on node/.append style={#1},
        x=\imagewidth,
        y=\imageheight
        ] on (#2) in node [anchor=north west] at ($(zoomboxes container.north west)+(\xpos pt,-\ypos pt)$);}
    \dospy
    \pgfmathtruncatemacro\pgfmathresult{ifthenelse(\columncount==\pgfkeysvalueof{/tikz/zoomboxarray columns},\rowcount+1,\rowcount)}
    \global\let\rowcount=\pgfmathresult
    \pgfmathtruncatemacro\pgfmathresult{ifthenelse(\columncount==\pgfkeysvalueof{/tikz/zoomboxarray columns},1,\columncount+1)}
    \global\let\columncount=\pgfmathresult
    \ifblackandwhitecycle
    \pgfmathtruncatemacro{\newpatternnumber}{\patternnumber+1}
    \global\edef\patternnumber{\newpatternnumber}
    \fi
\end{scope}
}

\newcommand{\matx}[1]{{\bm{#1}}} 
\newcommand{\vect}[1]{{\bm{#1}}} 
\newcommand{\set}[1]{{\mathbb{#1}}} 
\newcommand{\opt}[1]{{\mathcal{#1}}} 

\newcommand{\vp}{\vect{p}}
\newcommand{\vq}{\vect{q}}

\newcommand{\mE}{\matx{E}}

\newcommand{\mA}{\matx{A}}

\newcommand{\mN}{\matx{N}}

\newcommand{\mI}{\matx{I}}
\newcommand{\mJ}{\matx{J}}
\newcommand{\mR}{\matx{R}}

\newcommand{\mW}{\matx{W}}
\newcommand{\mGamma}{\matx{\Gamma}}
\newcommand{\mTheta}{\matx{\Theta}}
\newcommand{\mLambda}{\matx{\Lambda}}
\newcommand{\sR}{\set{R}}

\newcommand{\ie}{{\it i.e.}}
\newcommand{\eg}{{\it e.g.}}

\newcommand{\etal}{{\it et al.}}

\begin{document}

\title{Deep Bilateral Retinex for \\Low-Light Image Enhancement}


\author{
    Jinxiu Liang, Yong Xu, Yuhui Quan*, Jingwen Wang, Haibin Ling and Hui Ji
    \thanks{Jinxiu Liang, Yong Xu, and Yuhui Quan are with School of Computer Science and Engineering at South China University of Technology, Guangzhou, China. Yong Xu is also with Peng Cheng Laboratory, Shenzhen, China. Yuhui Quan is also with Guangdong Provincial Key Laboratory of Computational Intelligence and Cyberspace Information, China. (Email: cssherryliang@mail.scut.edu.cn; yxu@scut.edu.cn; csyhquan@scut.edu.cn)}
    \thanks{Jingwen Wang is with Tencent AI Lab, Shenzhen, China. (Email: jaywongjaywong@gmail.com)}
    \thanks{Haibin Ling is with the Department of Computer Science, Stony Brook University, Strony Brook, NY, USA. (Email: hling@cs.stonybrook.edu)}
    \thanks{Hui Ji is with Department of Mathematics at National University of Singapore, Singapore. (Email: matjh@nus.edu.sg)}
    \thanks{Asterisk indicates the corresponding author.}
}

\maketitle

\begin{abstract}
    Low-light images, \textbf{\textit{i.e.}} the images captured in low-light conditions, suffer from very poor visibility caused by low contrast, color distortion and significant measurement noise. Low-light image enhancement is about improving the visibility of low-light images. 
    As the measurement noise in low-light images is usually significant yet complex with spatially-varying characteristic,  how to handle the noise effectively is an important yet challenging problem in low-light image enhancement. 
    Based on the Retinex decomposition of natural images,  this paper proposes a deep learning method for low-light image enhancement with a particular focus on handling the measurement noise. 
    The basic idea is to train a neural network to generate a set of pixel-wise operators for simultaneously predicting the noise and the illumination layer, where the operators are defined in the bilateral space.
    Such an {integrated approach} allows us to have an accurate prediction of the reflectance layer in the presence of significant spatially-varying measurement noise.
    Extensive experiments on several benchmark datasets have shown that the proposed method is very competitive to the state-of-the-art methods, and has significant advantage over others when processing images captured in extremely low lighting conditions.
\end{abstract}


%
\IEEEpeerreviewmaketitle

\section{Introduction}
\label{sec:introduction}
\begin{figure*}
    \centering
    \setlength{\tabcolsep}{2pt}
        \begin{tabular}{ccccc}        	
            \includegraphics[width=0.19\linewidth]{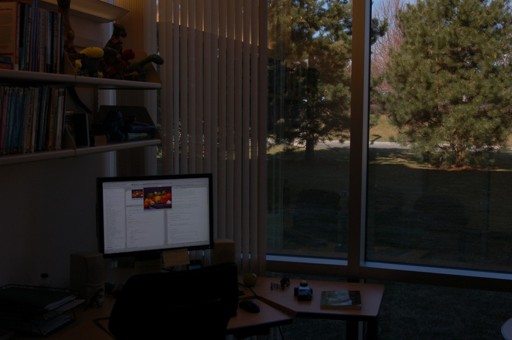} &
            \includegraphics[width=0.19\linewidth]{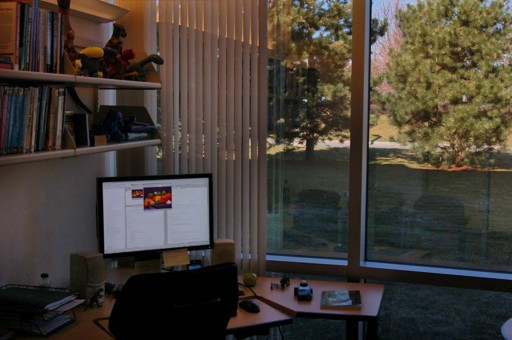} &
            \includegraphics[width=0.19\linewidth]{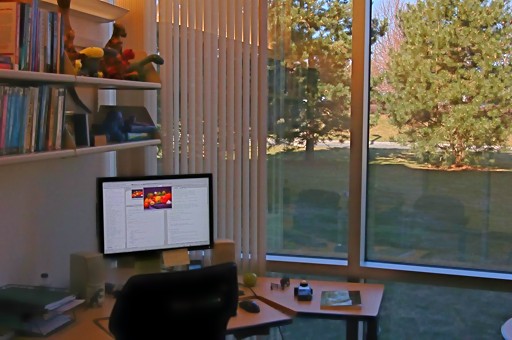} &
            \includegraphics[width=0.19\linewidth]{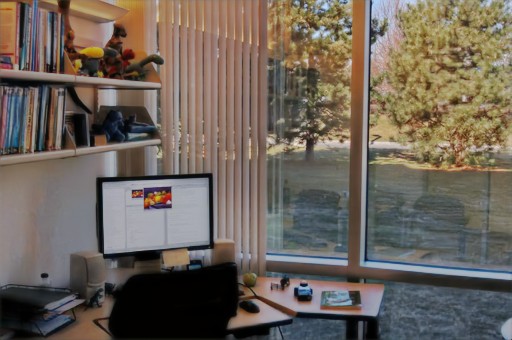} &
            \includegraphics[width=0.19\linewidth]{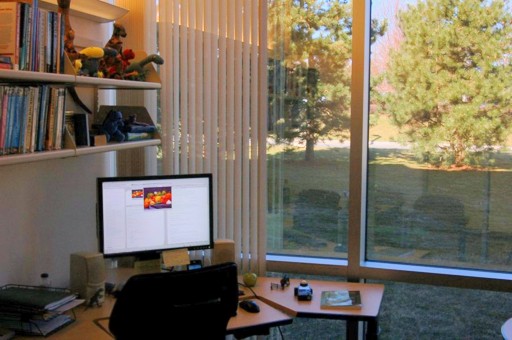} \\
            \includegraphics[width=0.19\linewidth]{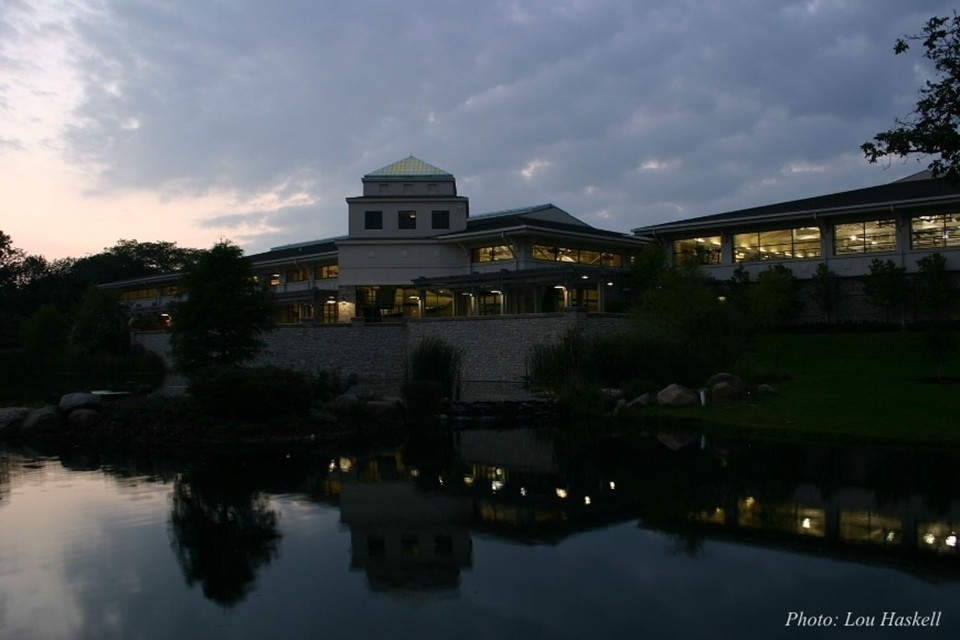} &
            \includegraphics[width=0.19\linewidth]{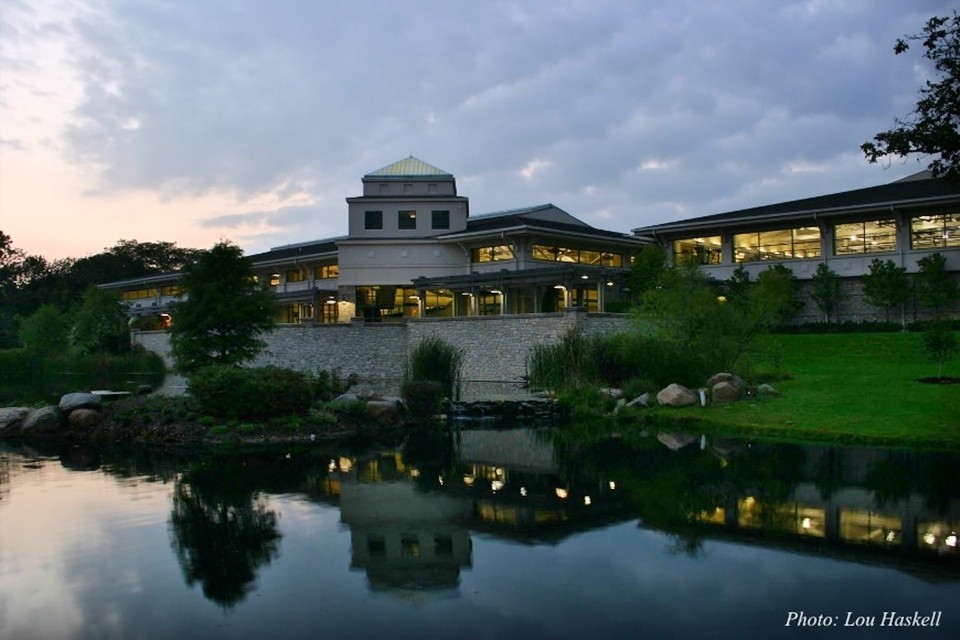} &
            \includegraphics[width=0.19\linewidth]{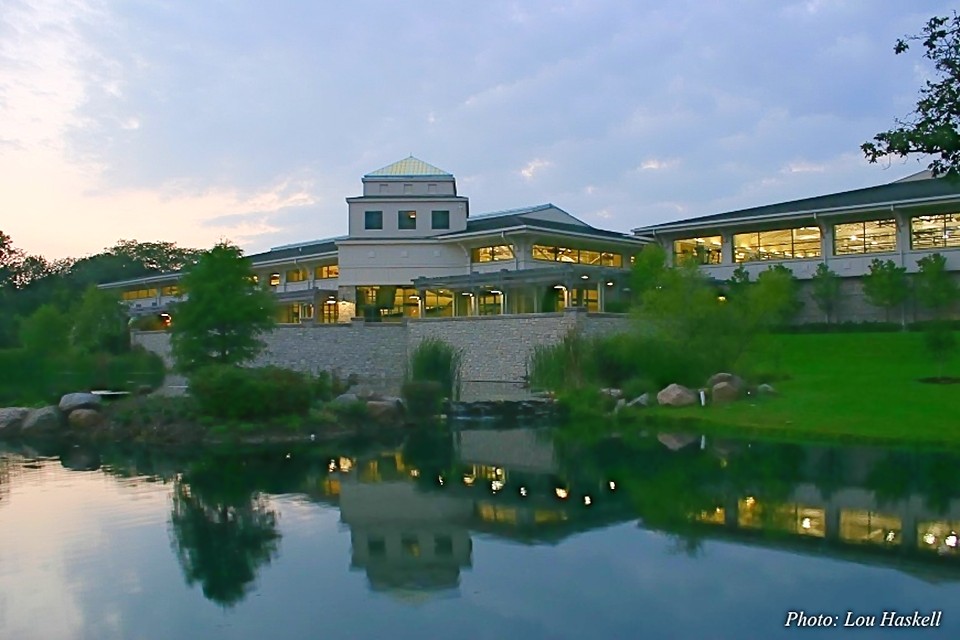} &
            \includegraphics[width=0.19\linewidth]{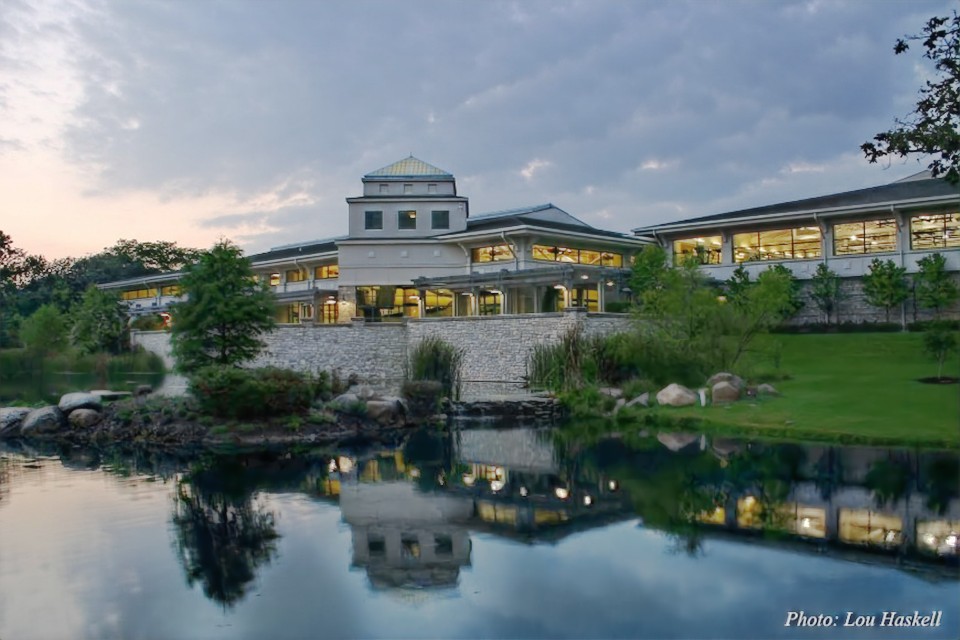} &
            \includegraphics[width=0.19\linewidth]{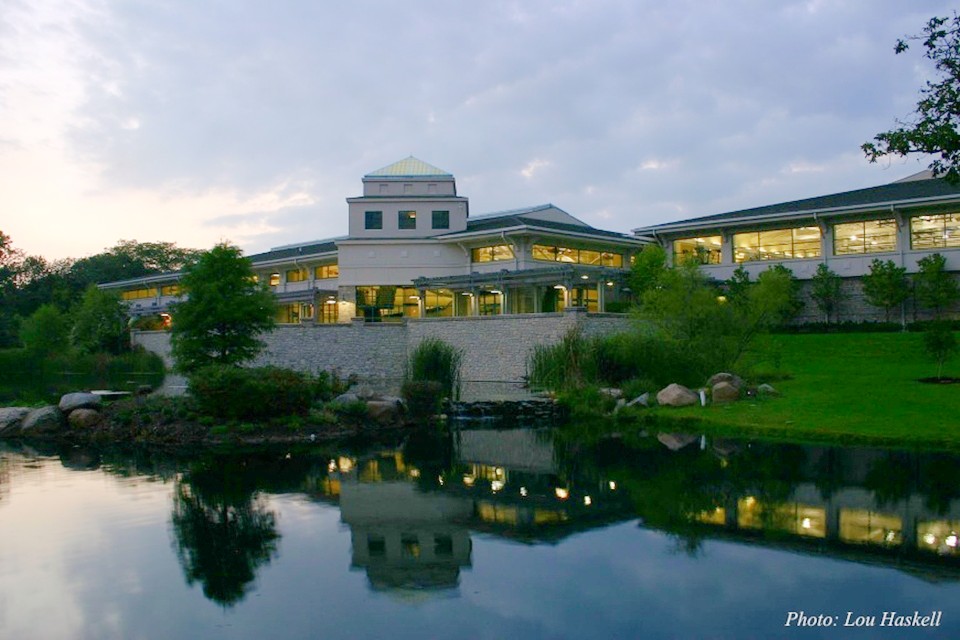}\\
            \includegraphics[width=0.19\linewidth]{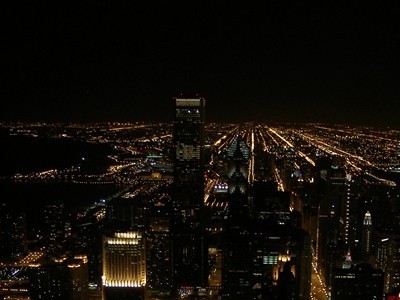} &
            \includegraphics[width=0.19\linewidth]{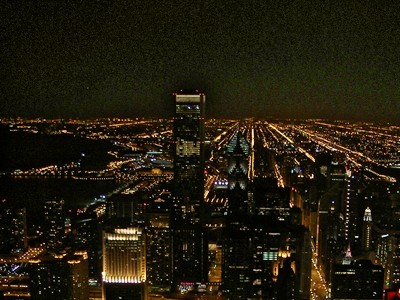} &
            \includegraphics[width=0.19\linewidth]{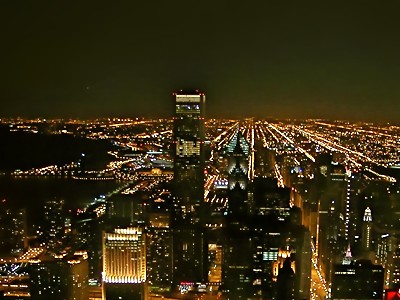} &
            \includegraphics[width=0.19\linewidth]{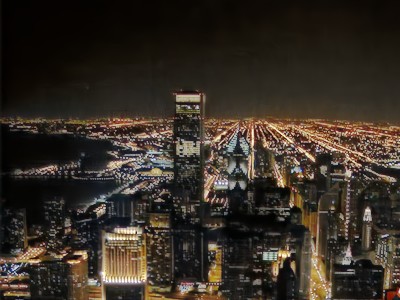} &
            \includegraphics[width=0.19\linewidth]{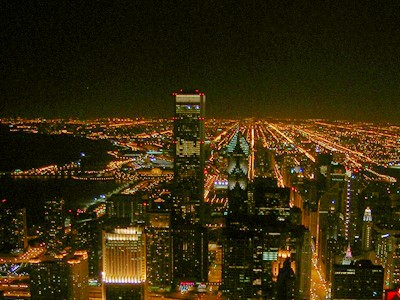}\\
            {\small (a) Low-light image } &  {\small (b) SRIE~\cite{fu2016weighted} } &  {\small (c) RRM~\cite{li2018structurerevealing} } &  {\small (d) KinD~\cite{zhang2019kindling}  } &  {\small (e) Ours}
        \end{tabular}
    \caption{\small Demonstration of existing low-light image enhancement methods and the proposed one. Please zoom-in for easier visual inspection.
    }
    \label{fig:teaser}
\end{figure*}

It often occurs in practice that one needs to capture images in low-light conditions, \eg~at dawn/twilight  and in dimly-lit indoor rooms.   
Images captured in low-light conditions, \ie~\emph{low-light images}, usually have poor visibility in terms of  low contrast, color distortion and low signal-to-noise-ratio (SNR). 
Low-light image enhancement is then about improving the visual quality of low-light images for better visibility of image details and  higher SNR. 
See Fig.~\ref{fig:teaser} for an illustration. 
Such a technique not only sees its practical values in digital photography, but also benefits many computer vision applications (\eg~surveillance and tracking) in low-light conditions.

There has been an enduring effort on developing effective techniques for low-light image enhancement, \eg~histogram equalization and gamma correction. 
In recent years, the Retinex model of images has been one prominent choice for developing more powerful low-light image enhancement techniques; see \eg~\cite{wang2013naturalness,fu2015probabilistic,fu2016fusionbased,fu2016weighted,guo2017lime,wei2018deep,zhang2019kindling,wang2019underexposed,li2018structurerevealing}. 
The Retinex model of images assumes that an image $\mI$ is composed of two different layers, the reflectance $\mR$ and the illumination $\mE$, in the following expression:
\begin{equation}\label{eqn:Retinex}
\mI = \mR\odot  \mE +\mN,
\end{equation}
where $\odot$ denotes element-wise multiplication, and $\mN$ denotes the measurement noise.
The layer $\mR$ denotes the reflectance map that encodes  inherent image structures, \ie~physical characteristics  of scenes/objects. 
The layer $\mE$ denotes the illumination map which is related to the light intensities of {scenes/objects} determined by the lighting condition.  
Once the Retinex decomposition of $\mI$ is done, one can  reconstruct a new image $\widetilde{\mI}$ with better visibility by replacing $\mE$ using another illumination layer $\widetilde{\mE}$:
\begin{equation}\label{eqn:Retinex_syn}
\widetilde{\mI}=\mR\odot \widetilde{\mE}.
\end{equation}
For instance, $\widetilde{\mE}$ can be defined using the gamma correction function $\widetilde{\mE}:=\mE^{\frac 1 \gamma}$.

It can be seen that the problem of low-light image enhancement can be recast as the Retinex decomposition problem~\eqref{eqn:Retinex}. It is an ill-posed inverse problem, and the low SNR of the input low-light image further {aggravates} the ill-posedness. Therefore, there are two main challenges for solving~\eqref{eqn:Retinex}:
\begin{enumerate}
    \item  How to resolve the ambiguities between the two maps, 
    \item  How to make the estimation robust to noise.
\end{enumerate}

Regarding the first question, the answer from most existing works is to impose certain prior on both the reflectance layer and the illumination layer. 
In the past, such priors usually are pre-defined based on empirical observations, \eg~spatial smoothness prior on the illumination layer~\cite{kimmel2003variational,wang2013naturalness, fu2016fusionbased,guo2017lime} and piece-wise smoothness prior on the reflectance layer~\cite{ma20111based,ng2011total,fu2016weighted}.
More recently, deep learning has become one promising tool of learning the priors for Retinex decomposition. 
It has been used either  for only estimating the illumination layer (\eg~\cite{wang2018gladnet,wang2019underexposed}) or for estimating both layers (\eg~\cite{wei2018deep,zhang2019kindling}). 

The answer to the second question also plays an important role in low-light image enhancement, as the  measurement noise will be noticeably amplified when taking a direct inversion.   The SNR of a low-light image is usually  much lower than its counterparts taken under normal lighting conditions. Recall that, as the light sensors of a camera  usually cannot receive adequate light in low-light conditions, the shot noise  caused by statistical quantum fluctuations will be much more prominent in a low-light image. Together with the necessity of an amplification of light sensitivity of sensors (\ie~a higher ISO) in low-light conditions, 
low-light images tend to have low SNRs. 
In other words, an effective denoising mechanism is another key component of  a Retinex-model-based low-light image enhancement method with good performance.

\subsection{Discussion on measurement noise of low-light images}\label{sec:dis}
As a low-light image often has rather low SNR, the treatment of measurement noise plays  an important role for the Retinex decomposition. Many existing methods ignore this issue, leading to noticeable noise magnification in the reflectance layer; see Fig.~\ref{fig:teaser} (b) and  Fig.~\ref{fig:LOL_146_decom} (b) for an illustration. 
Some other existing solutions deal with the magnified noise in the result by running a denoising post-processing. However, the noise after magnification has much more complex characteristic and is closely related to inherent image structures. As a result, the reflectance layer after post-processing often tends to be over-smoothed with many image details lost; see Fig.~\ref{fig:teaser} (d) and  Fig.~\ref{fig:LOL_146_decom} (d) for an illustration.

There exist profound connections between the noise $\mN$, the illumination layer $\mE$ as well as the reflectance layer $\mR$. 
The measurement noise $\mN$ spatially varies over different regions of a low-light image. It is not {\it i.i.d.} and thus cannot be easily distinguished from the structures of reflectance by off-the-shelf image denoisers or image denoisers with pre-defined regularizations. 
See Fig.~\ref{fig:teaser} (c) and Fig.~\ref{fig:LOL_146_decom} (c) for an illustration of the result using a pre-defined regularization model~from~\cite{li2018structurerevealing}. 
Indeed, the noise variance  is closely related to the illumination map $\mE$. In bright regions, $\mN$ is dominated  by the image-dependent shot noise caused by the randomness of light arrival. In dark regions, $\mN$ is dominated  by the image-independent read noise caused by the sensitivity of sensor readout.
In addition, there is also other noise from many sources, including dark current noise, thermal noise and quantization noise. Interested readers are referred to~\cite{wei2020physicsbased} for more details.

The treatment of measurement noise also plays a critical role for recovering the reflectance layer $\mR$. It can be seen from \eqref{eqn:Retinex} that once the illumination layer $\mE$ is estimated, one can estimate $\mR$ via a linear inversion. In a low-light image, many image structures of the reflectance layer, \eg~edges and textures, are of weak magnitude. Thus, it is challenging to distinguish noise from these weak structures. To effectively remove the noise during inversion, a powerful denoising scheme needs to be specifically designed for low-light images with low SNRs.

\subsection{Main idea}
Deep learning has emerged as a powerful tool in many image processing tasks. Based on the Retinex model \eqref{eqn:Retinex}, this paper aims at developing a powerful 
low-light image enhancement method with effective treatment on complex measurement noise. The proposed method takes a two-stage approach. 
Given an input low-light image $\mI$,  we first estimate the illumination layer $\mE$ and measurement noise $\mN$:
\begin{equation}\label{eqn:stage1}
\mI\rightarrow (\mN, \mE).
\end{equation}
Once $\mN$ and $\mE$ are estimated, the reflectance layer $\mR$ can be obtained by a linear inversion:
\begin{equation}\label{eqn:stage2}
\mR:=(\mI-\mN)\oslash \mE,
\end{equation}
where $\oslash$ denotes element-wise division.

In the procedure above, an accurate estimation of noise $\mN$ with spatially-varying characteristic is critical to the success of low-light image enhancement. As we discussed in Section~\ref{sec:dis}, there exists profound connection between the illumination layer $\mE$ and the noise $\mN$. Thus, we proposed a deep NN, called  \emph{Deep Bilateral Retinex} (DBR),  which is mainly an NN-based joint estimator of the measurement noise and the illumination layer. 

More specifically, in the proposed method, the interaction between the estimation of $\mE$ and $\mN$ is done by 
training a single NN  which takes the low-light image as input and outputs a pair of learnable pixel-wise linear transforms for predicting the two layers. The transform for predicting the illumination layer is simply a pixel-wise affine transform. For the noise, motivated by the bilateral filtering for image denoising and its NN extensions~\cite{tomasi1998bilateral,gharbi2017deep},  the pixel-wise linear transform learned for the noise estimation is defined in the so-called  \textit{bilateral space}, \ie, the spatial-range product space with an augmented dimension on  pixel color. 

Inside such a pair of learned pixel-wise linear transforms,
the module for estimating the noise $\mN$ is based on  the pixel-wise \emph{deformable convolution} which uses spatially-varying filtering kernels learned in the bilateral space. The module for estimating  the illumination layer $\mE$ is built on point-wise color transform matrices. Once the illumination layer $\mE$ and noise $\mN$ are estimated, the reflectance layer is predicted using \eqref{eqn:stage2}. Also, a loss function that encourages
the focus on image edges is adopted for further refinement on the separation between the noise and the reflectance layer.

%


\subsection{Contributions}
\label{sec:contributions}
The effective treatment on the measurement noise plays an important role in Retinex-decomposition-based low-light image enhancement. 
The measurement noise in low-light image is not only  significant in comparison to the magnitude of image structures, but also is spatially varying with complex statistical characteristics.  
This paper proposes a deep-learning-based method for low-light image enhancement with a particular focus on handling the measurement noise. 

By exploiting the inherent connections between the spatially-varying noise and the illumination layer, 
we develop a framework that enables the interaction between noise estimation and illumination layer estimation in the bilateral space. 
The effectiveness of the proposed method is extensively evaluated on several benchmarks. The experimental results show that the proposed method is very effective at handling measurement noise. 
For the images captured in very low-light conditions, the proposed method outperforms existing ones by a large margin. For the images captured in better lighting conditions whose measurement noise is relatively low, the proposed method still provides comparable performance to those state-of-the-art (SOTA) methods.


\section{Related Works}
\label{sec:related}
In the past, there have been extensive studies on low-light image enhancement. In the next, we give a brief discussion on existing low-light image enhancement methods, and focus more on Retinex-model-based methods.

\subsection{Non-Retinex-based methods}
Early works tackle the problem of low-light image enhancement by directly modifying the low-light image such that the resulting image has higher contrast.
The histogram equalization~\cite{arici2009histogram, celik2011contextual, lee2013contrast} improves the visibility of a low-light image by balancing its histogram.
The Gamma correction (power-law transformation)~\cite{yuan2012automatic, huang2013efficient} modifies the brightness of an image by increasing the brightness of dark regions and decreasing the brightness of bright regions. 
Multi-exposure sequence fusion is also exploited for the contrast enhancement in low-light images~\cite{ying2017bioinspired,cai2018learning}. 
Chen~\etal~\cite{chen2018learning} tackles the problem by directly modifying the raw data from image sensors using a learnable NN.

Since a direct contrast enhancement will magnify the measurement noise, much effort has been devoted to the noise reduction in contrast enhancement. 
Loza~\etal~\cite{loza2013automatic} performed wavelet-based noise reduction during contrast enhancement. 
Based on deep auto-encoder, Lore \etal~\cite{lore2017llnet} proposed a Low-Light Net (LLNet) to sequentially learn contrast enhancement and noise reduction.

\subsection{Retinex-based non-learning methods}
The Retinex image model \eqref{eqn:Retinex} proposed in \cite{land1977retinex} has been widely used for  image enhancement; see \eg~\cite{jobson1997properties,jobson1997multiscale,zhao2012closedform,wang2013naturalness,guo2017lime}. The majority of existing Retinex-based approaches assume the image being processed contains only negligible noise. The key of these methods is about how to resolve the ambiguities between the illumination and reflectance layers. Most existing non-learning methods resolve such ambiguities by imposing certain prior either on the illumination layer or the reflectance layer, or both.

Several methods proposed different priors on  the illumination layer. 
The smoothness prior is  first introduced to variational models by Kimmel~\etal~\cite{kimmel2003variational} which minimizes the squared $\ell_2$ norm of gradients of illumination layer.
Wang~\etal~\cite{wang2013naturalness} proposed a bright-pass filter for better preserving the naturalness of the illumination layer. Such an idea is further refined by Fu~\etal~\cite{fu2016fusionbased}  via fusing multiple derivatives of the illumination layer for better performance.
Guo~\etal~\cite{guo2017lime} proposed a structure-aware prior for the  illumination layer which is motivated from  relative total variation (RTV)~\cite{xu2012structure}. 
There is also some work  imposing the prior only on the reflectance layer. For instance, Ma~\etal~\cite{ma20111based} imposed a piece-wise smoothness prior on the reflectance layer. 

Another class of methods resolves the solution ambiguity by imposing the priors on both two layers. 
In Ng~\etal~\cite{ng2011total}, the TV prior is imposed on  both reflectance and illumination layers after applying the
logarithmic transformation on the input image.  
Instead of using logarithmic transform as a pre-processing, Fu \etal~\cite{fu2015probabilistic} introduced a probabilistic method for simultaneous illumination and reflectance estimation (SIRE) in the linear space rather than the logarithmic space. 
Another variation  comes form \cite{fu2016weighted} which proposes a weighted variational model to enhance the variation of derivative magnitudes in bright regions.

In addition to resolving the solution ambiguity, some methods are proposed to process low-light images with significant noise.
Elad~\etal~\cite{elad2005retinex} proposed to constrain the bilateral smoothness on pixel values of  both  illumination layer and reflectance  layer using two tailored bilateral filters.
A robust fidelity term with an explicit noise term is used in Ren~\etal~\cite{ren2018joint} and Li~\etal~\cite{li2018structurerevealing} to handle measurement noise. 
Nevertheless, the complex and spatially-varying characteristic of measurement noise 
makes these approaches not very effective.

\subsection{Retinex-based learning methods without noise handling}\label{sec:no-noise}
In recent years, deep learning has emerged as one prominent tool in image enhancement, including Retinex-based low-light image enhancement.
Wang~\etal~\cite{wang2018gladnet} proposed to estimate and adjust the illumination layer of a low-light image by an NN. 
Gharbi~\etal~\cite{gharbi2017deep} proposed a bilateral learning framework for photography enhancement, which trains an NN to predict point-wise color transform coefficients for the color vector at each pixel. 
The similar idea is used in~\cite{wang2019underexposed} that learns the image-to-illumination mapping for under-exposure correction. These methods do not take  the measurement noise into consideration. In the case of low SNR, the point-wise transform used in these methods is sensitive to noise, especially in the dark regions of low-light images. As a result, the visual quality of the results from these methods is not very satisfactory, especially for low-light images with low SNRs. 

\subsection{Retinex-based methods with noise handling}
The measurement noise of low-light images is often  quite significant.  Without appropriate noise treatment, those deep learning methods listed in Section~\ref{sec:no-noise} are likely to have  erroneous estimations of both  layers in dark regions.  
Recently, several deep learning methods have been proposed with the focus on better robustness to noise. 
Wei~\etal~\cite{wei2018deep} proposed to decompose a low/normal-light image into the corresponding reflectance and illumination layers by an NN and then adjust the illumination by another NN. 
An off-the-shelf denoiser was then used as a post-processing to remove the artifacts of the reflectance layer caused by noise. 
Zhang \etal~\cite{zhang2019kindling} trained a denoising NN for removing the artifacts of the reflectance layer, which leads to better visual quality of the result. 
However, as the artifacts caused by noise  have complex characteristic and are highly correlated to the reflectance layer, it is difficult to accurately separate artifacts and the truth reflectance layer. 
Often some details of the reflectance layer are wrongly removed as artifacts in these methods.

\section{Deep Bilateral Retinex}
\label{sec:method}


In this section, we aim at developing a deep learning method for Retinex-based low-light image enhancement with a built-in powerful denoising module.  Recall that the Retinex model of  a low-light image $\mI\in\sR^{P_x\times P_y\times 3}$ is expressed as
$$
\mI=\mR\odot \mE+\mN,
$$
where $\mR$ denotes the reflectance layer, $\mE$ denotes the illumination layer and $\mN$ denotes the measurement noise. Once ${\mN}$ and ${\mE}$ are estimated, 
the reflectance layer $\mR$ is obtained by
\begin{equation}
\label{eq:final}
\widetilde{\mR} = (\mI -{\mN})\oslash{\mE}.
\end{equation}
Following \cite{wang2019underexposed},
for a low-light image, we assume the illumination layer of its counterpart taken in the normal lighting condition  has the following  illumination layer: 
$$\widetilde{\mE}=\mE^{\frac{1}{\infty}}=\mathbf{1}.$$
In other words, the estimated reflectance layer $\widetilde \mR$ of the input low-light image is considered as 
the output of the proposed Retinex-based low-light image enhancement.

The focus of the proposed low-light image enhancement is then on how to estimate the noise layer $\mN$ and the illumination layer $\mE$. As we discussed in the previous section, the noise  characteristic of  $\mN$ is spatially varying and inherently related to the illumination layer $\mE$.  In the next, we introduce an NN architecture that enables a joint  prediction of the two layers with a built-in interaction mechanism.

\begin{figure*}[htbp!]
    \begin{center}
        \includegraphics[width=0.7\linewidth]{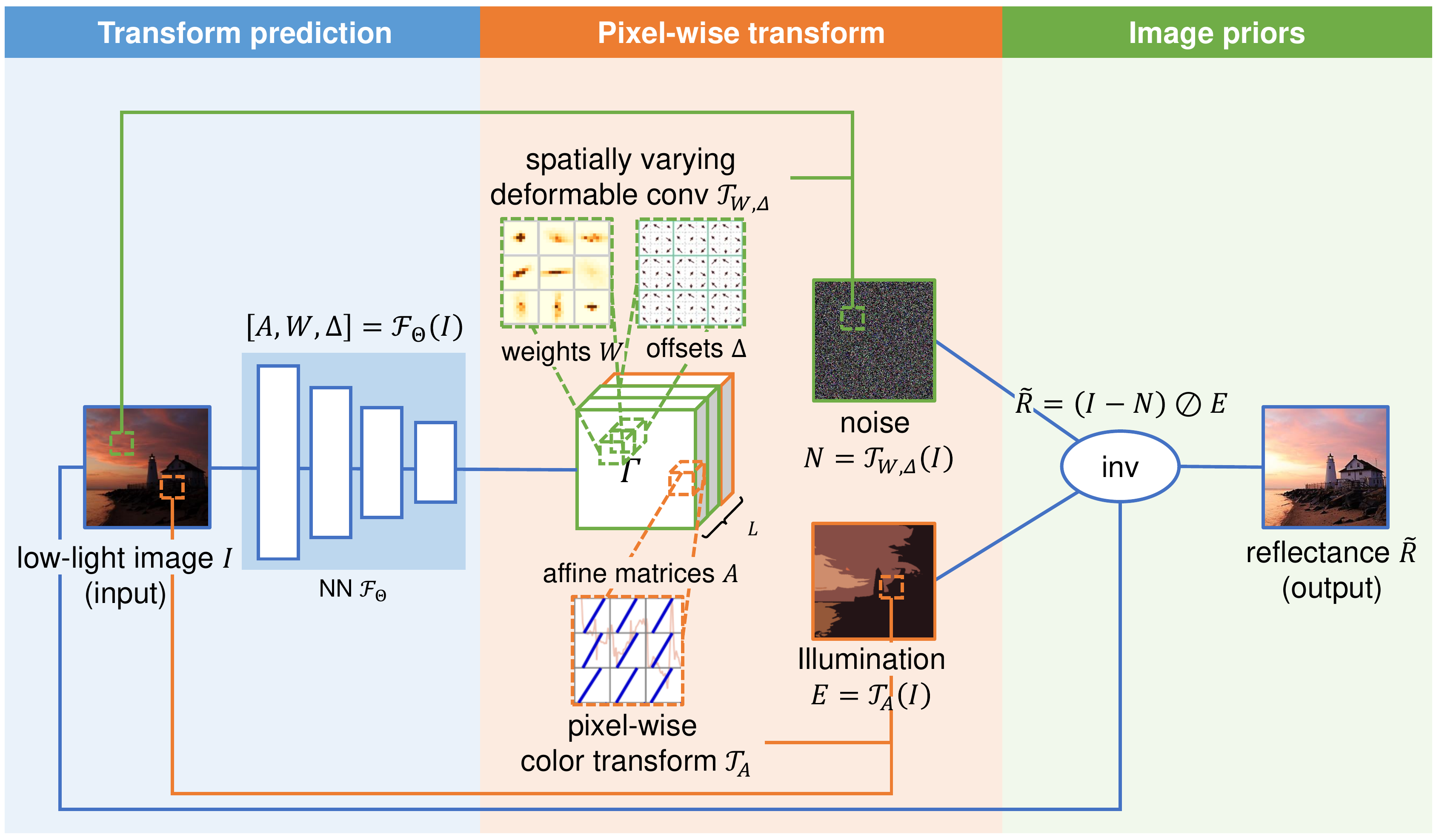}
        \caption{\small 
            Framework of the proposed method. 
            The NN $\opt{F}_\mTheta$ learns pixel-wise linear transforms $\mGamma$ including color transform $\opt{T}_{\mA}$ and deformable convolution $\opt{T}_{\mW,\Delta}$ to transform input $\mI$ pixel-wise into $\mE,\mN$ firstly. 
            Then the output $\widetilde{\mR}$ is obtained by performing inversion using \eqref{eq:final} with estimated $\mE,\mN$. 
        }
        \label{fig:overall}
    \end{center}
\end{figure*}


\subsection{Outline of the NN for joint estimation of $\mN$ and $\mE$ in bilateral space}
\label{sec:outline}
Recall that we need to have a joint estimation of $\mN$ and $\mE$ for exploiting their inherent correlation. Thus, instead of proposing an end-to-end network that directly maps the input image to these two layers, we propose to train an NN that learns a pair of linear transforms, which maps a low-light image $\mI$ to the noise layer $\mN$ and the illumination layer $\mE$, as shown in Fig.~\ref{fig:overall}.
More specifically, the proposed NN, denoted by $\opt{F}_\mTheta$, maps an input image $\mI$ to a pair of transforms:
\begin{equation}
\opt{F}_\mTheta: \ \ \mI\longrightarrow \{\underbrace{\opt{T}_{\mA}(\mI)}_{\mE}, \underbrace{\opt{T}_{\mW,\Delta}(\mI)}_{\mN}\}.
\label{eq:trans}
\end{equation}
Then, the noise $\mN$ and illumination layer $\mE$ are estimated by
\begin{equation}\label{eqn:mapping}
\begin{array}{ll}
\opt{T}_{\mA}: &\mI\longrightarrow \mE \\
\opt{T}_{\mW,\Delta}: &\mI\longrightarrow \mN.
\end{array}
\end{equation}

{It is shown in~\cite{chen2016bilateral} that many photographic transformations can be locally well-approximated by affine color transforms.} Therefore, for the illumination layer $\mE$, the transform $\opt{T}$ is defined by a set of affine transforms $\opt{A}=\{\mA_\vp\}_{\vp}\subseteq \mathbb{R}^{3,4}$. In other words, the operator $\opt{T}_\mA$ in
\eqref{eqn:mapping} is defined as
\begin{equation}
\label{eq:color_trans}
\opt{T}_{\mA}: \mI_\vp\longrightarrow \mE_\vp:= \mA_p[\mI_\vp,1]^\top,\quad\mbox{for each pixel $\vp$,}
\end{equation}
where the set $\opt{A}=\{\mA_\vp\}_\vp$ contain the coefficients predicted by the NN.



For the estimation of noise layer, we also need to learn a pixel-wise transform, as the noise $\mN$ has the spatially varying characteristic. Furthermore, the low SNR of low-light image makes it challenging to distinguish noise from the image edges with weak magnitude. In other words, we need to learn a pixel-wised transform with edge awareness. Motivated by the computational efficiency and edge adaptivity of bilateral filtering,  we propose to learn such a transform in the bilateral space, \ie~the space which treats each image pixel $\vp$ as a point $(\vp,\mI_\vp)$ in $\mathbb{R}^{5}$. The resulting transform can be expressed as a spatially-varying convolution:
\begin{equation}
\label{eq:deform_conv}
\opt{T}_{\mW,\Delta}: \mI_{\vp}\longrightarrow \mN_{\vp}:=\sum_{\vq \in \opt{N}_\vp}\mW_{\vp,{\vq+\Delta_{\vp,\vq}}} \mI_{\vq+\Delta_{\vp,\vq}}^\top,
\end{equation}
where $\opt{N}_\vp$ denotes a $K\times K$ regular neighborhood centered at pixel $\vp$ in image $\mI$. Each entry of \textit{kernel},  $\mW_{\vp,{\vq+\Delta_{\vp,\vq}}}\in\sR^{3 \times 3}$, is defined on the regular grid in the bilateral space, where 
$\Delta_{\vp,\vq}\in\set{R}^{2}$ denotes the associated \textit{offset}.
 
It can be seen that the family of coefficients, $\mGamma$, for defining the transform of estimating the noise layer is composed of
\begin{equation}\label{eqn:coeff}
\mGamma:= \ \ \{\mW_{\vp,{\vq+\Delta_{\vp,\vq}}}, \quad \Delta_{\vp,\vq}\}.
\end{equation}
Similarly, we propose to train an NN that takes the image $\mI$ as the input and outputs the prediction of the coefficients
above to obtain the transform~\eqref{eq:deform_conv}, which will then be applied to predicting the noise $\mN$.

\subsection{Detailed discussion on the transform}

\begin{figure}[htbp!]
    \begin{center}
        \includegraphics[width=1.0\linewidth]{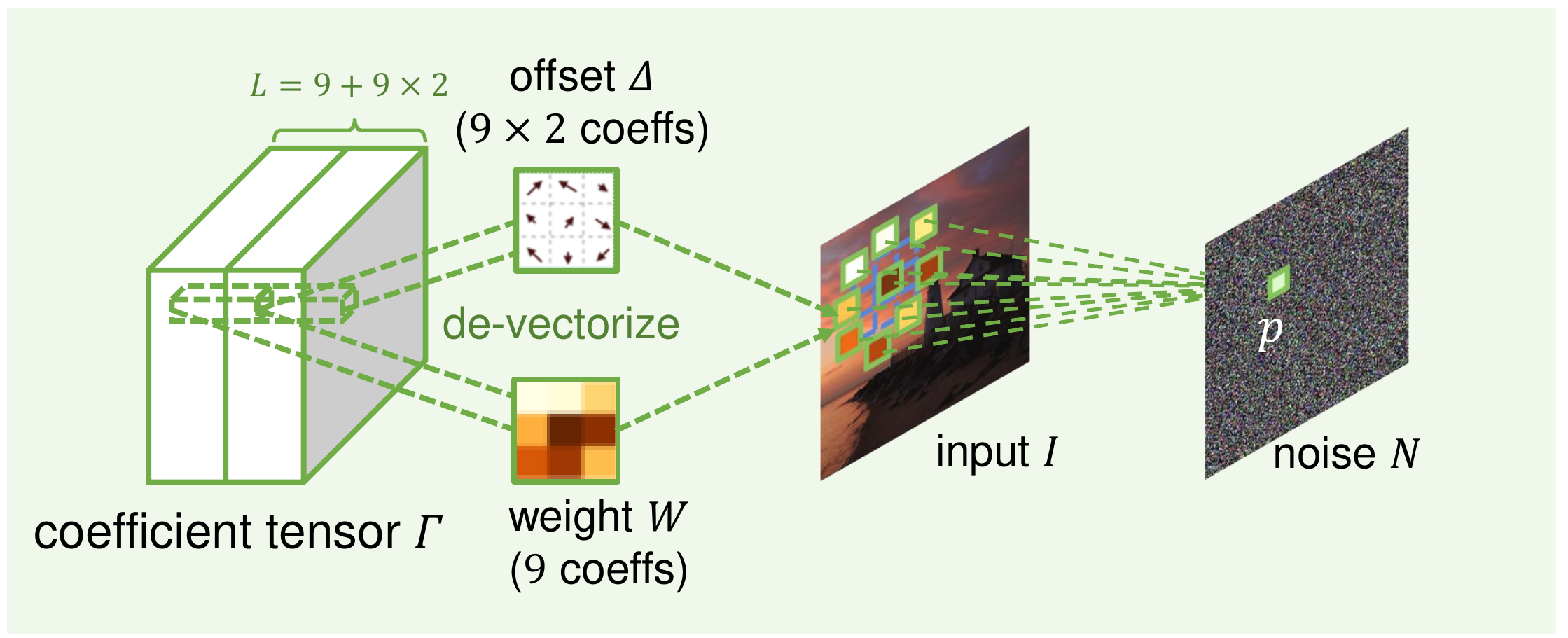}
        \caption{\small 
        Illustration of spatially-varying deformable convolution $\opt{T}_{\mW,\Delta}$ to estimate the noise $\mN$  from a low-light image $\mI$ .
        }
        \label{fig:abstract}
    \end{center}
\end{figure}

We give a more detailed discussion on the  spatially-varying convolution defined in \eqref{eq:deform_conv}:
\begin{equation*}
\opt{T}_{\mW,\Delta}: \mI_{\vp}\longrightarrow \mN_{\vp}:=\sum_{\vq \in \opt{N}_\vp}\mW_{\vp,{\vq+\Delta_{\vp,\vq}}} \mI_{\vq+\Delta_{\vp,\vq}}^\top,
\end{equation*}
which is used for predicting the noise $\mN$. 
See Fig.~\ref{fig:abstract} for the illustration of the transform.
The offsets $\{\Delta_{\vp,\vq}\}_{\vp,\vq}$ used in the proposed method are chosen from a larger neighborhood $W\times W$ where scalar $W  (W\ge K)$ denotes the window size, which are firstly scaled to $[0,1]$ by a sigmoid function and then linearly scaled to $[-W, W]$.
Notice that these offsets do not form a regular grid, and we use the bi-linear interpolation to generate $\mI_{\vq+\Delta_{\vp,\vq}}$.

For spatially-varying kernels $\{\mW_{\vp,{\vq+\Delta_{\vp,\vq}}}\}_{\vp,\vq}$, we need to use them to estimate the noise of  a low-light image. As the energy of noise is typically concentrated on high-frequency channels, we impose that $\{\mW_{\vp,{\vq+\Delta_{\vp,\vq}}}\}_{\vp,\vq}$ should be  high-pass filters, which is done by normalizing the kernels to be zero mean\footnote{When these kernels are used to estimate the noise-free image rather than noise in ablation study in Sec.~\ref{sec:experiments}, we add a softmax layer to ensure that the kernels with positive values and sums to $1$.}.

\subsection{Transform prediction in bilateral space}
\label{sec:network}
\begin{figure}[htbp!]
    \begin{center}
        \includegraphics[width=1.0\linewidth]{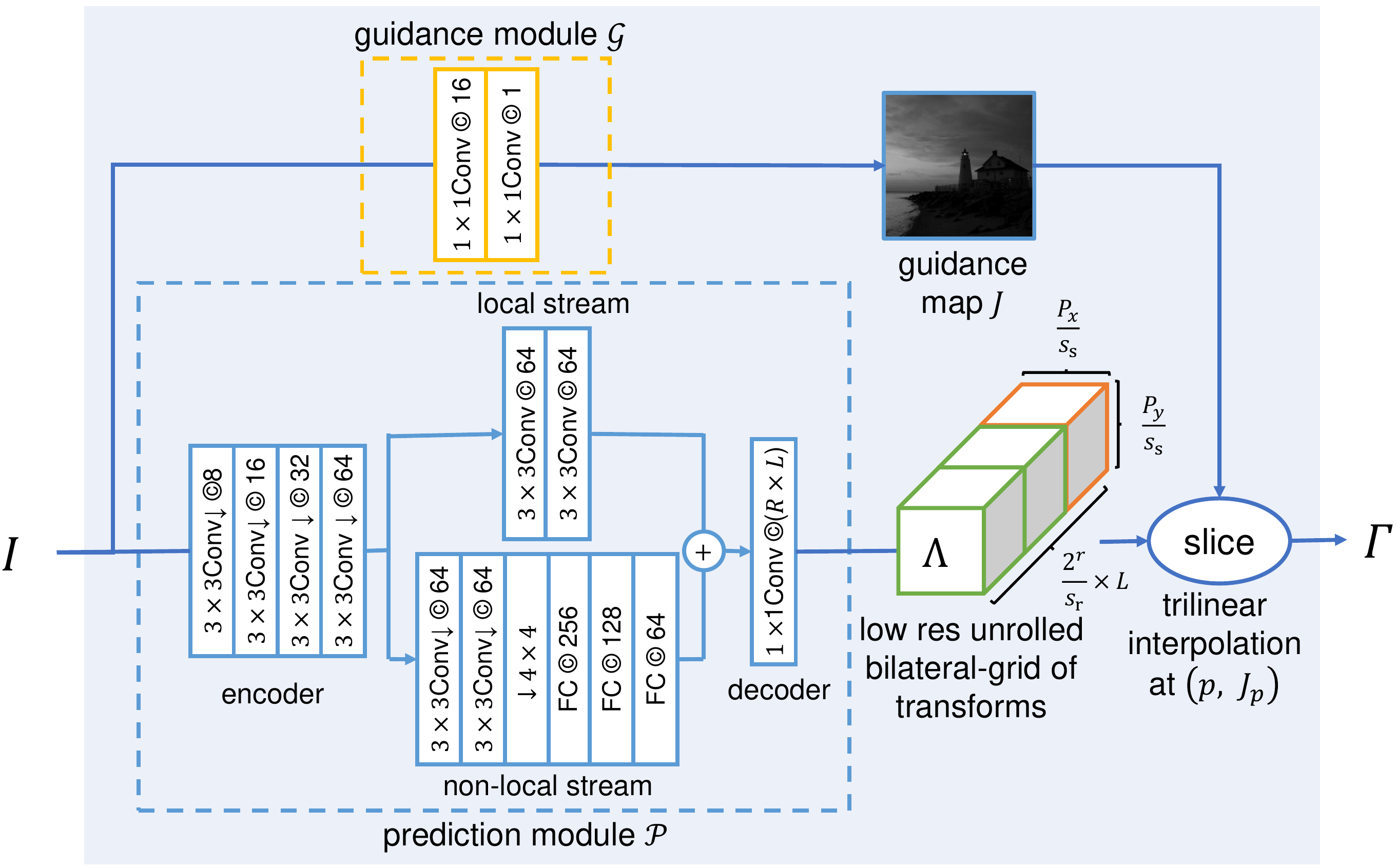}
        \caption{\small 
        Architecture of the NN $\opt{F}_\mTheta$ to predict transform coefficients $\mGamma$ from a low-light image $\mI$ in bilateral space.
        }
        \label{fig:network}
    \end{center}
\end{figure}

The prediction of per-pixel kernels is not a new idea. It has been exploited in denoising~\cite{mildenhall2018burst, bako2017kernelpredicting,xu2019learning}, video interpolation~\cite{niklaus2017videoa,niklaus2017video} and joint image filtering~\cite{kim2019deformable}. All of them are learned in the image space, which is not suitable for separating noise and weak image gradients of a low-light image. 
In this section, we give a detailed discussion on the NN for predicting the transform coefficients \eqref{eqn:coeff} of the spatial varying convolution in the bilateral space. See
Fig.~\ref{fig:network} for the outline of the NN, where there are three main modules: guidance module $\opt{G}$, prediction module $\opt{P}$, and slicing module.

The guidance module $\opt{G}$ produces a single-channel image $\mJ \in \set{R}^{P_x\times P_y}$ whose edges are likely to be kept in the resulting image,
\begin{equation}
\label{eq:guidance}
    \opt{G}:\mI \longrightarrow \mJ,
\end{equation}
where $\mJ_\vp \in \{1,\cdots,2^r\}$ (typically $r=8$) denotes the gray scale value at pixel $\vp=(\vp_x,\vp_y)$.

Recall that the bilateral space refers to the spatial-range product space with an augmented dimension on  pixel color compared to the pixel space. 
In our method, the transform coefficients are predicted and thus efficiently embedded in the reduced bilateral space such that the produced per-pixel spatial varying convolutions have the edge-aware properties that is critical for our task.
Specifically, the prediction module $\opt{P}$ predicts a low-resolution bilateral grid of transform coefficients $\mLambda$:
\begin{equation}
\label{eq:prediction}
\opt{P}:\mI \longrightarrow \mLambda,
\end{equation}
where $\mLambda$ indexed by $(i,j,k) \in \set{R}^{[P_x/s_{\rm s}] \times[ P_y/s_{\rm s}] \times [2^r/s_{\rm r}]}$ is downsampled by $\opt{P}$ with sampling rates $s_{\rm s}, s_{\rm r}$ for \underline{s}patial and \underline{r}ange domain respectively.
The  module $\opt{P}$ has a two-stream structure. The one with fully-connected layers encodes non-local information, and the other with only convolutional layers captures local information. The features from  the local and non-local streams are fused eventually to generate $\mLambda$. 

The parameter 
$\mLambda$ is then rolled into a bilateral grid and sliced into the coefficient tensor $\mGamma$ of full resolution $P_x \times P_y$:
\begin{equation}
\label{eq:slice}
\mGamma_\vp := \sum_{i,j,k}{\delta(s_{\rm s} \vp_x-i)\delta(s_{\rm s} \vp_y-j)\delta(s_{\rm r} \mJ_\vp-k)\mLambda_{i,j,k}},
\end{equation}
 which is a 3D tensor of the same spatial resolution as $\mI$ with $L$ channels in the third dimension, where $\delta(\cdot)=\max{(1-\lvert \cdot \rvert,0)}$ is a linear interpolation kernel.
Thanks to the slicing operation, the resulting coefficient $\mGamma_\vp$ to define the transforms that will then map the input to output is smooth in the bilateral space and keep the discontinuities of $\mJ$. 
Such a design regularizes the output towards edge-aware solutions even though edge preservation is not explicitly handled.

\subsection{Cost function with regularizations}
The NN $\opt{F}_\mTheta$ is trained for predicting  pixel-wise transforms, which will be used to estimate the illumination layer $\mE$ and the noise $\mN$ from the input $\mI$. Then, the reflectance layer $\widetilde{\mR}$ will be estimated by \eqref{eq:final}. See  Fig.~\ref{fig:overall} for the pipeline of the method.

Consider a dataset of $N$ image pairs $\{(\mI_i, \mR_i)\}_{i=1}^N$, where $\mR_i$ denotes the ground truth of the reflectance layer of an input image $\mI_i$.
Several regularizations are imposed on the loss function in order to separate the two layers $\mE$ and $\mR$, in the presence of significant noise.
The loss function $\opt{L}$ is defined as the summation of three terms:
\begin{equation}
\opt{L}:={{\opt{L}_r}(\mR, \widetilde{\mR})  + \lambda_n {\opt{L}_n}(\mN) + \lambda_e {\opt{L}_e}(\mE,\mI)},
\end{equation}
where $\lambda_n$ and $\lambda_e$ are two regularization parameters. The term ${\opt{L}_{r}}(\mR, \widetilde{\mR})$ measures the fidelity on the estimated reflectance layer, the term ${\opt{L}_n}(\mN)$ denotes the regularization on the estimate of noise, and the term ${\opt{L}_e}(\mE,\mI)$  denotes the regularization on the estimate of illumination layer.

The fidelity term on the reflectance layer is defined by
\begin{equation}
{\opt{L}}_r = \sum_{\vp}({\lVert\widetilde{\mR}_\vp-\mR_\vp \rVert}_1+\lambda_g{\lVert\nabla\widetilde{\mR}_\vp-\nabla\mR_\vp \rVert}_1),
\end{equation}
where $\nabla$ denotes the first order difference operator, and $\lambda_g$ is a weighting parameter.
The fidelity is measured  in both intensity and gradient domains using the $\ell_1$-norm metric. Such a loss function is helpful to enhance the  robustness to noise and keep sharp edges in the estimate of the reflectance layer.

Motivated by relative total variation for separating  cartoon structure and textures in~\cite{xu2012structure}, we propose the following regularization on the estimate of noise:
\begin{equation}
\opt{L}_n = \sum_{\vp}\lVert \sum_{\vq\in\opt{N}_\vp}
G_\sigma(\Vert \vp-\vq\rVert_2)\nabla\mN_\vq \rVert_1
\end{equation}
where $G_\sigma$ denoted the 2D Gaussian kernel $G_\sigma (x)=\frac{1}{2\pi \sigma^2}\exp{({-\frac{x^2}{2\sigma^2}})}$
($\sigma=1$ is used in the implementation). It can be seen that such a regularization alleviates possible attenuation of image edges so as to keep sharp edges in the reflectance layer.

The third term $L_e$ is about the regularization on the illumination layer $\mE$. In this paper, we consider a piece-wise smoothness prior for the illumination layer, which is formulated as a re-weighted $\ell_1$-norm on the gradients of $\mE$:
\begin{equation}
\opt{L}_e = \sum_{\vp}   \frac{\lVert \nabla \mE_\vp\rVert_1}{\lVert \nabla \mI_\vp\rVert_1^\theta+\epsilon},
\end{equation}
where the weights are inversely proportional to the magnitude of 
image gradients. In other words, the larger the magnitude of low-light image gradient is, the more likely it indicates the discontinuity of the illumination layer. The exponential parameter $\theta=1.2$ is to control the likeliness and   $\epsilon$ is a small constant for avoiding the  division by zero. 
In addition, we impose physical constraints  on $\mE$: $\mI \le \mE \le \mathbf{1}$.
In training,
all input-target images are normalized from original n-bit RGB color channels to $[0, 1]$.
We set $\mI$ as the lower bound of  $\mE$ to ensure that the obtained $\widetilde{\mR}$ is bounded by $\mathbf{1}$, whereas setting $\mathbf{1}$ as the upper bound of $\mE$ to avoid mistakenly darkening the low-light images.


\section{Experiments}
\label{sec:experiments}

\subsection{Datasets}
The proposed method is trained on the LOL dataset~\cite{wei2018deep}, which includes 1500 low/normal-light image pairs.
Concretely, there are $500$ image pairs of size $400\times 600$ captured  in \textit{real} scenes and 1000 image pairs  of size $384\times 384$ synthesized from raw data.
We use 1000 synthesis pairs and 485 real pairs for training and the remaining 15 real pairs for test as suggested in~\cite{wei2018deep}.
Since the images in the test set of LOL are taken in extreme low-light conditions (as shown in the topleft of Fig.~\ref{fig:LOL_493}), the dark regions of the images are full of intensive noise.
The results on this dataset reveal the performance in challenging low-light conditions.

In addition to the LOL dataset, we also evaluate the proposed method on other four widely-adopted benchmarks for low-light image enhancement that contain underexposed or low-light images without corresponding normal-light reference:
(i) DICM contains 69 captured images from commercial digital cameras collected by~\cite{lee2013contrast}.
(ii) MEF contains 17 high-quality image sequences including natural scenarios, indoor and outdoor views, and man-made architectures provided by~\cite{ma2015perceptual}.
Each image sequence has several multi-exposure images, and we select one of poor-exposed images as input to perform evaluation.
(iii) LIME contains 10 low-light images used in~\cite{guo2017lime}.
(iv) NPE contains 8 outdoor natural scene images which are used in~\cite{wang2013naturalness}.

\subsection{Metric for evaluation}
For all datasets, four quality metrics are adopted for evaluation:
(i) Lightness Order Error (LOE)~\cite{wang2013naturalness} is designed for objectively measuring the lightness distortion.
The computation requires only the low-light images as references.
However, as pointed out in~\cite{guo2017lime}, using the low-light input as reference might be problematic. Therefore, for dataset which provides normal-light reference, we additionally measure LOE$_{\rm ref}$ which uses the normal-light image as the reference.
(ii) Blind Image Spatial Quality Evaluator (BRISQUE)~\cite{mittal2012noreference} correlates subjective quality scores and can measure the quality of images with common distortion such as compression artifacts, blurring, and noise.
(iii) Natural Image Quality Evaluator (NIQE)~\cite{mittal2013making} does not relate to subjective quality scores and can measure the quality of images with arbitrary distortion.
(iv) Perception based Image Quality Evaluator (PIQE)~\cite{venkatanathn2015blind} measures the block-wise quality of images with arbitrary distortion.
Lower values of the four metrics reflect better perceptual quality.

For the LOL dataset which provides reference normal-light images, two extra full-reference metrics are used, \ie~Peak Signal-to-Noise Ratio (PSNR) and Structural Similarity (SSIM) Index~\cite{wang2004image}, with higher value for better quality.

\subsection{Implementation details}
Our approach is implemented using PyTorch~\cite{paszke2019pytorch} and trained on an Nvidia Titan RTX GPU and Intel i7-7700K 4.20GHz CPU.
We use Adam optimizer~\cite{kingma2015adam} with a fixed learning rate of $10^{-4}$ and $\ell_2$ weight decay of $10^{-8}$.
Other hyper parameters are set as default (\ie, $\beta_1=0.9$, $\beta_2=0.999$ and $\epsilon=10^{-8}$).
Totally 2000 epochs are used for training.
During training, each image is normalized to the range $[0, 1]$.
The batch size is set to 16.
For data augmentation, we randomly cropped $256\times 256$ patches followed by random mirroring, resizing and rotation for all patches.
The weights for the convolutional and fully-connected layers are initialized according to~\cite{he2015delving} and the biases are initialized to 0.
In our experiments, the parameter setting for deformable convolutions are: kernel size $K=3$ and window size $W=15$.
The sampling rates in the prediction module are $s_{\rm s}=16$ and $s_{\rm r}=32$ for \underline{s}patial and \underline{r}ange domain respectively.
As for the loss function, we set $\lambda_g=0.1$, $\lambda_n=1$, $\lambda_e=1$, $\epsilon=10^{-4}$.

\begin{table*}[htbp!]
    \centering
    \caption{\small Quantitative results on the LOL~\cite{wei2018deep} test set.}
    \sisetup{detect-all=true,detect-weight=true,detect-inline-weight=math}
    \begin{tabular}{cS[table-format=2.4]S[table-format=1.4]S[table-format=4.2]S[table-format=4.2]S[table-format=1.4]S[table-format=2.4]S[table-format=2.4]}
        \toprule
        {Method} & \multicolumn{1}{c}{{PSNR(dB)$\uparrow$}} & \multicolumn{1}{c}{{SSIM}$\uparrow$} & \multicolumn{1}{c}{{LOE$_{\rm ref}\downarrow$}} & \multicolumn{1}{c}{{LOE$\downarrow$}} & \multicolumn{1}{c}{{NIQE$\downarrow$}} & \multicolumn{1}{c}{{BRISQUE$\downarrow$}} & \multicolumn{1}{c}{{PIQE$\downarrow$}} \\
        \midrule
        \midrule
        HE    & 14.8006  & 0.4112  & 485.67  & 250.61  & 8.4763  & 37.8472  & 51.7788  \\
        MSR~\cite{jobson1997multiscale}   & 13.1728  & 0.4787  & 1428.35  & 1414.38  & 8.1136  & 32.6192  & 45.9488  \\
        Dong~\cite{dong2010fast}  & 16.7165  & 0.5824  & 409.18  & 89.32  & 8.3157  & 34.4683  & 46.6898  \\
        NPE~\cite{wang2013naturalness}   & 16.9697  & 0.5894  & 681.42  & 479.72  & 8.4390  & 35.1236  & 47.6651  \\
        SRIE~\cite{fu2016weighted}  & 11.8552  & 0.4979  & \BF 370.44 & \BF 83.81 & \BF 7.2869 & \BF 27.6113 & \BF 27.7037 \\
        MF~\cite{fu2016fusionbased}    & 16.9662  & 0.6049  & 429.20  & 211.19  & 8.8770  & 34.7835  & 47.1974  \\
        BIMEF~\cite{ying2017bioinspired} & 13.8753  & 0.5771  & 387.83  & 141.16  & 7.5150  & 27.6511  & 28.1861  \\
        LIME~\cite{guo2017lime}  & 16.7586  & 0.5644  & 800.34  & 695.50  & 8.3777  & 36.0971  & 49.9709  \\
        NPIE-MLLS~\cite{wang2018naturalness} & 16.6972  & 0.5945  & 548.25  & 317.40  & 8.1588  & 33.8586  & 44.0897  \\
        RetinexNet~\cite{wei2018deep} & 16.7740  & 0.5594  & 1059.27  & 993.29  & 8.8792  & 39.5860  & 57.6731  \\
        DeepUPE~\cite{wang2019underexposed}  & \BF 20.3736  & \BF 0.6379  & 444.51  & 284.92  & 7.8642  & 32.8006  & 40.6565  \\
        \midrule
        \midrule
        JED~\cite{ren2018joint}    & 13.6857  & 0.6280  & 398.14  & 432.89  & 5.3767  & 29.0680  & 40.7568  \\
        RRM~\cite{li2018structurerevealing}    & 13.8765  & 0.6577  & 453.30  & 460.87  & 5.8100  & 34.9902  & 47.3000  \\
        KinD~\cite{zhang2019kindling} & 17.6476  & 0.7601  & 549.03  & 493.00  & 4.7100  & 26.6443  & 45.2404  \\
        KinD-nonblind~\cite{zhang2019kindling} & 20.3792  & \BF 0.8045  & 449.01  & 406.45  & 5.3546  & 32.4347  & 65.6035  \\
        Ours  & \BF 22.5156  & 0.7864  & \BF 289.05  & \BF 277.35  & \BF 3.6354  & \BF 21.7781  & \BF 21.0840  \\
        \bottomrule
    \end{tabular}%
    \label{tab:LOL}%
\end{table*}%

\begin{figure*}[htbp!]\centering
    \begin{subfigure}[b]{0.24\textwidth}
        \begin{tikzpicture}[zoomboxarray, zoomboxes below, zoomboxarray inner gap=0.25cm, zoomboxarray columns=2, zoomboxarray rows=1]
        \node [image node] { \includegraphics[width=\textwidth]{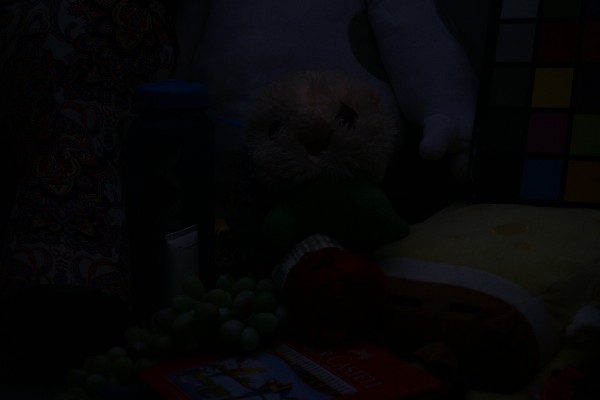} };
        \zoombox[magnification=4.63, color code=red]{0.35,0.93}
        \zoombox[magnification=2.67, color code=blue]{0.56,0.24}
        \end{tikzpicture}
        \caption{\small Input}
    \end{subfigure}
    \begin{subfigure}[b]{0.24\textwidth}
        \begin{tikzpicture}[zoomboxarray, zoomboxes below, zoomboxarray inner gap=0.25cm, zoomboxarray columns=2, zoomboxarray rows=1]
        \node [image node] { \includegraphics[width=\textwidth]{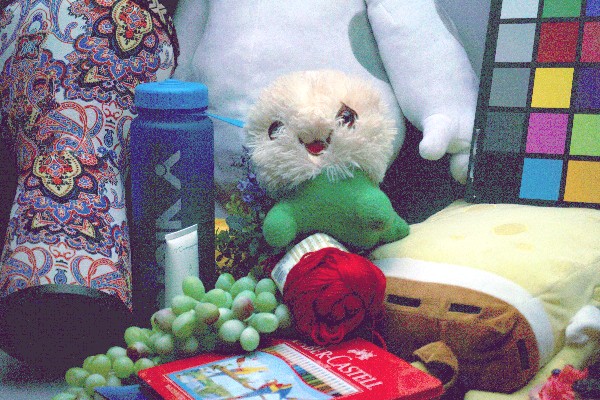} };
        \zoombox[magnification=4.63, color code=red]{0.35,0.93}
        \zoombox[magnification=2.67, color code=blue]{0.56,0.24}
        \end{tikzpicture}
        \caption{\small HE}
    \end{subfigure}
    \begin{subfigure}[b]{0.24\textwidth}
        \begin{tikzpicture}[zoomboxarray, zoomboxes below, zoomboxarray inner gap=0.25cm, zoomboxarray columns=2, zoomboxarray rows=1]
        \node [image node] { \includegraphics[width=\textwidth]{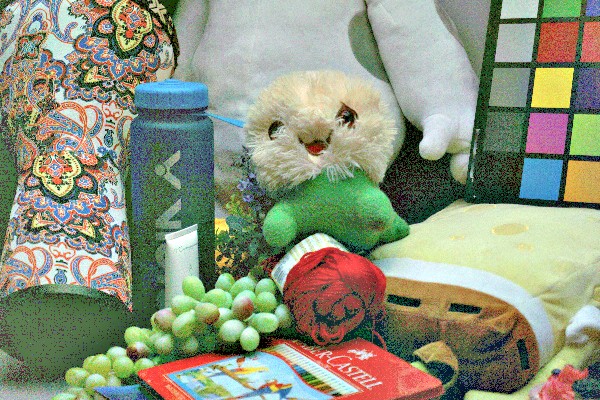} };
        \zoombox[magnification=4.63, color code=red]{0.35,0.93}
        \zoombox[magnification=2.67, color code=blue]{0.56,0.24}
        \end{tikzpicture}
        \caption{\small MSR~\cite{jobson1997multiscale}}
    \end{subfigure}
    \begin{subfigure}[b]{0.24\textwidth}
        \begin{tikzpicture}[zoomboxarray, zoomboxes below, zoomboxarray inner gap=0.25cm, zoomboxarray columns=2, zoomboxarray rows=1]
        \node [image node] { \includegraphics[width=\textwidth]{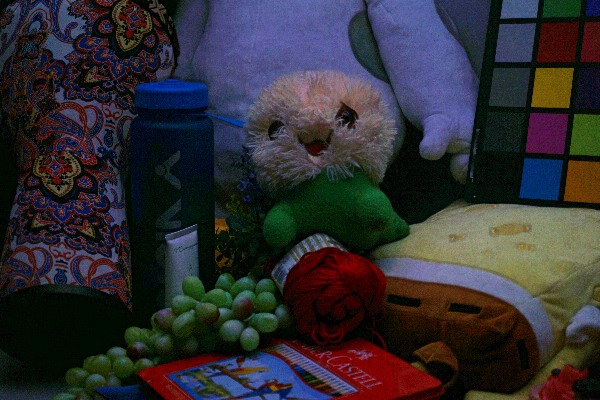} };
        \zoombox[magnification=4.63, color code=red]{0.35,0.93}
        \zoombox[magnification=2.67, color code=blue]{0.56,0.24}
        \end{tikzpicture}
        \caption{\small Dong~\cite{dong2010fast}}
    \end{subfigure}

    \begin{subfigure}[b]{0.24\textwidth}
        \begin{tikzpicture}[zoomboxarray, zoomboxes below, zoomboxarray inner gap=0.25cm, zoomboxarray columns=2, zoomboxarray rows=1]
        \node [image node] { \includegraphics[width=\textwidth]{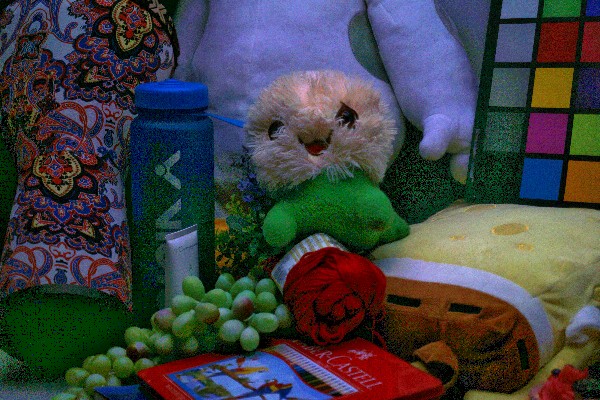} };
        \zoombox[magnification=4.63, color code=red]{0.35,0.93}
        \zoombox[magnification=2.67, color code=blue]{0.56,0.24}
        \end{tikzpicture}
        \caption{\small NPE~\cite{wang2013naturalness}}
    \end{subfigure}
    \begin{subfigure}[b]{0.24\textwidth}
        \begin{tikzpicture}[zoomboxarray, zoomboxes below, zoomboxarray inner gap=0.25cm, zoomboxarray columns=2, zoomboxarray rows=1]
        \node [image node] { \includegraphics[width=\textwidth]{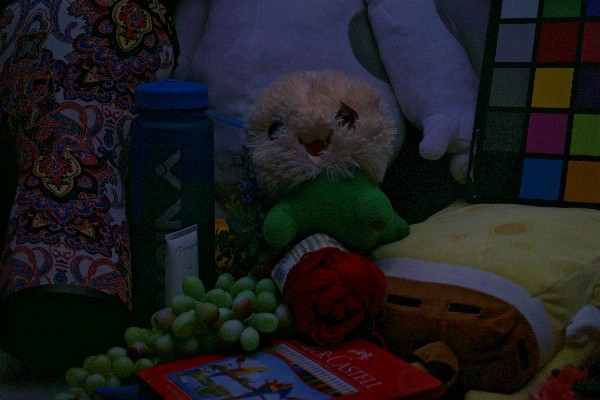} };
        \zoombox[magnification=4.63, color code=red]{0.35,0.93}
        \zoombox[magnification=2.67, color code=blue]{0.56,0.24}
        \end{tikzpicture}
        \caption{\small SRIE~\cite{fu2016weighted}}
    \end{subfigure}
    \begin{subfigure}[b]{0.24\textwidth}
        \begin{tikzpicture}[zoomboxarray, zoomboxes below, zoomboxarray inner gap=0.25cm, zoomboxarray columns=2, zoomboxarray rows=1]
        \node [image node] { \includegraphics[width=\textwidth]{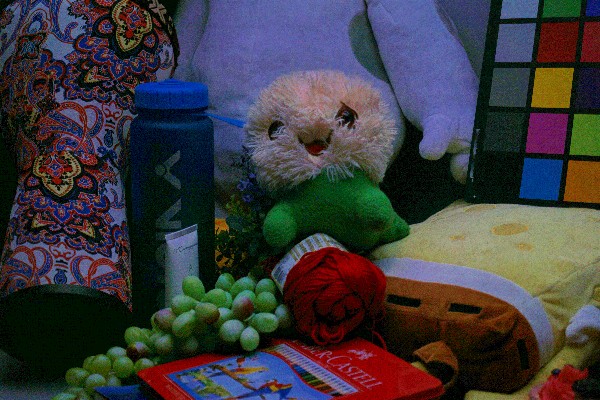} };
        \zoombox[magnification=4.63, color code=red]{0.35,0.93}
        \zoombox[magnification=2.67, color code=blue]{0.56,0.24}
        \end{tikzpicture}
        \caption{\small MF~\cite{fu2016fusionbased}}
    \end{subfigure}
    \begin{subfigure}[b]{0.24\textwidth}
        \begin{tikzpicture}[zoomboxarray, zoomboxes below, zoomboxarray inner gap=0.25cm, zoomboxarray columns=2, zoomboxarray rows=1]
        \node [image node] { \includegraphics[width=\textwidth]{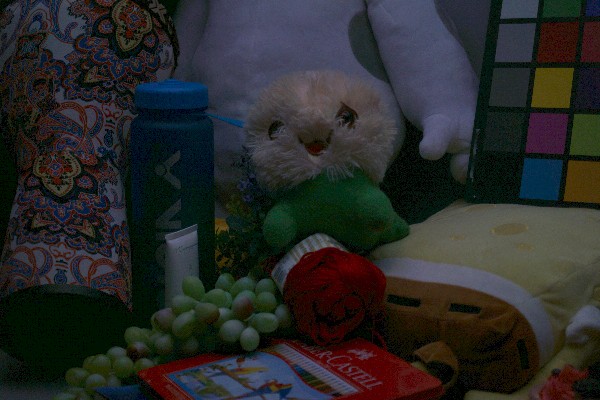} };
        \zoombox[magnification=4.63, color code=red]{0.35,0.93}
        \zoombox[magnification=2.67, color code=blue]{0.56,0.24}
        \end{tikzpicture}
        \caption{\small BIMEF~\cite{ying2017bioinspired}}
    \end{subfigure}
    
    \begin{subfigure}[b]{0.24\textwidth}
        \begin{tikzpicture}[zoomboxarray, zoomboxes below, zoomboxarray inner gap=0.25cm, zoomboxarray columns=2, zoomboxarray rows=1]
        \node [image node] { \includegraphics[width=\textwidth]{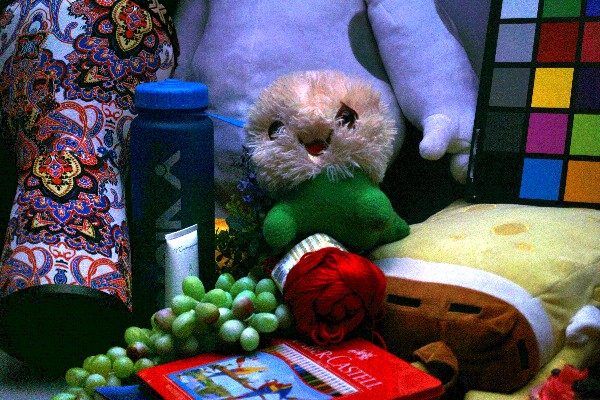} };
        \zoombox[magnification=4.63, color code=red]{0.35,0.93}
        \zoombox[magnification=2.67, color code=blue]{0.56,0.24}
        \end{tikzpicture}
        \caption{\small LIME~\cite{guo2017lime}}
    \end{subfigure}
    \begin{subfigure}[b]{0.24\textwidth}
        \begin{tikzpicture}[zoomboxarray, zoomboxes below, zoomboxarray inner gap=0.25cm, zoomboxarray columns=2, zoomboxarray rows=1]
        \node [image node] { \includegraphics[width=\textwidth]{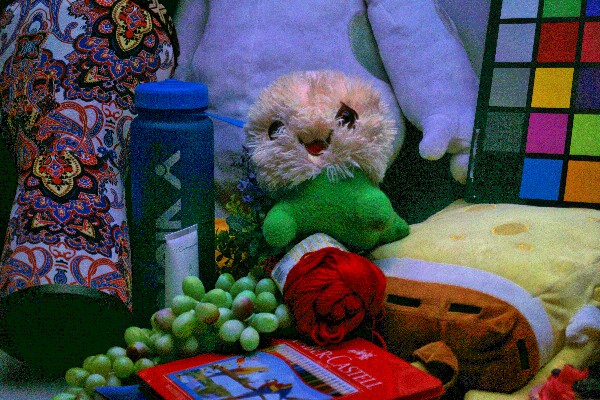} };
        \zoombox[magnification=4.63, color code=red]{0.35,0.93}
        \zoombox[magnification=2.67, color code=blue]{0.56,0.24}
        \end{tikzpicture}
        \caption{\small NPIE-MLLS~\cite{wang2018naturalness}}
    \end{subfigure}
    \begin{subfigure}[b]{0.24\textwidth}
        \begin{tikzpicture}[zoomboxarray, zoomboxes below, zoomboxarray inner gap=0.25cm, zoomboxarray columns=2, zoomboxarray rows=1]
        \node [image node] { \includegraphics[width=\textwidth]{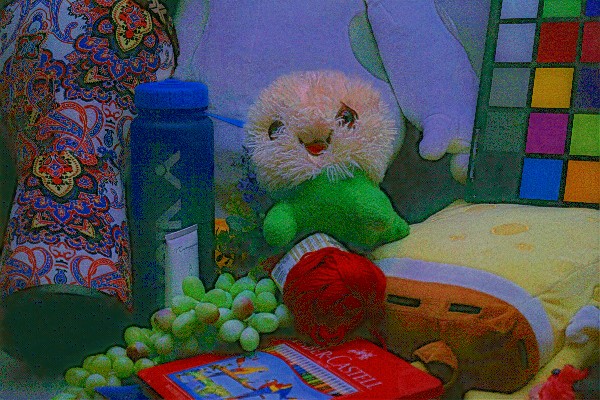} };
        \zoombox[magnification=4.63, color code=red]{0.35,0.93}
        \zoombox[magnification=2.67, color code=blue]{0.56,0.24}
        \end{tikzpicture}
        \caption{\small RetinexNet~\cite{wei2018deep}}
    \end{subfigure}
    \begin{subfigure}[b]{0.24\textwidth}
        \begin{tikzpicture}[zoomboxarray, zoomboxes below, zoomboxarray inner gap=0.25cm, zoomboxarray columns=2, zoomboxarray rows=1]
        \node [image node] { \includegraphics[width=\textwidth]{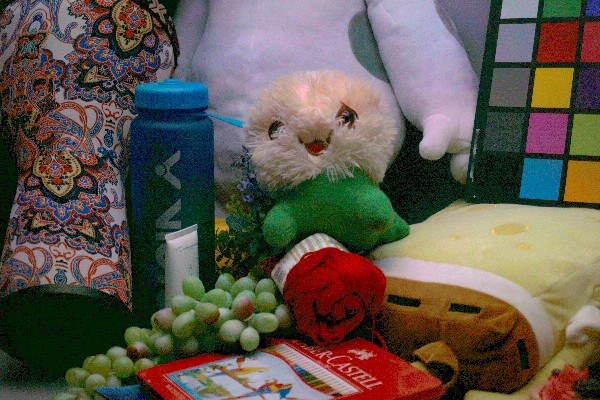} };
        \zoombox[magnification=4.63, color code=red]{0.35,0.93}
        \zoombox[magnification=2.67, color code=blue]{0.56,0.24}
        \end{tikzpicture}
        \caption{\small DeepUPE~\cite{wang2019underexposed}}
    \end{subfigure}
    
    \begin{subfigure}[b]{0.24\textwidth}
        \begin{tikzpicture}[zoomboxarray, zoomboxes below, zoomboxarray inner gap=0.25cm, zoomboxarray columns=2, zoomboxarray rows=1]
        \node [image node] { \includegraphics[width=\textwidth]{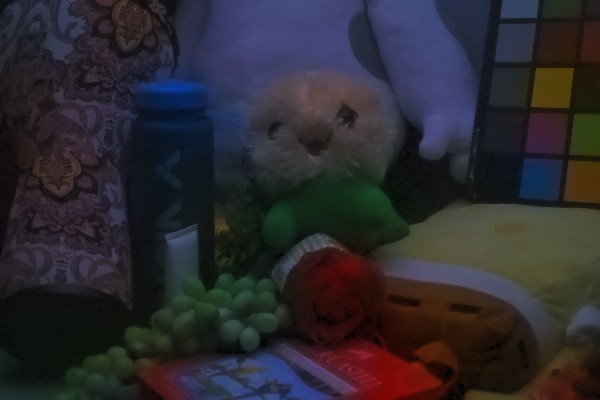} };
        \zoombox[magnification=4.63, color code=red]{0.35,0.93}
        \zoombox[magnification=2.67, color code=blue]{0.56,0.24}
        \end{tikzpicture}
        \caption{\small JED~\cite{ren2018joint}}
    \end{subfigure}
    \begin{subfigure}[b]{0.24\textwidth}
        \begin{tikzpicture}[zoomboxarray, zoomboxes below, zoomboxarray inner gap=0.25cm, zoomboxarray columns=2, zoomboxarray rows=1]
        \node [image node] { \includegraphics[width=\textwidth]{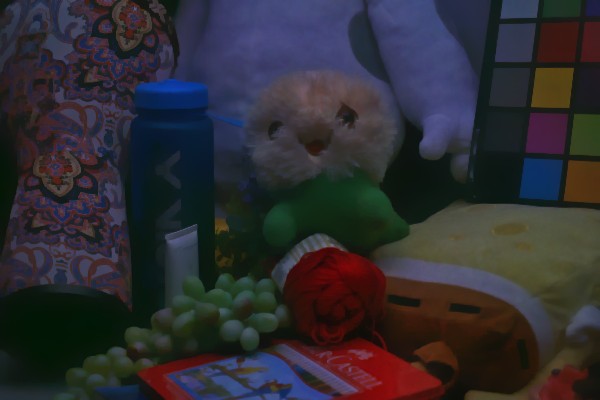} };
        \zoombox[magnification=4.63, color code=red]{0.35,0.93}
        \zoombox[magnification=2.67, color code=blue]{0.56,0.24}
        \end{tikzpicture}
        \caption{\small RRM~\cite{li2018structurerevealing}}
    \end{subfigure}
    \begin{subfigure}[b]{0.24\textwidth}
        \begin{tikzpicture}[zoomboxarray, zoomboxes below, zoomboxarray inner gap=0.25cm, zoomboxarray columns=2, zoomboxarray rows=1]
        \node [image node] { \includegraphics[width=\textwidth]{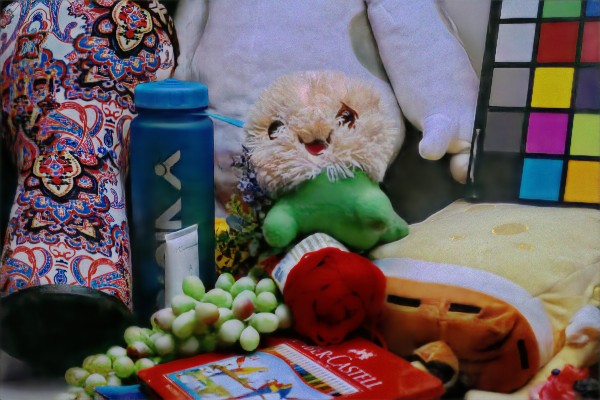} };
        \zoombox[magnification=4.63, color code=red]{0.35,0.93}
        \zoombox[magnification=2.67, color code=blue]{0.56,0.24}
        \end{tikzpicture}
        \caption{\small KinD~\cite{zhang2019kindling}}
    \end{subfigure}
    \begin{subfigure}[b]{0.24\textwidth}
        \begin{tikzpicture}[zoomboxarray, zoomboxes below, zoomboxarray inner gap=0.25cm, zoomboxarray columns=2, zoomboxarray rows=1]
        \node [image node] { \includegraphics[width=\textwidth]{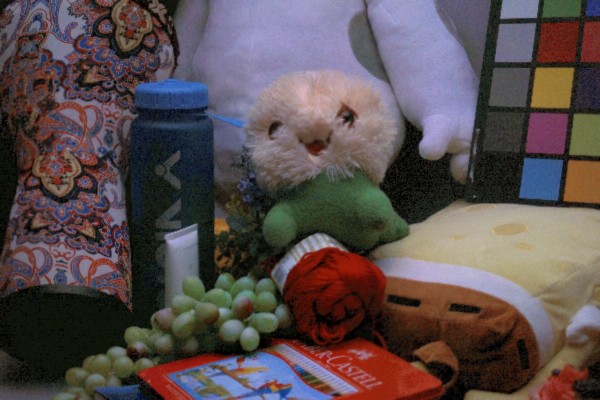} };
        \zoombox[magnification=4.63, color code=red]{0.35,0.93}
        \zoombox[magnification=2.67, color code=blue]{0.56,0.24}
        \end{tikzpicture}
        \caption{\small Ours}
    \end{subfigure}
    \caption{\small Visual comparisons of different methods on low-light images from the LOL test set.}
    \label{fig:LOL_493}
\end{figure*}

\begin{figure*}[htbp!]\centering
\begin{subfigure}[b]{0.24\textwidth}
    \begin{tikzpicture}[zoomboxarray, zoomboxes below, zoomboxarray inner gap=0.25cm, zoomboxarray columns=2, zoomboxarray rows=1]
    \node [image node] { \includegraphics[width=\textwidth]{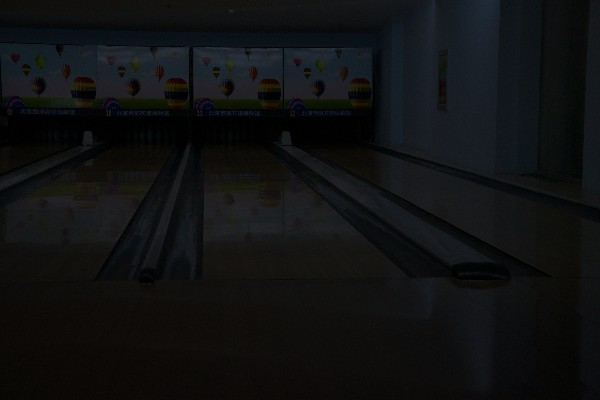} };
    \zoombox[magnification=6.82, color code=red]{0.79,0.71}
    \zoombox[magnification=5.56, color code=blue]{0.44,0.77}
    \end{tikzpicture}
    \caption{Input}
\end{subfigure}
\begin{subfigure}[b]{0.24\textwidth}
    \begin{tikzpicture}[zoomboxarray, zoomboxes below, zoomboxarray inner gap=0.25cm, zoomboxarray columns=2, zoomboxarray rows=1]
    \node [image node] { \includegraphics[width=\textwidth]{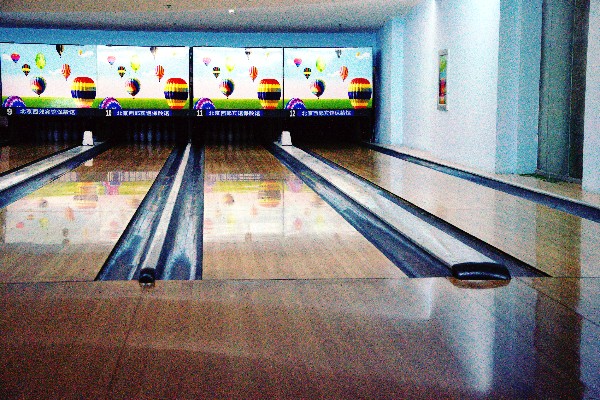} };
    \zoombox[magnification=6.82, color code=red]{0.79,0.71}
    \zoombox[magnification=5.56, color code=blue]{0.44,0.77}
    \end{tikzpicture}
    \caption{HE}
\end{subfigure}
\begin{subfigure}[b]{0.24\textwidth}
    \begin{tikzpicture}[zoomboxarray, zoomboxes below, zoomboxarray inner gap=0.25cm, zoomboxarray columns=2, zoomboxarray rows=1]
    \node [image node] { \includegraphics[width=\textwidth]{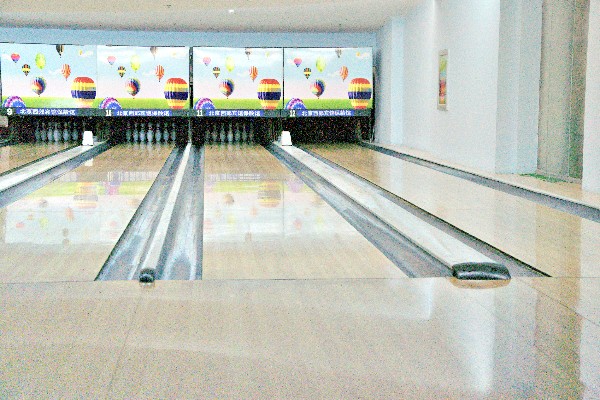} };
    \zoombox[magnification=6.82, color code=red]{0.79,0.71}
    \zoombox[magnification=5.56, color code=blue]{0.44,0.77}
    \end{tikzpicture}
    \caption{MSR~\cite{jobson1997multiscale}}
\end{subfigure}
\begin{subfigure}[b]{0.24\textwidth}
    \begin{tikzpicture}[zoomboxarray, zoomboxes below, zoomboxarray inner gap=0.25cm, zoomboxarray columns=2, zoomboxarray rows=1]
    \node [image node] { \includegraphics[width=\textwidth]{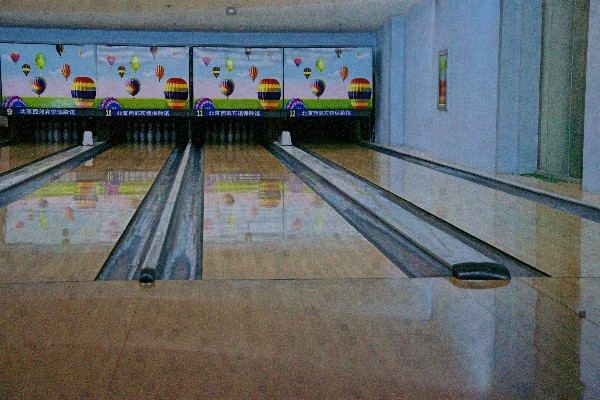} };
    \zoombox[magnification=6.82, color code=red]{0.79,0.71}
    \zoombox[magnification=5.56, color code=blue]{0.44,0.77}
    \end{tikzpicture}
    \caption{Dong~\cite{dong2010fast}}
\end{subfigure}

\begin{subfigure}[b]{0.24\textwidth}
    \begin{tikzpicture}[zoomboxarray, zoomboxes below, zoomboxarray inner gap=0.25cm, zoomboxarray columns=2, zoomboxarray rows=1]
    \node [image node] { \includegraphics[width=\textwidth]{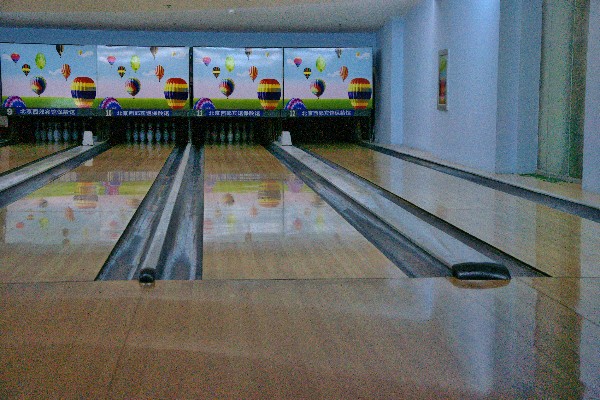} };
    \zoombox[magnification=6.82, color code=red]{0.79,0.71}
    \zoombox[magnification=5.56, color code=blue]{0.44,0.77}
    \end{tikzpicture}
    \caption{NPE~\cite{wang2013naturalness}}
\end{subfigure}
\begin{subfigure}[b]{0.24\textwidth}
    \begin{tikzpicture}[zoomboxarray, zoomboxes below, zoomboxarray inner gap=0.25cm, zoomboxarray columns=2, zoomboxarray rows=1]
    \node [image node] { \includegraphics[width=\textwidth]{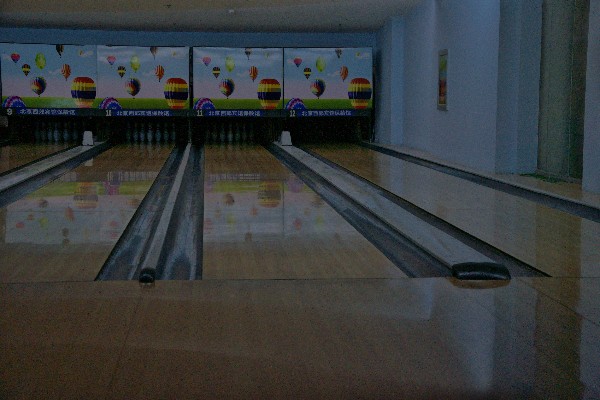} };
    \zoombox[magnification=6.82, color code=red]{0.79,0.71}
    \zoombox[magnification=5.56, color code=blue]{0.44,0.77}
    \end{tikzpicture}
    \caption{SRIE~\cite{fu2016weighted}}
\end{subfigure}
\begin{subfigure}[b]{0.24\textwidth}
    \begin{tikzpicture}[zoomboxarray, zoomboxes below, zoomboxarray inner gap=0.25cm, zoomboxarray columns=2, zoomboxarray rows=1]
    \node [image node] { \includegraphics[width=\textwidth]{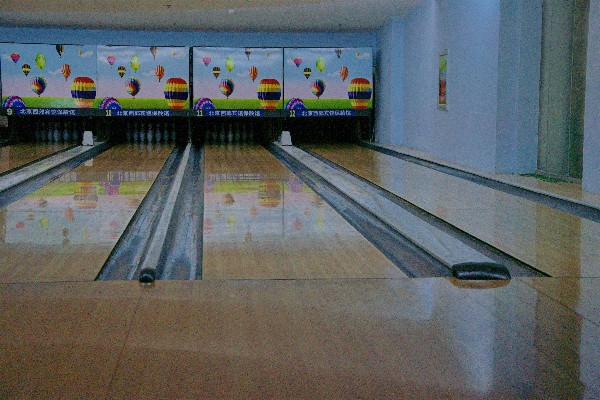} };
    \zoombox[magnification=6.82, color code=red]{0.79,0.71}
    \zoombox[magnification=5.56, color code=blue]{0.44,0.77}
    \end{tikzpicture}
    \caption{MF~\cite{fu2016fusionbased}}
\end{subfigure}
\begin{subfigure}[b]{0.24\textwidth}
    \begin{tikzpicture}[zoomboxarray, zoomboxes below, zoomboxarray inner gap=0.25cm, zoomboxarray columns=2, zoomboxarray rows=1]
    \node [image node] { \includegraphics[width=\textwidth]{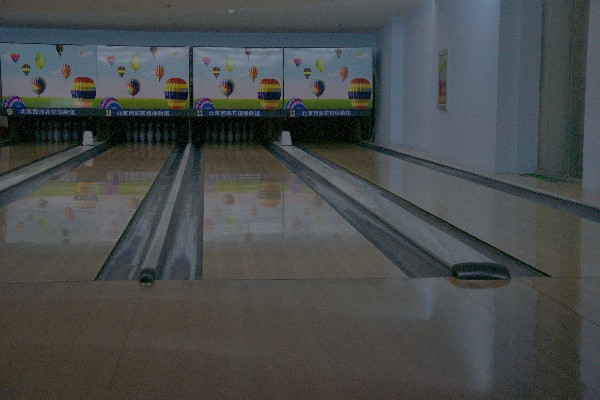} };
    \zoombox[magnification=6.82, color code=red]{0.79,0.71}
    \zoombox[magnification=5.56, color code=blue]{0.44,0.77}
    \end{tikzpicture}
    \caption{BIMEF~\cite{ying2017bioinspired}}
\end{subfigure}

\begin{subfigure}[b]{0.24\textwidth}
    \begin{tikzpicture}[zoomboxarray, zoomboxes below, zoomboxarray inner gap=0.25cm, zoomboxarray columns=2, zoomboxarray rows=1]
    \node [image node] { \includegraphics[width=\textwidth]{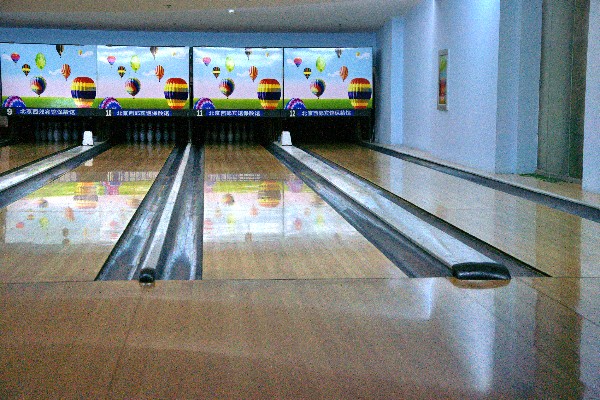} };
    \zoombox[magnification=6.82, color code=red]{0.79,0.71}
    \zoombox[magnification=5.56, color code=blue]{0.44,0.77}
    \end{tikzpicture}
    \caption{LIME~\cite{guo2017lime}}
\end{subfigure}
\begin{subfigure}[b]{0.24\textwidth}
    \begin{tikzpicture}[zoomboxarray, zoomboxes below, zoomboxarray inner gap=0.25cm, zoomboxarray columns=2, zoomboxarray rows=1]
    \node [image node] { \includegraphics[width=\textwidth]{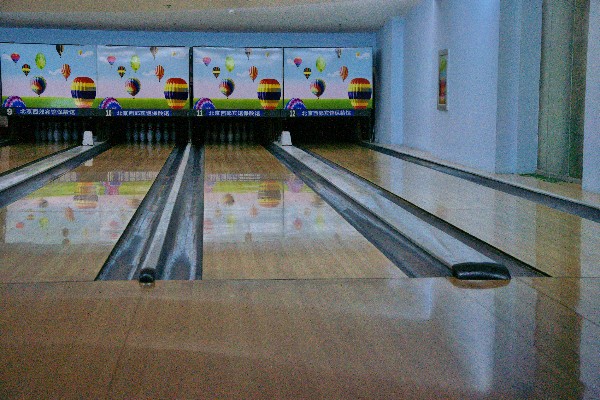} };
    \zoombox[magnification=6.82, color code=red]{0.79,0.71}
    \zoombox[magnification=5.56, color code=blue]{0.44,0.77}
    \end{tikzpicture}
    \caption{NPIE-MLLS~\cite{wang2018naturalness}}
\end{subfigure}
\begin{subfigure}[b]{0.24\textwidth}
    \begin{tikzpicture}[zoomboxarray, zoomboxes below, zoomboxarray inner gap=0.25cm, zoomboxarray columns=2, zoomboxarray rows=1]
    \node [image node] { \includegraphics[width=\textwidth]{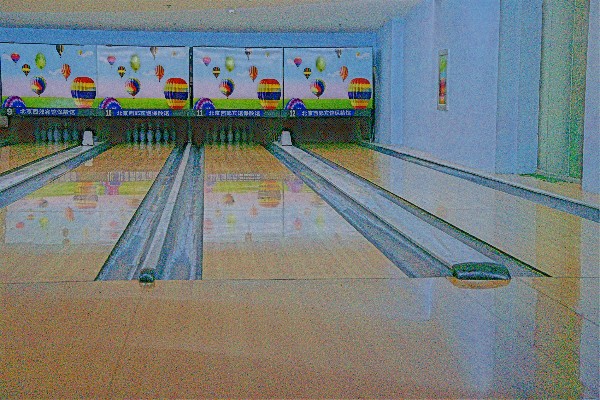} };
    \zoombox[magnification=6.82, color code=red]{0.79,0.71}
    \zoombox[magnification=5.56, color code=blue]{0.44,0.77}
    \end{tikzpicture}
    \caption{RetinexNet~\cite{wei2018deep}}
\end{subfigure}
\begin{subfigure}[b]{0.24\textwidth}
    \begin{tikzpicture}[zoomboxarray, zoomboxes below, zoomboxarray inner gap=0.25cm, zoomboxarray columns=2, zoomboxarray rows=1]
    \node [image node] { \includegraphics[width=\textwidth]{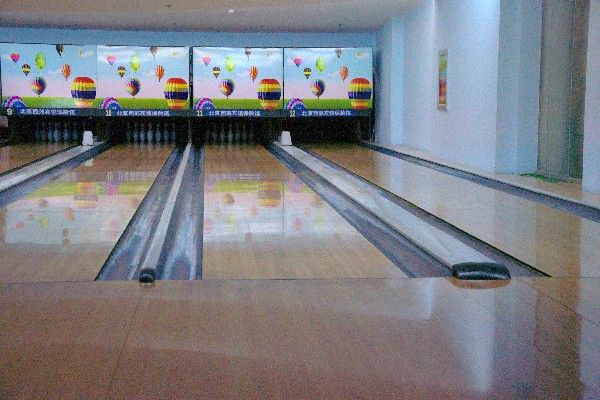} };
    \zoombox[magnification=6.82, color code=red]{0.79,0.71}
    \zoombox[magnification=5.56, color code=blue]{0.44,0.77}
    \end{tikzpicture}
    \caption{DeepUPE~\cite{wang2019underexposed}}
\end{subfigure}

\begin{subfigure}[b]{0.24\textwidth}
    \begin{tikzpicture}[zoomboxarray, zoomboxes below, zoomboxarray inner gap=0.25cm, zoomboxarray columns=2, zoomboxarray rows=1]
    \node [image node] { \includegraphics[width=\textwidth]{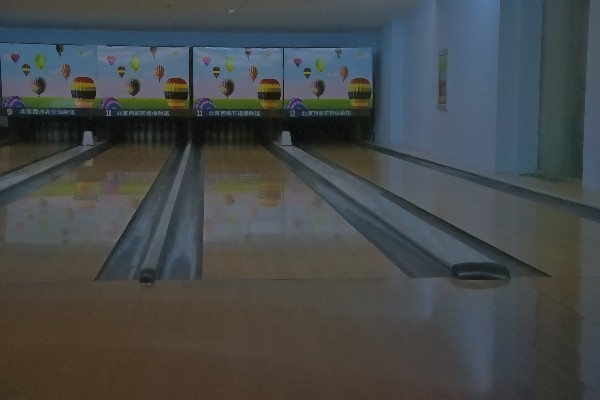} };
    \zoombox[magnification=6.82, color code=red]{0.79,0.71}
    \zoombox[magnification=5.56, color code=blue]{0.44,0.77}
    \end{tikzpicture}
    \caption{JED~\cite{ren2018joint}}
\end{subfigure}
\begin{subfigure}[b]{0.24\textwidth}
    \begin{tikzpicture}[zoomboxarray, zoomboxes below, zoomboxarray inner gap=0.25cm, zoomboxarray columns=2, zoomboxarray rows=1]
    \node [image node] { \includegraphics[width=\textwidth]{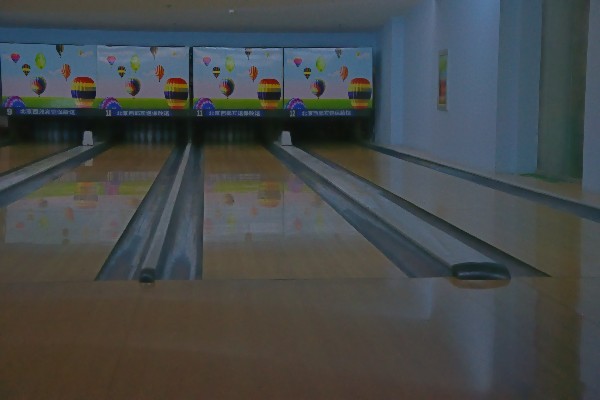} };
    \zoombox[magnification=6.82, color code=red]{0.79,0.71}
    \zoombox[magnification=5.56, color code=blue]{0.44,0.77}
    \end{tikzpicture}
    \caption{RRM~\cite{li2018structurerevealing}}
\end{subfigure}
\begin{subfigure}[b]{0.24\textwidth}
    \begin{tikzpicture}[zoomboxarray, zoomboxes below, zoomboxarray inner gap=0.25cm, zoomboxarray columns=2, zoomboxarray rows=1]
    \node [image node] { \includegraphics[width=\textwidth]{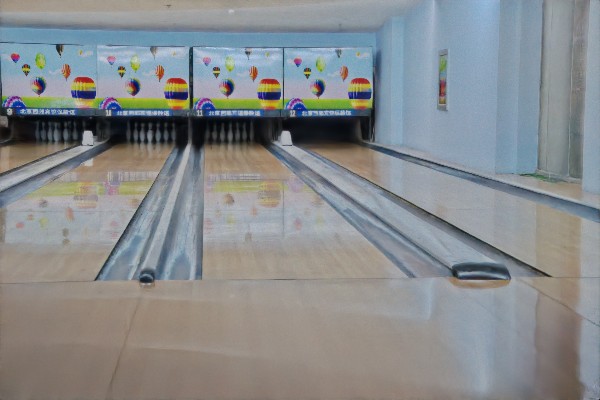} };
    \zoombox[magnification=6.82, color code=red]{0.79,0.71}
    \zoombox[magnification=5.56, color code=blue]{0.44,0.77}
    \end{tikzpicture}
    \caption{KinD~\cite{zhang2019kindling}}
\end{subfigure}
\begin{subfigure}[b]{0.24\textwidth}
    \begin{tikzpicture}[zoomboxarray, zoomboxes below, zoomboxarray inner gap=0.25cm, zoomboxarray columns=2, zoomboxarray rows=1]
    \node [image node] { \includegraphics[width=\textwidth]{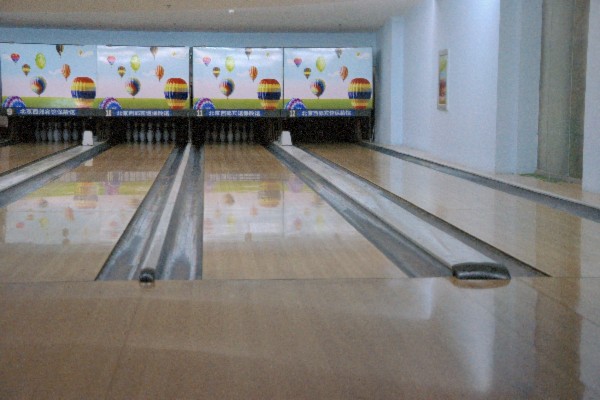} };
    \zoombox[magnification=6.82, color code=red]{0.79,0.71}
    \zoombox[magnification=5.56, color code=blue]{0.44,0.77}
    \end{tikzpicture}
    \caption{Ours}
\end{subfigure}
\caption{Visual comparisons of different methods on low-light images from the LOL test set.}
\label{fig:LOL_669}
\end{figure*}

\begin{figure*}[htbp!]\centering
\setlength{\tabcolsep}{2pt}
\begin{tabular}{ccccc}
        \begin{tikzpicture}[zoomboxarray, zoomboxes below, zoomboxarray inner gap=0.25cm, zoomboxarray columns=2, zoomboxarray rows=1]
        \node [image node] { \includegraphics[width=0.19\textwidth]{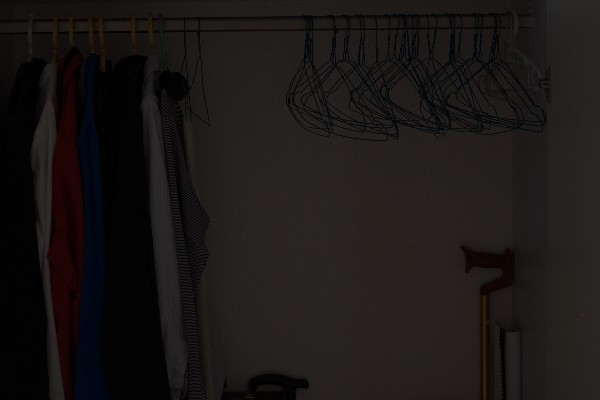} };
        \zoombox[magnification=4.29, color code=red]{0.11,0.61}
        \zoombox[magnification=2.78, color code=blue]{0.88,0.52}
        \end{tikzpicture}
        &
        \begin{tikzpicture}[zoomboxarray, zoomboxes below, zoomboxarray inner gap=0.25cm, zoomboxarray columns=2, zoomboxarray rows=1]
        \node [image node] { \includegraphics[width=0.19\textwidth]{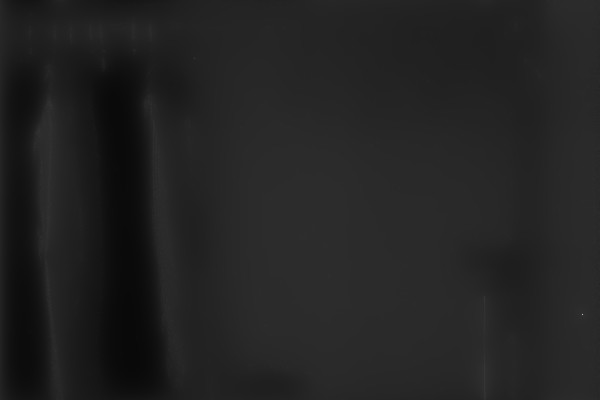} };
        \zoombox[magnification=4.29, color code=red]{0.11,0.61}
        \zoombox[magnification=2.78, color code=blue]{0.88,0.52}
        \end{tikzpicture}
        &
        \begin{tikzpicture}[zoomboxarray, zoomboxes below, zoomboxarray inner gap=0.25cm, zoomboxarray columns=2, zoomboxarray rows=1]
        \node [image node] { \includegraphics[width=0.19\textwidth]{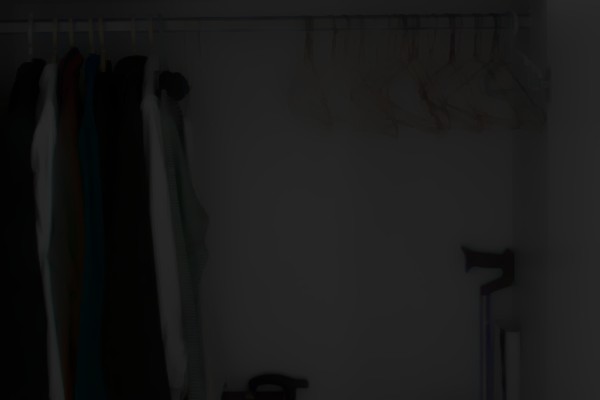} };
        \zoombox[magnification=4.29, color code=red]{0.11,0.61}
        \zoombox[magnification=2.78, color code=blue]{0.88,0.52}
        \end{tikzpicture}
        &
        \begin{tikzpicture}[zoomboxarray, zoomboxes below, zoomboxarray inner gap=0.25cm, zoomboxarray columns=2, zoomboxarray rows=1]
        \node [image node] { \includegraphics[width=0.19\textwidth]{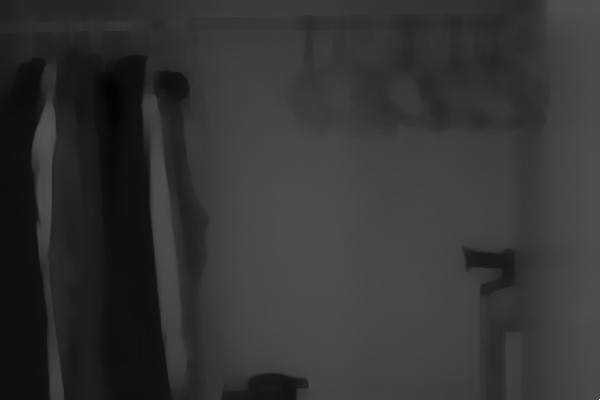} };
        \zoombox[magnification=4.29, color code=red]{0.11,0.61}
        \zoombox[magnification=2.78, color code=blue]{0.88,0.52}
        \end{tikzpicture}
        &
        \begin{tikzpicture}[zoomboxarray, zoomboxes below, zoomboxarray inner gap=0.25cm, zoomboxarray columns=2, zoomboxarray rows=1]
        \node [image node] { \includegraphics[width=0.19\textwidth]{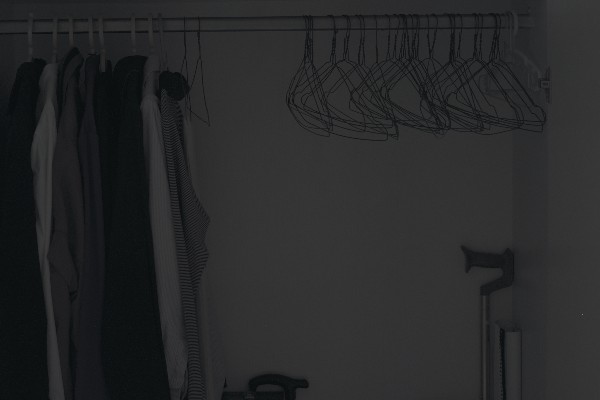} };
        \zoombox[magnification=4.29, color code=red]{0.11,0.61}
        \zoombox[magnification=2.78, color code=blue]{0.88,0.52}
        \end{tikzpicture}
        \\ & 
        \begin{tikzpicture}[zoomboxarray, zoomboxes below, zoomboxarray inner gap=0.25cm, zoomboxarray columns=2, zoomboxarray rows=1]
        \node [image node] { \includegraphics[width=0.19\textwidth]{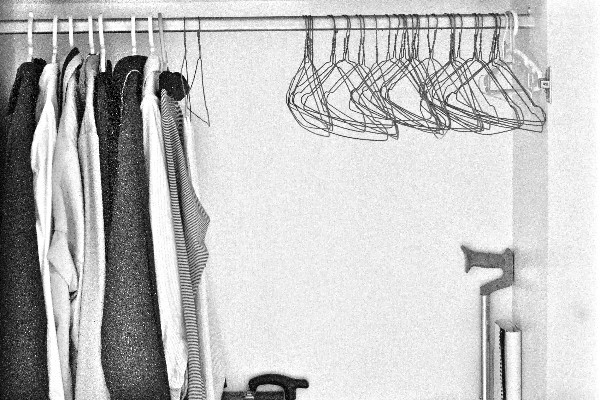} };
        \zoombox[magnification=4.29, color code=red]{0.11,0.61}
        \zoombox[magnification=2.78, color code=blue]{0.88,0.52}
        \end{tikzpicture}
        &
        \begin{tikzpicture}[zoomboxarray, zoomboxes below, zoomboxarray inner gap=0.25cm, zoomboxarray columns=2, zoomboxarray rows=1]
        \node [image node] { \includegraphics[width=0.19\textwidth]{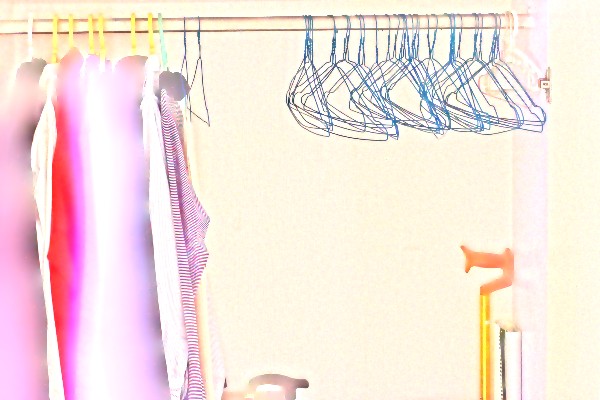} };
        \zoombox[magnification=4.29, color code=red]{0.11,0.61}
        \zoombox[magnification=2.78, color code=blue]{0.88,0.52}
        \end{tikzpicture}
        &
        \begin{tikzpicture}[zoomboxarray, zoomboxes below, zoomboxarray inner gap=0.25cm, zoomboxarray columns=2, zoomboxarray rows=1]
        \node [image node] { \includegraphics[width=0.19\textwidth]{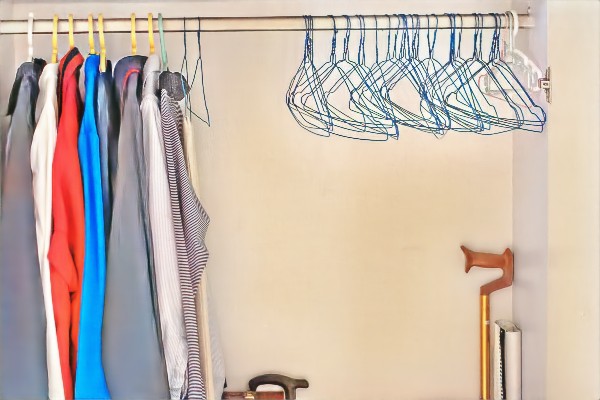} };
        \zoombox[magnification=4.29, color code=red]{0.11,0.61}
        \zoombox[magnification=2.78, color code=blue]{0.88,0.52}
        \end{tikzpicture}
        &
        \begin{tikzpicture}[zoomboxarray, zoomboxes below, zoomboxarray inner gap=0.25cm, zoomboxarray columns=2, zoomboxarray rows=1]
        \node [image node] { \includegraphics[width=0.19\textwidth]{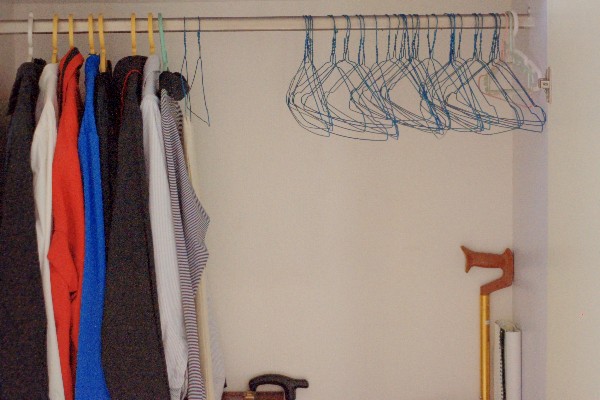} };
        \zoombox[magnification=4.29, color code=red]{0.11,0.61}
        \zoombox[magnification=2.78, color code=blue]{0.88,0.52}
        \end{tikzpicture}
        \\
        {\small (a) Input } &  {\small (b) SRIE~\cite{fu2016fusionbased} } &  {\small (c) RRM~\cite{li2018structurerevealing} } &  {\small (d) KinD~\cite{zhang2019kindling} } &  {\small (e) Ours}
    \end{tabular}
    \caption{\small Illustration of Retinex decomposition results. (b)-(e): Top: illumination; Bottom: reflectance.}
    \label{fig:LOL_146_decom}
\end{figure*}

\subsection{Results and comparisons}
We compare the proposed method with 3 existing state-of-the-art methods that explicitly consider how to deal with severe noise in low-light images, including
Joint Enhancement and Denoising (JED)~\cite{ren2018joint},
Robust Retinex Model (RRM)~\cite{li2018structurerevealing},
and Kindling the Darkness (KinD)~\cite{zhang2019kindling}.
JED and RRM are variational methods, while KinD is a deep learning-based method which uses an NN separately trained for denoising.

In addition, 12 classic low-light image enhancement methods, though without considering the existence of noise, is also evaluated for comparison.
They are histogram equalization (HE),
Multi-Scale Retinex with Color Restoration (MSRCR)~\cite{jobson1997multiscale},
dehazing based method (Dong)~\cite{dong2010fast},
naturalness preserved enhancement algorithm (NPE)~\cite{wang2013naturalness}, SRIE~\cite{fu2016weighted},
Multi-deviation Fusion method (MF)~\cite{fu2016fusionbased},
low-light image enhancement via illumination map estimation (LIME)~\cite{guo2017lime},
Bio-Inspired Multi-Exposure Fusion (BIMEF)~\cite{ying2017bioinspired},
Naturalness Preserved Image Enhancement Using a Priori Multi-Layer Lightness Statistics (NPIE-MLLS)~\cite{wang2018naturalness},
RetinexNet~\cite{wei2018deep},
and Deep Underexposed Photo Enhancement (DeepUPE)~\cite{wang2019underexposed}.
Among them, RetinexNet and DeepUPE are methods based on deep learning.
HE is performed by using the MATLAB built-in function \textit{histeq}.
The results of other methods are generated by the codes released by the authors, with recommended experiment settings.


\subsubsection{Results on LOL}

The quantitative results on the LOL dataset are listed in Table~\ref{tab:LOL}. It can be seen that the proposed method significantly outperforms all other compared methods for all seven metrics except for SSIM, of which our score is only slightly lower than KinD-nonblind.
It is noted that, KinD are fed with an illumination ratio computed from the ground truth data, which is denoted as ``nonblind''. In contrast, our method does not require such nonblind information.
The results of KinD without a ground truth illumination ratio is also reported for fair comparison. 
The proposed method outperform this blind version of KinD in terms of all metrics in this setting.
All above noticeable performance improvement has demonstrated the effectiveness of the proposed approach.

Please see some visual comparisons in Fig.~\ref{fig:LOL_493}.
It can be seen from the first three rows that the methods without noise treatment mechanisms produce noisy results, especially in the dark areas of the original image, although some of them such as LIME and DeepUPE do produce relative vivid colors.
For instance, as shown in Fig.~\ref{fig:LOL_493} and ~\ref{fig:LOL_669}, the hands of the white doll are full of noise and artifacts in the enhanced results on first three rows.
In contrast, the methods with noise treatment mechanisms, \ie, JED, RRM, KinD and the proposed method, suppress noise well. 
Among them, the former three over-smooth the image details and textures, while the proposed method not only produces pleasing colors, but also trades off well between noise suppression and preservation of details.

Please see Fig.~\ref{fig:LOL_146_decom} for the illumination and reflectance layers estimated by several Retinex decomposition methods.
Thanks to the edge-aware technique of bilateral learning, the proposed method produced edges of larger multitudes.


\subsubsection{Results on the other datasets}
Table~\ref{tab:MEF},~\ref{tab:DICM},~\ref{tab:LIME},~\ref{tab:NPE} summarize the results on the MEF, DICM, LIME, NPE datasets respectively.
On these datasets, the methods without denoising mechanisms achieved the best  quantitative results.
On MEF, BIMEF outperforms other methods in terms of all metrics except for NIQE.
As for DICM, LIME, and NPE, there are no methods outperforming others in terms of all metrics.
Specifically, SRIE outperforms other methods in terms of LOE.
It is not surprising that none of methods with denoising mechanisms is among the best ones on these datasets, since images from these datasets are most underexposed with absence of noise.
In order to deal with noise, the methods with denoising mechanisms in low-light images inevitably brings smooth artifacts to the images, which have negative effects on the quality evaluation.
We present results on these datasets  more to evaluate whether the methods designed for severe noise can generalize well for underexposed images.

\begin{figure*}[htbp!]\centering

\begin{subfigure}[b]{0.24\textwidth}
    \begin{tikzpicture}[zoomboxarray, zoomboxes below, zoomboxarray inner gap=0.25cm, zoomboxarray columns=2, zoomboxarray rows=1]
    \node [image node] { \includegraphics[width=\textwidth]{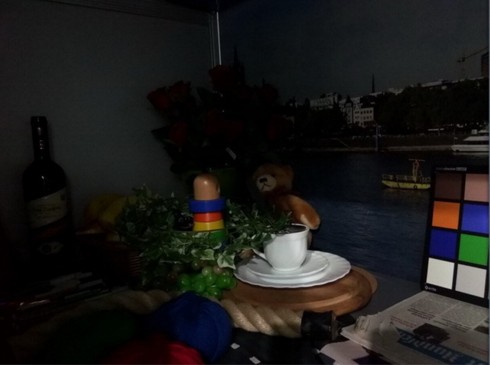} };
    \zoombox[magnification=2.58, color code=red]{0.21,0.45}
    \zoombox[magnification=1.76, color code=blue]{0.44,0.70}
    \end{tikzpicture}
    \caption{Input}
\end{subfigure}
\begin{subfigure}[b]{0.24\textwidth}
    \begin{tikzpicture}[zoomboxarray, zoomboxes below, zoomboxarray inner gap=0.25cm, zoomboxarray columns=2, zoomboxarray rows=1]
    \node [image node] { \includegraphics[width=\textwidth]{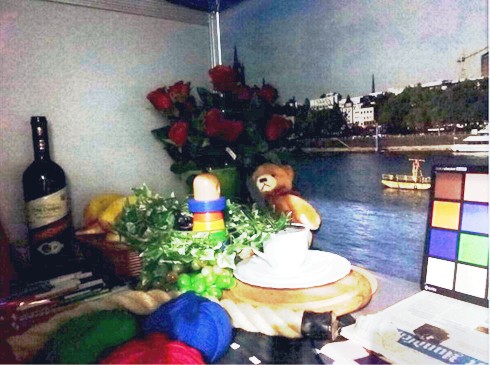} };
    \zoombox[magnification=2.58, color code=red]{0.21,0.45}
    \zoombox[magnification=1.76, color code=blue]{0.44,0.70}
    \end{tikzpicture}
    \caption{HE}
\end{subfigure}
\begin{subfigure}[b]{0.24\textwidth}
    \begin{tikzpicture}[zoomboxarray, zoomboxes below, zoomboxarray inner gap=0.25cm, zoomboxarray columns=2, zoomboxarray rows=1]
    \node [image node] { \includegraphics[width=\textwidth]{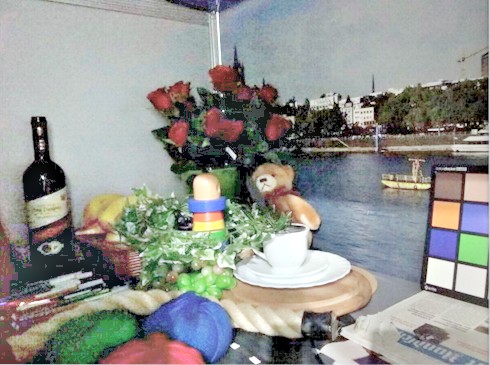} };
    \zoombox[magnification=2.58, color code=red]{0.21,0.45}
    \zoombox[magnification=1.76, color code=blue]{0.44,0.70}
    \end{tikzpicture}
    \caption{MSR~\cite{jobson1997multiscale}}
\end{subfigure}
\begin{subfigure}[b]{0.24\textwidth}
    \begin{tikzpicture}[zoomboxarray, zoomboxes below, zoomboxarray inner gap=0.25cm, zoomboxarray columns=2, zoomboxarray rows=1]
    \node [image node] { \includegraphics[width=\textwidth]{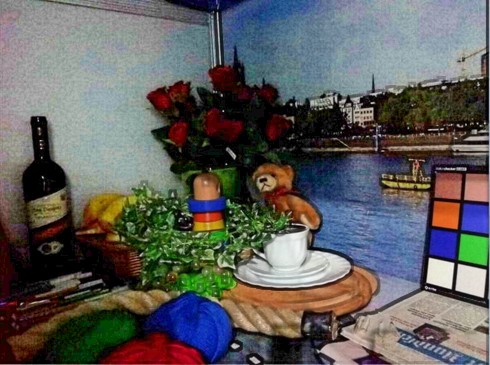} };
    \zoombox[magnification=2.58, color code=red]{0.21,0.45}
    \zoombox[magnification=1.76, color code=blue]{0.44,0.70}
    \end{tikzpicture}
    \caption{Dong~\cite{dong2010fast}}
\end{subfigure}

\begin{subfigure}[b]{0.24\textwidth}
    \begin{tikzpicture}[zoomboxarray, zoomboxes below, zoomboxarray inner gap=0.25cm, zoomboxarray columns=2, zoomboxarray rows=1]
    \node [image node] { \includegraphics[width=\textwidth]{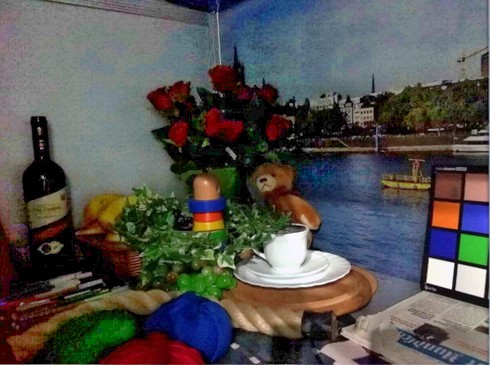} };
    \zoombox[magnification=2.58, color code=red]{0.21,0.45}
    \zoombox[magnification=1.76, color code=blue]{0.44,0.70}
    \end{tikzpicture}
    \caption{NPE~\cite{wang2013naturalness}}
\end{subfigure}
\begin{subfigure}[b]{0.24\textwidth}
    \begin{tikzpicture}[zoomboxarray, zoomboxes below, zoomboxarray inner gap=0.25cm, zoomboxarray columns=2, zoomboxarray rows=1]
    \node [image node] { \includegraphics[width=\textwidth]{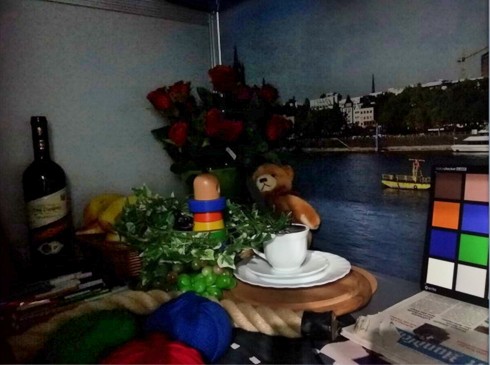} };
    \zoombox[magnification=2.58, color code=red]{0.21,0.45}
    \zoombox[magnification=1.76, color code=blue]{0.44,0.70}
    \end{tikzpicture}
    \caption{SRIE~\cite{fu2016weighted}}
\end{subfigure}
\begin{subfigure}[b]{0.24\textwidth}
    \begin{tikzpicture}[zoomboxarray, zoomboxes below, zoomboxarray inner gap=0.25cm, zoomboxarray columns=2, zoomboxarray rows=1]
    \node [image node] { \includegraphics[width=\textwidth]{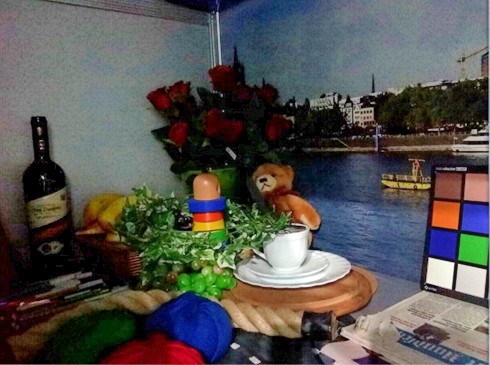} };
    \zoombox[magnification=2.58, color code=red]{0.21,0.45}
    \zoombox[magnification=1.76, color code=blue]{0.44,0.70}
    \end{tikzpicture}
    \caption{MF~\cite{fu2016fusionbased}}
\end{subfigure}
\begin{subfigure}[b]{0.24\textwidth}
    \begin{tikzpicture}[zoomboxarray, zoomboxes below, zoomboxarray inner gap=0.25cm, zoomboxarray columns=2, zoomboxarray rows=1]
    \node [image node] { \includegraphics[width=\textwidth]{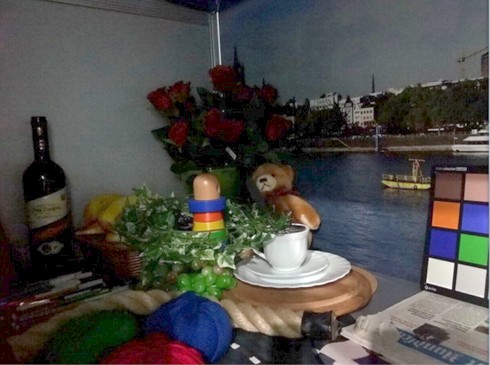} };
    \zoombox[magnification=2.58, color code=red]{0.21,0.45}
    \zoombox[magnification=1.76, color code=blue]{0.44,0.70}
    \end{tikzpicture}
    \caption{BIMEF~\cite{ying2017bioinspired}}
\end{subfigure}

\begin{subfigure}[b]{0.24\textwidth}
    \begin{tikzpicture}[zoomboxarray, zoomboxes below, zoomboxarray inner gap=0.25cm, zoomboxarray columns=2, zoomboxarray rows=1]
    \node [image node] { \includegraphics[width=\textwidth]{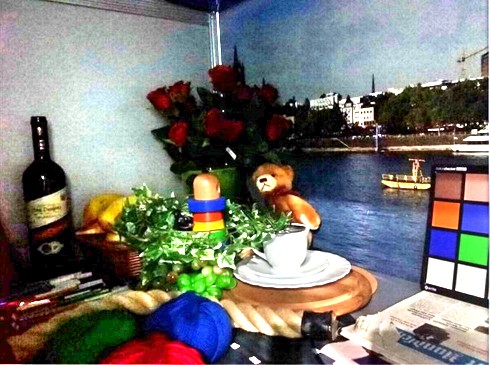} };
    \zoombox[magnification=2.58, color code=red]{0.21,0.45}
    \zoombox[magnification=1.76, color code=blue]{0.44,0.70}
    \end{tikzpicture}
    \caption{LIME~\cite{guo2017lime}}
\end{subfigure}
\begin{subfigure}[b]{0.24\textwidth}
    \begin{tikzpicture}[zoomboxarray, zoomboxes below, zoomboxarray inner gap=0.25cm, zoomboxarray columns=2, zoomboxarray rows=1]
    \node [image node] { \includegraphics[width=\textwidth]{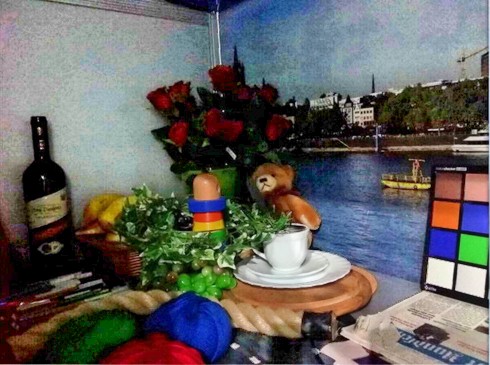} };
    \zoombox[magnification=2.58, color code=red]{0.21,0.45}
    \zoombox[magnification=1.76, color code=blue]{0.44,0.70}
    \end{tikzpicture}
    \caption{NPIE-MLLS~\cite{wang2018naturalness}}
\end{subfigure}
\begin{subfigure}[b]{0.24\textwidth}
    \begin{tikzpicture}[zoomboxarray, zoomboxes below, zoomboxarray inner gap=0.25cm, zoomboxarray columns=2, zoomboxarray rows=1]
    \node [image node] { \includegraphics[width=\textwidth]{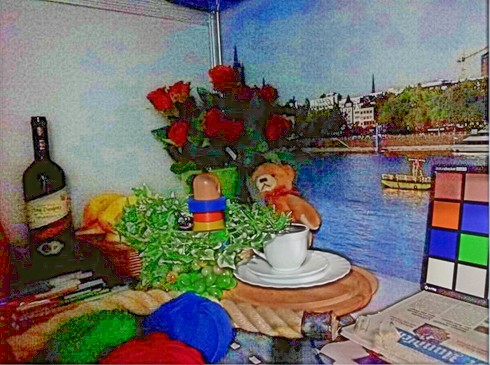} };
    \zoombox[magnification=2.58, color code=red]{0.21,0.45}
    \zoombox[magnification=1.76, color code=blue]{0.44,0.70}
    \end{tikzpicture}
    \caption{RetinexNet~\cite{wei2018deep}}
\end{subfigure}
\begin{subfigure}[b]{0.24\textwidth}
    \begin{tikzpicture}[zoomboxarray, zoomboxes below, zoomboxarray inner gap=0.25cm, zoomboxarray columns=2, zoomboxarray rows=1]
    \node [image node] { \includegraphics[width=\textwidth]{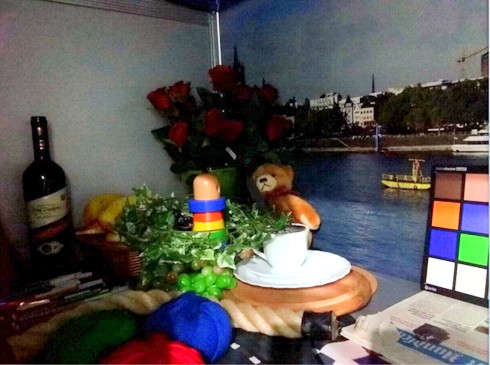} };
    \zoombox[magnification=2.58, color code=red]{0.21,0.45}
    \zoombox[magnification=1.76, color code=blue]{0.44,0.70}
    \end{tikzpicture}
    \caption{DeepUPE~\cite{wang2019underexposed}}
\end{subfigure}

\begin{subfigure}[b]{0.24\textwidth}
    \begin{tikzpicture}[zoomboxarray, zoomboxes below, zoomboxarray inner gap=0.25cm, zoomboxarray columns=2, zoomboxarray rows=1]
    \node [image node] { \includegraphics[width=\textwidth]{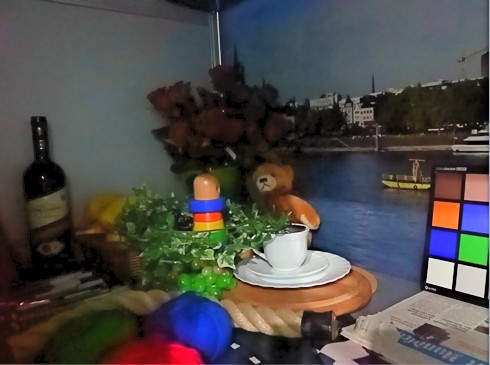} };
    \zoombox[magnification=2.58, color code=red]{0.21,0.45}
    \zoombox[magnification=1.76, color code=blue]{0.44,0.70}
    \end{tikzpicture}
    \caption{JED~\cite{ren2018joint}}
\end{subfigure}
\begin{subfigure}[b]{0.24\textwidth}
    \begin{tikzpicture}[zoomboxarray, zoomboxes below, zoomboxarray inner gap=0.25cm, zoomboxarray columns=2, zoomboxarray rows=1]
    \node [image node] { \includegraphics[width=\textwidth]{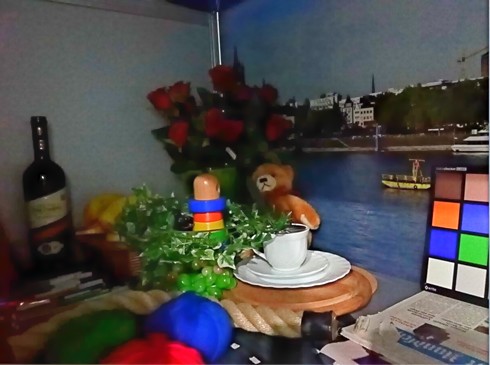} };
    \zoombox[magnification=2.58, color code=red]{0.21,0.45}
    \zoombox[magnification=1.76, color code=blue]{0.44,0.70}
    \end{tikzpicture}
    \caption{RRM~\cite{li2018structurerevealing}}
\end{subfigure}
\begin{subfigure}[b]{0.24\textwidth}
    \begin{tikzpicture}[zoomboxarray, zoomboxes below, zoomboxarray inner gap=0.25cm, zoomboxarray columns=2, zoomboxarray rows=1]
    \node [image node] { \includegraphics[width=\textwidth]{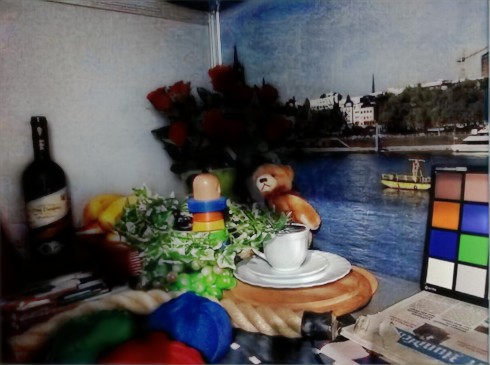} };
    \zoombox[magnification=2.58, color code=red]{0.21,0.45}
    \zoombox[magnification=1.76, color code=blue]{0.44,0.70}
    \end{tikzpicture}
    \caption{KinD~\cite{zhang2019kindling}}
\end{subfigure}
\begin{subfigure}[b]{0.24\textwidth}
    \begin{tikzpicture}[zoomboxarray, zoomboxes below, zoomboxarray inner gap=0.25cm, zoomboxarray columns=2, zoomboxarray rows=1]
    \node [image node] { \includegraphics[width=\textwidth]{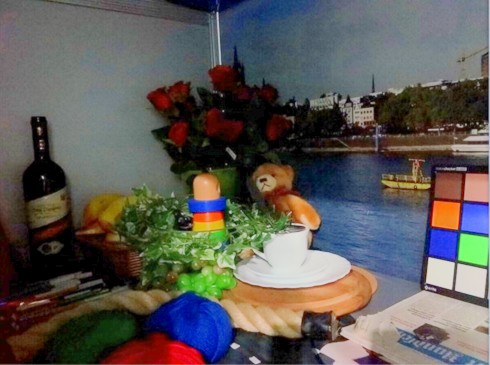} };
    \zoombox[magnification=2.58, color code=red]{0.21,0.45}
    \zoombox[magnification=1.76, color code=blue]{0.44,0.70}
    \end{tikzpicture}
    \caption{Ours}
\end{subfigure}
\caption{Visual comparisons of different methods on low-light images from the LIME dataset.}
\label{fig:LIME_8}
\end{figure*}

It can be seen that, the proposed method still obtained good results the on MEF and DICM datasets.
On the LIME dataset, the score of the proposed method is only lower than JED in terms of NIQE.
In the NPE dataset, the score of the proposed method is only lower than KinD.
The qualitative results are presented in the supplementary material.
Please see Fig.~\ref{fig:MEF_ChineseGarden} and ~\ref{fig:LIME_8} for visual comparison.
\begin{table}[htbp!]
    \centering
    \caption{\small Quantitative results on the MEF~\cite{ma2015perceptual} dataset.}
    \addtolength{\tabcolsep}{-2.5pt}
    \sisetup{detect-all=true,detect-weight=true,detect-inline-weight=math}
    \begin{tabular}{cS[table-format=4.2]S[table-format=1.4]S[table-format=2.4]S[table-format=2.4]}
        \toprule
        {Method} & \multicolumn{1}{c}{{LOE$\downarrow$}} & \multicolumn{1}{c}{{NIQE$\downarrow$}} & \multicolumn{1}{c}{{BRISQUE$\downarrow$}} & \multicolumn{1}{c}{{PIQE$\downarrow$}} \\
        \midrule
        \midrule
        HE    & 236.15  & 3.6455  & 26.3202  & 42.0736  \\
        MSR~\cite{jobson1997multiscale}   & 1011.88  & \BF 3.3090  & 23.0138  & 38.4410  \\
        Dong~\cite{dong2010fast}  & 221.72  & 4.0989  & 26.4946  & 37.0443  \\
        NPE~\cite{wang2013naturalness}   & 422.32  & 3.5295  & 23.7685  & 35.8246  \\
        SRIE~\cite{fu2016weighted}  & 210.26  & 3.4741  & 22.0880  & 37.7866  \\
        MF~\cite{fu2016fusionbased}    & 208.13  & 3.4923  & 22.7244  & 34.1056  \\
        BIMEF~\cite{ying2017bioinspired} & \BF 155.62  & 3.3290  & \BF 20.2203  & \BF 33.7428  \\
        LIME~\cite{guo2017lime}  & 939.11  & 3.7023  & 24.0577  & 40.0240  \\
        NPIE-MLLS~\cite{wang2018naturalness} & 344.95  & 3.3370  & 22.3205  & 34.8772  \\
        RetinexNet~\cite{wei2018deep} & 708.25  & 4.4099  & 26.0365  & 41.2155  \\
        DeepUPE~\cite{wang2019underexposed}  & 214.77  & 3.3717  & 22.6911  & 37.2156  \\
        \midrule
        \midrule
        JED~\cite{ren2018joint}    & 294.15  & 4.5209  & 30.7195  & 47.9936  \\
        RRM~\cite{li2018structurerevealing}    & 311.39  & 5.0621  & 32.9379  & 52.8415  \\
        KinD~\cite{zhang2019kindling} & 275.47  & 3.8767  & 30.4408  & 53.9012  \\
        Ours  & \BF 172.80  & \BF 3.4673  & \BF 22.2387  & \BF 35.6976  \\
        \bottomrule
    \end{tabular}%
    \label{tab:MEF}%
\end{table}%

\begin{table}[htbp!]
    \centering
    \caption{\small Quantitative results on the DICM~\cite{lee2013contrast} dataset.}
    \addtolength{\tabcolsep}{-2.5pt}
    \sisetup{detect-all=true,detect-weight=true,detect-inline-weight=math}
    \begin{tabular}{cS[table-format=4.2]S[table-format=1.4]S[table-format=2.4]S[table-format=2.4]}
        \toprule
        {Method} & \multicolumn{1}{c}{{LOE$\downarrow$}} & \multicolumn{1}{c}{{NIQE$\downarrow$}} & \multicolumn{1}{c}{{BRISQUE$\downarrow$}} & \multicolumn{1}{c}{{PIQE$\downarrow$}} \\
        \midrule
        \midrule
        HE    & 270.31  & 3.8635  & 25.4847  & 41.2639  \\
        MSR~\cite{jobson1997multiscale}   & 1127.82  & \BF 3.6766  & 26.0095  & 40.1992  \\
        Dong~\cite{dong2010fast}  & 276.97  & 4.1191  & 26.7334  & 36.0868  \\
        NPE~\cite{wang2013naturalness}   & 209.00  & 3.7601  & \BF 25.3145  & 37.1805  \\
        SRIE~\cite{fu2016weighted}  & \BF 162.22  & 3.8987  & 27.6980  & 39.3550  \\
        MF~\cite{fu2016fusionbased}    & 321.18  & 3.8441  & 25.6823  & 35.9124  \\
        BIMEF~\cite{ying2017bioinspired} & 239.27  & 3.8459  & 26.8110  & 36.6763  \\
        LIME~\cite{guo2017lime}  & 1107.77  & 3.8588  & 26.8838  & 41.3392  \\
        NPIE-MLLS~\cite{wang2018naturalness} & 264.60  & 3.7360  & 25.4938  & 37.0152  \\
        RetinexNet~\cite{wei2018deep} & 636.16  & 4.4154  & 26.6565  & 37.9280  \\
        DeepUPE~\cite{wang2019underexposed}  & 193.38  & 3.9229  & 26.9957  & \BF 35.4541  \\
        \midrule
        \midrule
        JED~\cite{ren2018joint}    & 483.89  & 4.2605  & 27.4469  & 41.3614  \\
        RRM~\cite{li2018structurerevealing}    & 518.37  & 4.5970  & 30.1769  & 42.0824  \\
        KinD~\cite{zhang2019kindling} & 261.77  & 4.1505  & 30.6981  & 47.2272  \\
        Ours  & \BF 235.23  & \BF 3.7409  & \BF 26.4639  & \BF 32.1176  \\
        \bottomrule
    \end{tabular}%
    \label{tab:DICM}%
\end{table}%

\begin{table}[htbp!]
    \centering
    \caption{\small Quantitative results on the LIME~\cite{guo2017lime} dataset.}
    \addtolength{\tabcolsep}{-2.5pt}
    \sisetup{detect-all=true,detect-weight=true,detect-inline-weight=math}
    \begin{tabular}{cS[table-format=4.2]S[table-format=1.4]S[table-format=2.4]S[table-format=2.4]}
        \toprule
        {Method} & \multicolumn{1}{c}{{LOE$\downarrow$}} & \multicolumn{1}{c}{{NIQE$\downarrow$}} & \multicolumn{1}{c}{{BRISQUE$\downarrow$}} & \multicolumn{1}{c}{{PIQE$\downarrow$}} \\
        \midrule
        \midrule
        HE    & 322.22  & 4.3966  & 22.6083  & 42.6408  \\
        MSR~\cite{jobson1997multiscale}   & 864.68  & 3.7642  & 22.6987  & 40.3416  \\
        Dong~\cite{dong2010fast}  & 241.40  & 4.0516  & 26.2226  & 38.4488  \\
        NPE~\cite{wang2013naturalness}   & 334.92  & 3.9048  & \BF 22.1569  & 36.4812  \\
        SRIE~\cite{fu2016weighted}  & \BF 106.31  & 3.7863  & 24.1806  & \BF 34.7993  \\
        MF~\cite{fu2016fusionbased}    & 183.25  & 4.0673  & 22.2979  & 36.4530  \\
        BIMEF~\cite{ying2017bioinspired} & 136.90  & 3.8596  & 23.1353  & 35.7954  \\
        LIME~\cite{guo2017lime}  & 793.90  & 4.1549  & 22.3094  & 41.0324  \\
        NPIE-MLLS~\cite{wang2018naturalness} & 300.51  & \BF 3.5788  & 22.5060  & 37.5379  \\
        RetinexNet~\cite{wei2018deep} & 539.64  & 4.5983  & 26.1007  & 42.7741  \\
        DeepUPE~\cite{wang2019underexposed}  & 185.88  & 3.8982  & 25.6375  & 37.4441  \\
        \midrule
        \midrule
        JED~\cite{ren2018joint}    & 270.76  & 4.7180  & 28.5736  & 42.8337  \\
        RRM~\cite{li2018structurerevealing}    & 276.12  & 4.6426  & 29.1015  & 41.3977  \\
        KinD~\cite{zhang2019kindling} & 255.80  & 4.7632  & 26.7730  & 45.3084  \\
        Ours  & \BF 153.99  & \BF 4.0431  & \BF 23.1246  & \BF 37.9949  \\
        \bottomrule
    \end{tabular}%
    \label{tab:LIME}%
\end{table}%

\begin{table}[htbp!]
    \centering
    \caption{\small Quantitative results on the NPE~\cite{wang2013naturalness} dataset.}
    \addtolength{\tabcolsep}{-2.5pt}
    \sisetup{detect-all=true,detect-weight=true,detect-inline-weight=math}
    \begin{tabular}{cS[table-format=4.2]S[table-format=1.4]S[table-format=2.4]S[table-format=2.4]}
        \toprule
        {Method} & \multicolumn{1}{c}{{LOE$\downarrow$}} & \multicolumn{1}{c}{{NIQE$\downarrow$}} & \multicolumn{1}{c}{{BRISQUE$\downarrow$}} & \multicolumn{1}{c}{{PIQE$\downarrow$}} \\
        \midrule
        \midrule
        HE    & 212.75  & 3.9780  & 24.0995  & 38.6459  \\
        MSR~\cite{jobson1997multiscale}   & 1270.14  & 4.3663  & 24.2051  & 35.7164  \\
        Dong~\cite{dong2010fast}  & 181.15  & 4.1263  & \BF 23.1679  & \BF 31.9268  \\
        NPE~\cite{wang2013naturalness}   & 204.65  & \BF 3.9520  & 24.4615  & 31.9680  \\
        SRIE~\cite{fu2016weighted}  & \BF 159.34  & 3.9795  & 25.6351  & 35.3823  \\
        MF~\cite{fu2016fusionbased}    & 303.82  & 4.1048  & 26.2524  & 34.3280  \\
        BIMEF~\cite{ying2017bioinspired} & 233.27  & 4.1328  & 24.4256  & 33.4234  \\
        LIME~\cite{guo2017lime}  & 1174.92  & 4.2629  & 26.1337  & 36.8072  \\
        NPIE-MLLS~\cite{wang2018naturalness} & 219.79  & 4.0245  & 25.1294  & 32.1868  \\
        RetinexNet~\cite{wei2018deep} & 653.85  & 4.5669  & 26.8732  & 34.5105  \\
        DeepUPE~\cite{wang2019underexposed}  & 173.33  & 4.3201  & 23.8473  & 35.1940  \\
        \midrule
        \midrule
        JED~\cite{ren2018joint}    & 595.71  & 4.9958  & 26.5272  & 41.1340  \\
        RRM~\cite{li2018structurerevealing}    & 634.86  & 4.8452  & 24.7838  & 38.6205  \\
        KinD~\cite{zhang2019kindling} & \BF 180.91  & \BF 4.1607  & \BF 24.2792  & 43.1331  \\
        Ours  & 275.00  & 4.3807  & 24.8661  & \BF 37.0634  \\
        \bottomrule
    \end{tabular}%
    \label{tab:NPE}%
\end{table}%
\begin{figure*}[htbp!]\centering
\begin{subfigure}[b]{0.24\textwidth}
    \begin{tikzpicture}[zoomboxarray, zoomboxes below, zoomboxarray inner gap=0.25cm, zoomboxarray columns=2, zoomboxarray rows=1]
    \node [image node] { \includegraphics[width=\textwidth]{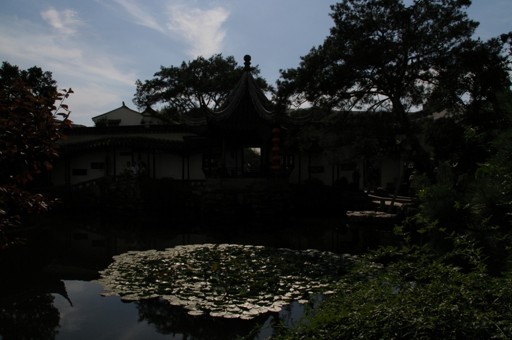} };
    \zoombox[magnification=3.01, color code=red]{0.22,0.56}
    \zoombox[magnification=2.78, color code=blue]{0.48,0.70}
    \end{tikzpicture}
    \caption{  Input}
\end{subfigure}
\begin{subfigure}[b]{0.24\textwidth}
    \begin{tikzpicture}[zoomboxarray, zoomboxes below, zoomboxarray inner gap=0.25cm, zoomboxarray columns=2, zoomboxarray rows=1]
    \node [image node] { \includegraphics[width=\textwidth]{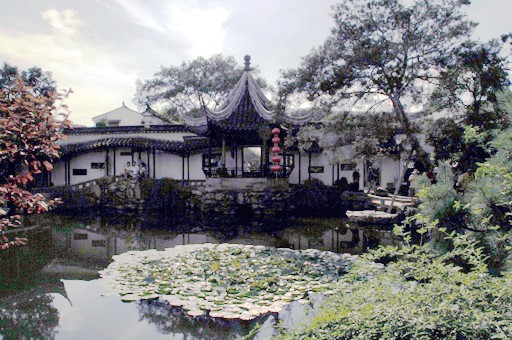} };
    \zoombox[magnification=3.01, color code=red]{0.22,0.56}
    \zoombox[magnification=2.78, color code=blue]{0.48,0.70}
    \end{tikzpicture}
    \caption{  HE}
\end{subfigure}
\begin{subfigure}[b]{0.24\textwidth}
    \begin{tikzpicture}[zoomboxarray, zoomboxes below, zoomboxarray inner gap=0.25cm, zoomboxarray columns=2, zoomboxarray rows=1]
    \node [image node] { \includegraphics[width=\textwidth]{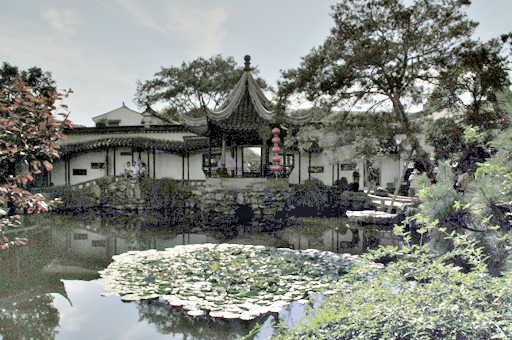} };
    \zoombox[magnification=3.01, color code=red]{0.22,0.56}
    \zoombox[magnification=2.78, color code=blue]{0.48,0.70}
    \end{tikzpicture}
    \caption{  MSR~\cite{jobson1997multiscale}}
\end{subfigure}
\begin{subfigure}[b]{0.24\textwidth}
    \begin{tikzpicture}[zoomboxarray, zoomboxes below, zoomboxarray inner gap=0.25cm, zoomboxarray columns=2, zoomboxarray rows=1]
    \node [image node] { \includegraphics[width=\textwidth]{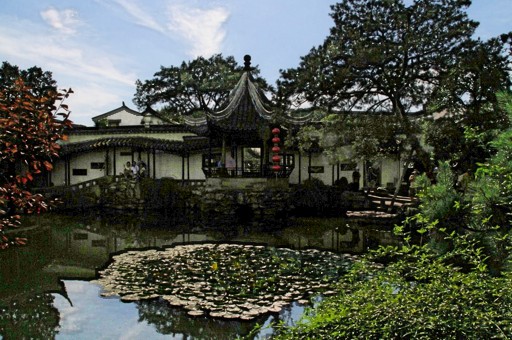} };
    \zoombox[magnification=3.01, color code=red]{0.22,0.56}
    \zoombox[magnification=2.78, color code=blue]{0.48,0.70}
    \end{tikzpicture}
    \caption{  Dong~\cite{dong2010fast}}
\end{subfigure}

\begin{subfigure}[b]{0.24\textwidth}
    \begin{tikzpicture}[zoomboxarray, zoomboxes below, zoomboxarray inner gap=0.25cm, zoomboxarray columns=2, zoomboxarray rows=1]
    \node [image node] { \includegraphics[width=\textwidth]{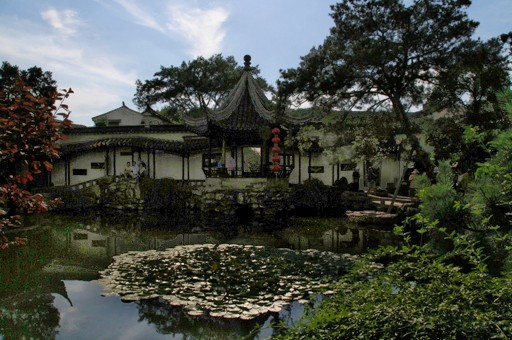} };
    \zoombox[magnification=3.01, color code=red]{0.22,0.56}
    \zoombox[magnification=2.78, color code=blue]{0.48,0.70}
    \end{tikzpicture}
    \caption{  NPE~\cite{wang2013naturalness}}
\end{subfigure}
\begin{subfigure}[b]{0.24\textwidth}
    \begin{tikzpicture}[zoomboxarray, zoomboxes below, zoomboxarray inner gap=0.25cm, zoomboxarray columns=2, zoomboxarray rows=1]
    \node [image node] { \includegraphics[width=\textwidth]{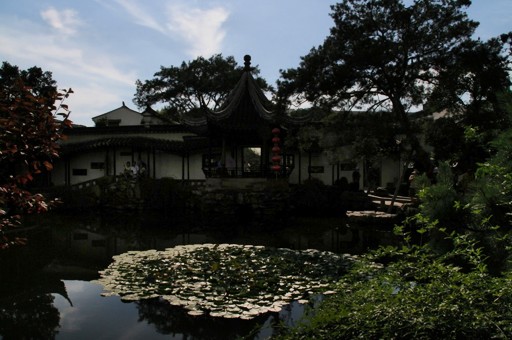} };
    \zoombox[magnification=3.01, color code=red]{0.22,0.56}
    \zoombox[magnification=2.78, color code=blue]{0.48,0.70}
    \end{tikzpicture}
    \caption{  SRIE~\cite{fu2016weighted}}
\end{subfigure}    
\begin{subfigure}[b]{0.24\textwidth}
    \begin{tikzpicture}[zoomboxarray, zoomboxes below, zoomboxarray inner gap=0.25cm, zoomboxarray columns=2, zoomboxarray rows=1]
    \node [image node] { \includegraphics[width=\textwidth]{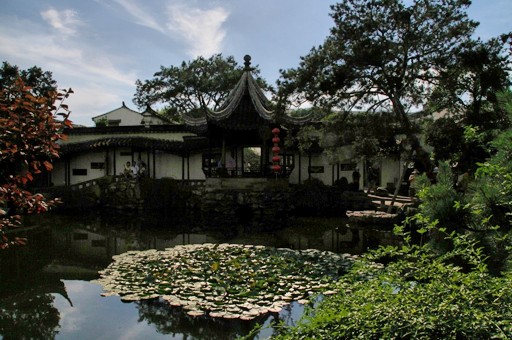} };
    \zoombox[magnification=3.01, color code=red]{0.22,0.56}
    \zoombox[magnification=2.78, color code=blue]{0.48,0.70}
    \end{tikzpicture}
    \caption{  MF~\cite{fu2016fusionbased}}
\end{subfigure}
\begin{subfigure}[b]{0.24\textwidth}
    \begin{tikzpicture}[zoomboxarray, zoomboxes below, zoomboxarray inner gap=0.25cm, zoomboxarray columns=2, zoomboxarray rows=1]
    \node [image node] { \includegraphics[width=\textwidth]{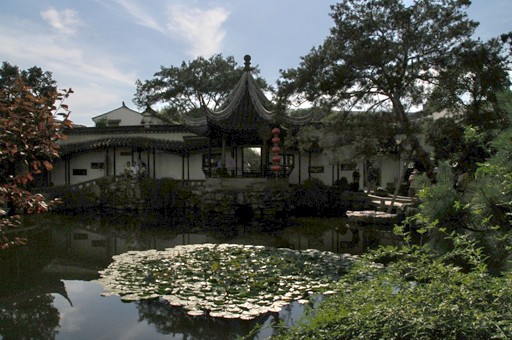} };
    \zoombox[magnification=3.01, color code=red]{0.22,0.56}
    \zoombox[magnification=2.78, color code=blue]{0.48,0.70}
    \end{tikzpicture}
    \caption{  BIMEF~\cite{ying2017bioinspired}}
\end{subfigure}    

\begin{subfigure}[b]{0.24\textwidth}
    \begin{tikzpicture}[zoomboxarray, zoomboxes below, zoomboxarray inner gap=0.25cm, zoomboxarray columns=2, zoomboxarray rows=1]
    \node [image node] { \includegraphics[width=\textwidth]{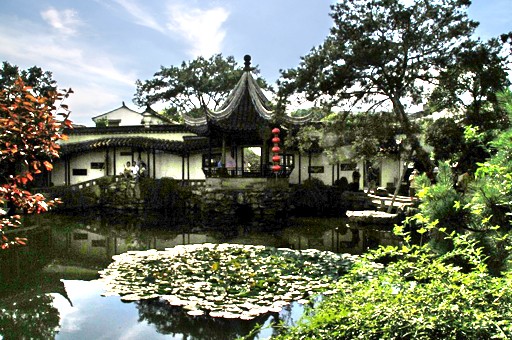} };
    \zoombox[magnification=3.01, color code=red]{0.22,0.56}
    \zoombox[magnification=2.78, color code=blue]{0.48,0.70}
    \end{tikzpicture}
    \caption{  LIME~\cite{guo2017lime}}
\end{subfigure}
\begin{subfigure}[b]{0.24\textwidth}
    \begin{tikzpicture}[zoomboxarray, zoomboxes below, zoomboxarray inner gap=0.25cm, zoomboxarray columns=2, zoomboxarray rows=1]
    \node [image node] { \includegraphics[width=\textwidth]{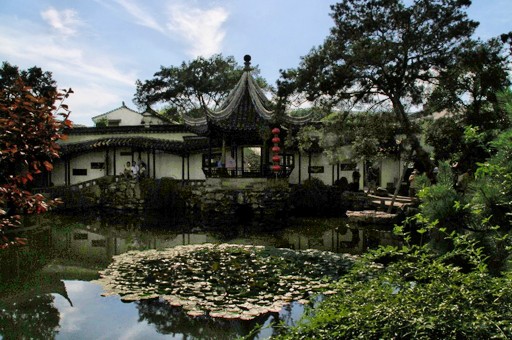} };
    \zoombox[magnification=3.01, color code=red]{0.22,0.56}
    \zoombox[magnification=2.78, color code=blue]{0.48,0.70}
    \end{tikzpicture}
    \caption{  NPIE-MLLS~\cite{wang2018naturalness}}
\end{subfigure}
\begin{subfigure}[b]{0.24\textwidth}
    \begin{tikzpicture}[zoomboxarray, zoomboxes below, zoomboxarray inner gap=0.25cm, zoomboxarray columns=2, zoomboxarray rows=1]
    \node [image node] { \includegraphics[width=\textwidth]{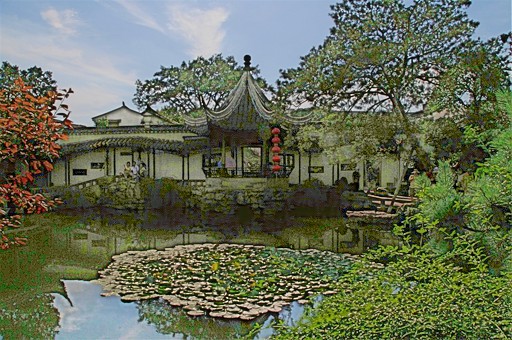} };
    \zoombox[magnification=3.01, color code=red]{0.22,0.56}
    \zoombox[magnification=2.78, color code=blue]{0.48,0.70}
    \end{tikzpicture}
    \caption{  RetinexNet~\cite{wei2018deep}}
\end{subfigure}
\begin{subfigure}[b]{0.24\textwidth}
    \begin{tikzpicture}[zoomboxarray, zoomboxes below, zoomboxarray inner gap=0.25cm, zoomboxarray columns=2, zoomboxarray rows=1]
    \node [image node] { \includegraphics[width=\textwidth]{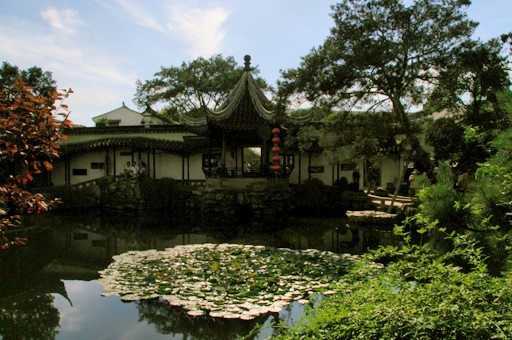} };
    \zoombox[magnification=3.01, color code=red]{0.22,0.56}
    \zoombox[magnification=2.78, color code=blue]{0.48,0.70}
    \end{tikzpicture}
    \caption{  DeepUPE~\cite{wang2019underexposed}}
\end{subfigure}

\begin{subfigure}[b]{0.24\textwidth}
    \begin{tikzpicture}[zoomboxarray, zoomboxes below, zoomboxarray inner gap=0.25cm, zoomboxarray columns=2, zoomboxarray rows=1]
    \node [image node] { \includegraphics[width=\textwidth]{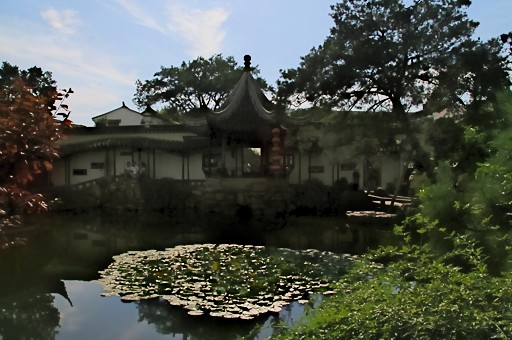} };
    \zoombox[magnification=3.01, color code=red]{0.22,0.56}
    \zoombox[magnification=2.78, color code=blue]{0.48,0.70}
    \end{tikzpicture}
    \caption{  JED~\cite{ren2018joint}}
\end{subfigure}
\begin{subfigure}[b]{0.24\textwidth}
    \begin{tikzpicture}[zoomboxarray, zoomboxes below, zoomboxarray inner gap=0.25cm, zoomboxarray columns=2, zoomboxarray rows=1]
    \node [image node] { \includegraphics[width=\textwidth]{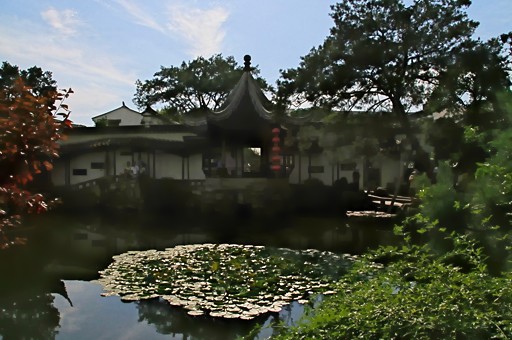} };
    \zoombox[magnification=3.01, color code=red]{0.22,0.56}
    \zoombox[magnification=2.78, color code=blue]{0.48,0.70}
    \end{tikzpicture}
    \caption{  RRM~\cite{li2018structurerevealing}}
\end{subfigure}
\begin{subfigure}[b]{0.24\textwidth}
    \begin{tikzpicture}[zoomboxarray, zoomboxes below, zoomboxarray inner gap=0.25cm, zoomboxarray columns=2, zoomboxarray rows=1]
    \node [image node] { \includegraphics[width=\textwidth]{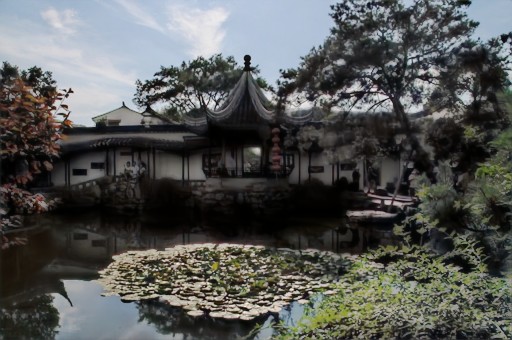} };
    \zoombox[magnification=3.01, color code=red]{0.22,0.56}
    \zoombox[magnification=2.78, color code=blue]{0.48,0.70}
    \end{tikzpicture}
    \caption{  KinD~\cite{zhang2019kindling}}
\end{subfigure}
\begin{subfigure}[b]{0.24\textwidth}
    \begin{tikzpicture}[zoomboxarray, zoomboxes below, zoomboxarray inner gap=0.25cm, zoomboxarray columns=2, zoomboxarray rows=1]
    \node [image node] { \includegraphics[width=\textwidth]{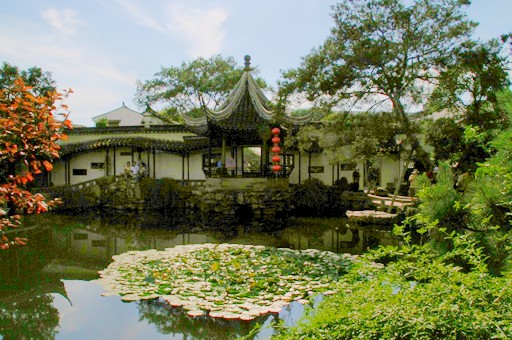} };
    \zoombox[magnification=3.01, color code=red]{0.22,0.56}
    \zoombox[magnification=2.78, color code=blue]{0.48,0.70}
    \end{tikzpicture}
    \caption{  Ours}
\end{subfigure}
\caption{  Visual comparisons of different methods on low-light images from the MEF dataset.}
\label{fig:MEF_ChineseGarden}
\end{figure*}

\subsection{Ablation study}

\subsubsection{Ablation study on the transforms}

\begin{table}[htbp!]
    \centering
    \caption{\small Ablation study on the transforms for estimating $\mN$.}
    \addtolength{\tabcolsep}{-2.5pt}
    \sisetup{detect-all=true,detect-weight=true,detect-inline-weight=math}
    \begin{tabular}{cS[table-format=2.4]S[table-format=1.4]}
        \toprule
        {transforms} & \multicolumn{1}{c}{{PSNR(dB)}} & \multicolumn{1}{c}{{SSIM}} \\
        \midrule
        another set of affine matrices & 20.5639  & 0.6718  \\
        non-deformable (rigid) kernels   & 21.2876  & 0.7579  \\
        deformable kernels with W=7   & 21.6337  & 0.7760  \\
        deformable kernels with W=15 & \BF 22.5156  & \BF 0.7864 \\
        deformable kernels with W=31   & 21.2314  & 0.7741  \\
        \bottomrule
    \end{tabular}%
    \label{tab:kernel}%
\end{table}%

\begin{figure*}[htbp!]\centering
\setlength{\tabcolsep}{2pt}
\begin{tabular}{ccccccc}
\vspace{3pt}
    {\small $\widetilde{\mR}$}&
    \adjincludegraphics[valign=m,width=0.155\linewidth,trim={{.58\width} {.24\height} {.12\width} {.31\height}},clip]{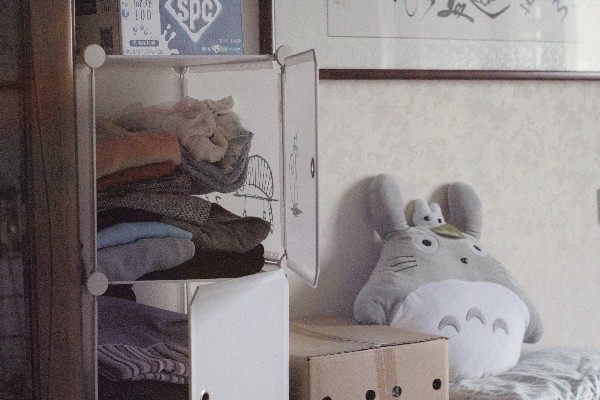}&
    \adjincludegraphics[valign=m,width=0.155\linewidth,trim={{.58\width} {.24\height} {.12\width} {.31\height}},clip]{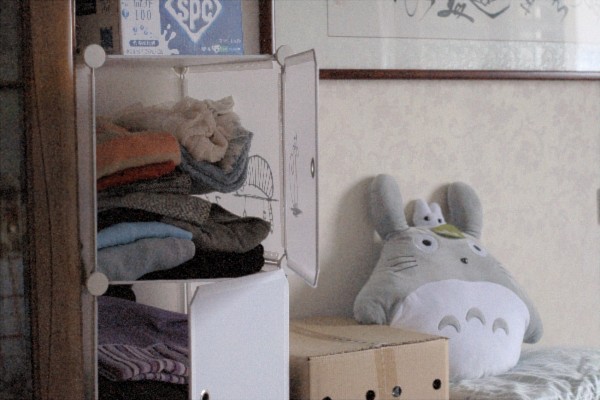}&
    \adjincludegraphics[valign=m,width=0.155\linewidth,trim={{.58\width} {.24\height} {.12\width} {.31\height}},clip]{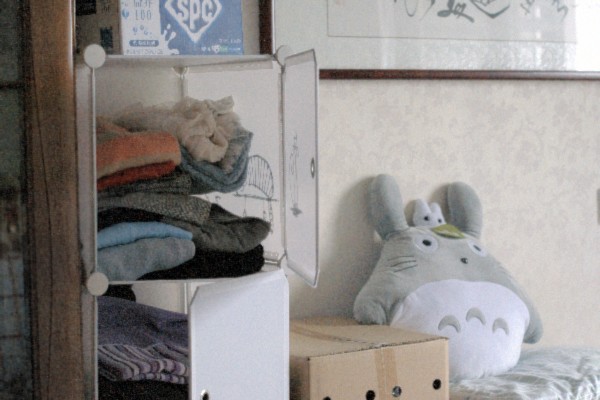}&
    \adjincludegraphics[valign=m,width=0.155\linewidth,trim={{.58\width} {.24\height} {.12\width} {.31\height}},clip]{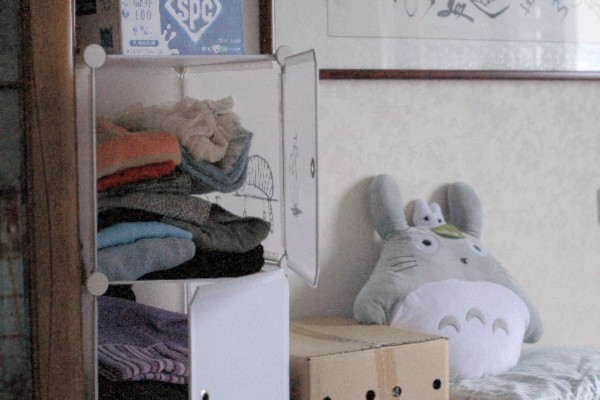}&
    \adjincludegraphics[valign=m,width=0.155\linewidth,trim={{.58\width} {.24\height} {.12\width} {.31\height}},clip]{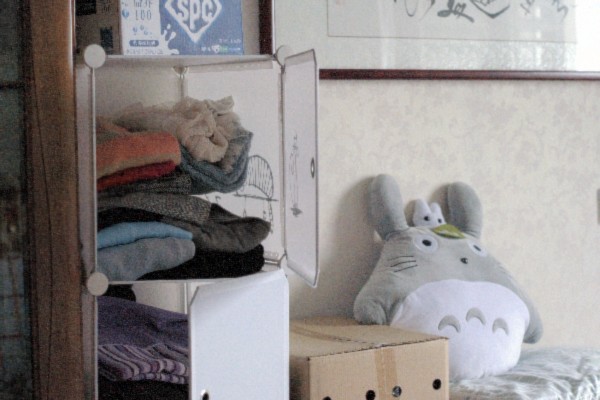}&
    \adjincludegraphics[valign=m,width=0.155\linewidth,trim={{.58\width} {.24\height} {.12\width} {.31\height}},clip]{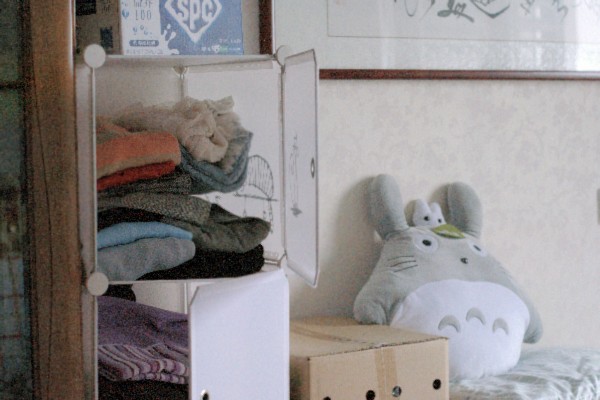}\\
\vspace{3pt}
    {\small $\mE$}&
    \adjincludegraphics[valign=m,width=0.155\linewidth,trim={{.58\width} {.24\height} {.12\width} {.31\height}},clip]{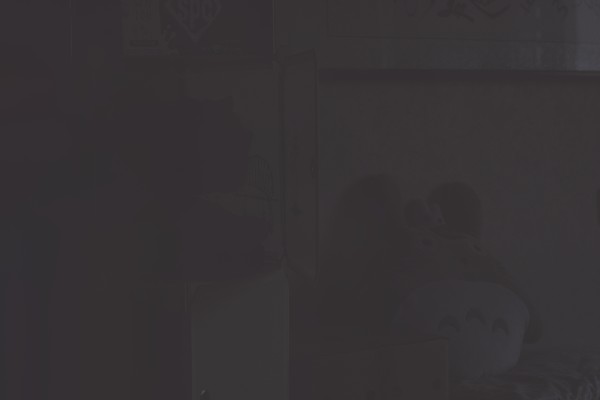}&
    \adjincludegraphics[valign=m,width=0.155\linewidth,trim={{.58\width} {.24\height} {.12\width} {.31\height}},clip]{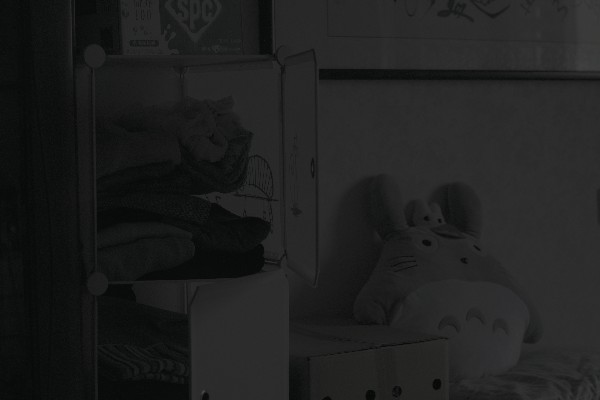}&
    \adjincludegraphics[valign=m,width=0.155\linewidth,trim={{.58\width} {.24\height} {.12\width} {.31\height}},clip]{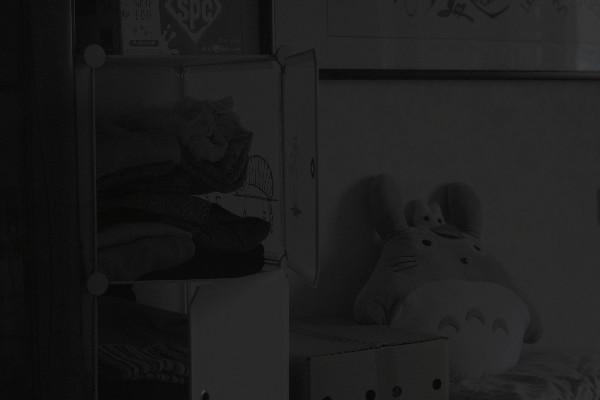}&
    \adjincludegraphics[valign=m,width=0.155\linewidth,trim={{.58\width} {.24\height} {.12\width} {.31\height}},clip]{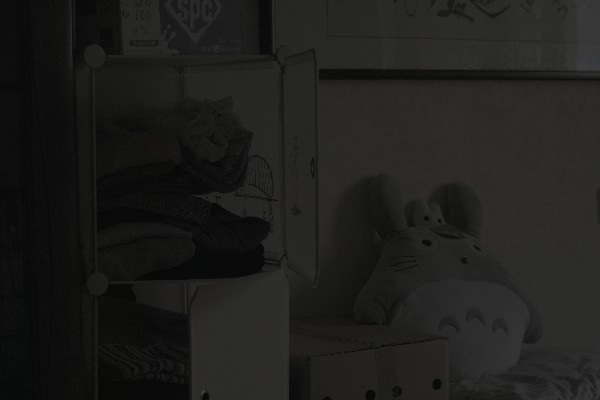}&
    \adjincludegraphics[valign=m,width=0.155\linewidth,trim={{.58\width} {.24\height} {.12\width} {.31\height}},clip]{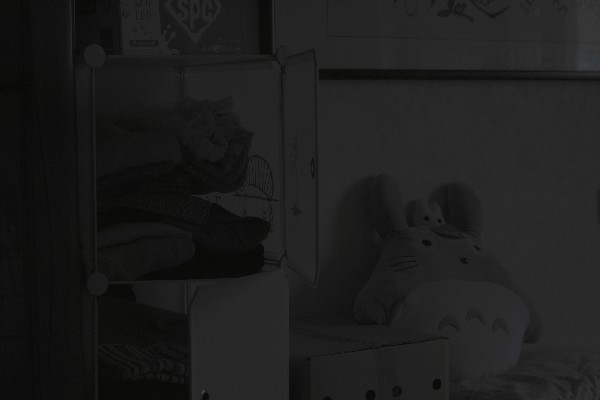}&
    \adjincludegraphics[valign=m,width=0.155\linewidth,trim={{.58\width} {.24\height} {.12\width} {.31\height}},clip]{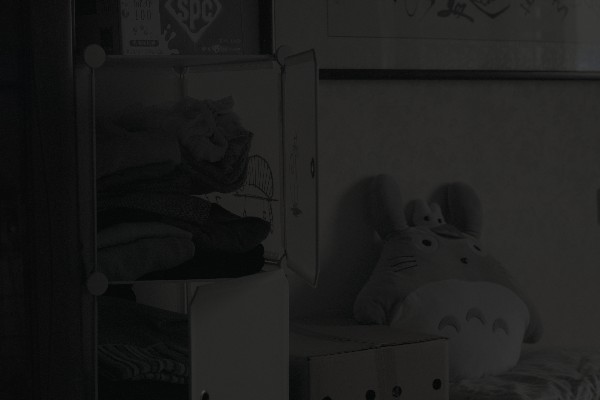}
    \\
\vspace{3pt}
    {\small $\mN$}&
    \adjincludegraphics[valign=m,width=0.155\linewidth,trim={{.58\width} {.24\height} {.12\width} {.31\height}},clip]{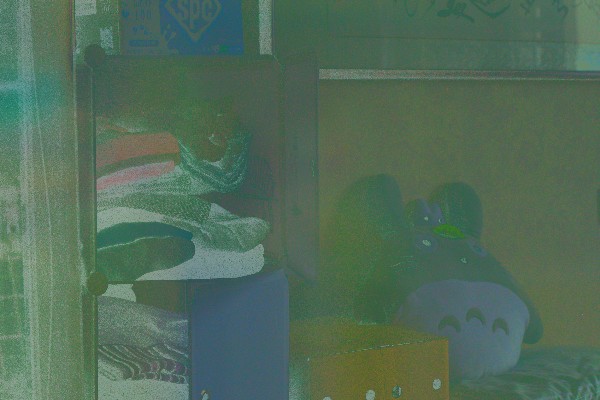}&
    \adjincludegraphics[valign=m,width=0.155\linewidth,trim={{.58\width} {.24\height} {.12\width} {.31\height}},clip]{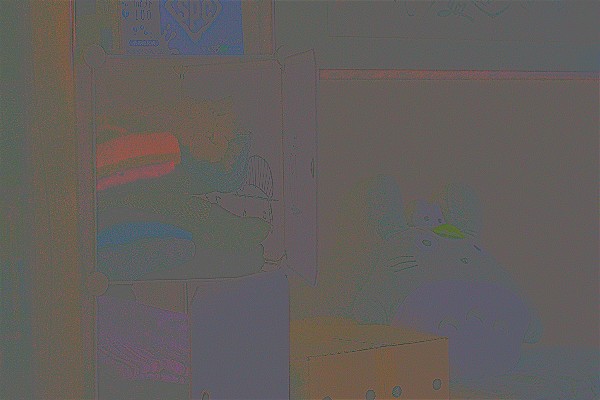}&
    \adjincludegraphics[valign=m,width=0.155\linewidth,trim={{.58\width} {.24\height} {.12\width} {.31\height}},clip]{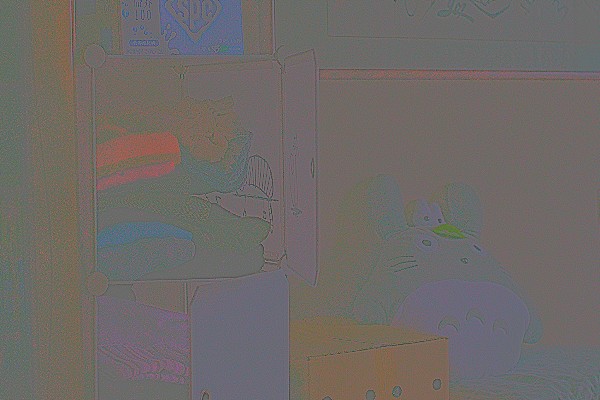}&
    \adjincludegraphics[valign=m,width=0.155\linewidth,trim={{.58\width} {.24\height} {.12\width} {.31\height}},clip]{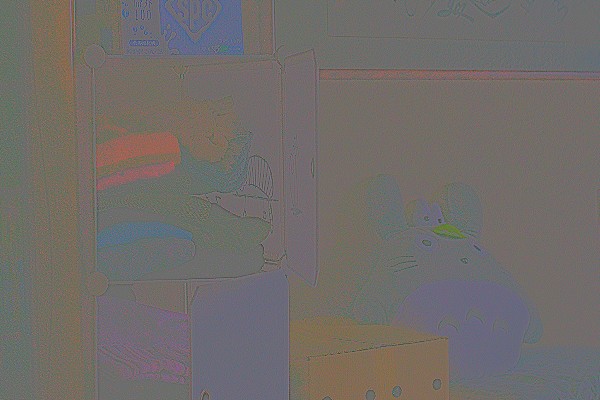}&
    \adjincludegraphics[valign=m,width=0.155\linewidth,trim={{.58\width} {.24\height} {.12\width} {.31\height}},clip]{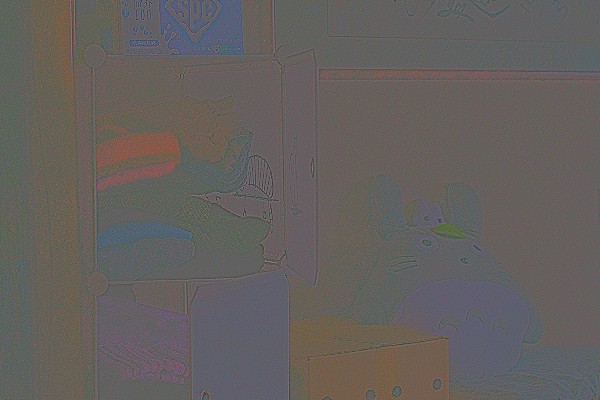}&
    \adjincludegraphics[valign=m,width=0.155\linewidth,trim={{.58\width} {.24\height} {.12\width} {.31\height}},clip]{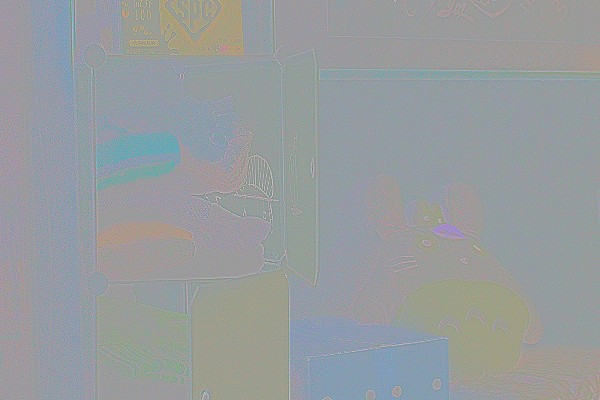}
    \\
  &  {\small (a) w/o $\opt{T}_{\mW,\Delta}$  } &  {\small (b) $\opt{T}_{\mW}$ } &  {\small (c) $\opt{T}_{\mW,\Delta}\ W=7$ } &  {\small (d) $\opt{T}_{\mW,\Delta}\ W=15$ } &  {\small (e) $\opt{T}_{\mW,\Delta}\ W=31$ } &  {\small (f) learning $\mI-\mN$}\\
\end{tabular}
\caption{\small 
Qualitative comparison of the results using different transforms.
}
\label{fig:trans}
\end{figure*}

One component to distinguish our method from existing ones is the spatially varying deformable convolution learned in the bilateral space. To verify its effectiveness of the large receptive field and irregular sampling positions,
we conduct controlled experiments as listed in Table~\ref{tab:kernel}.
The corresponding results generated by transforms with different receptive fields are shown in Fig~\ref{fig:trans}.
The result in the first row of Table~\ref{tab:kernel} is obtained by using the point operation for image-to-noise mapping in the same form of that for image-to-illumination mapping.
In this setting, the receptive field of the learned kernels is $1\times 1$.
As shown in Fig~\ref{fig:trans} (a), the produced results with only the point operations are dominated by amplified noise as it is hard to distinguish those noise without neighborhood information. 
The second row shows results using learned kernels sampling on a rigid square neighborhood.
The next three rows show results using deformable convolution with both learned kernels and learned sampling positions from windows of size $W\times W$, which can better distinguish the noise from texture, as evidenced by the noise component $\mN$ in Fig~\ref{fig:trans} (c), (d) and (e).
It is noted that the rows from top to bottom show results obtained by transforms with increasingly larger sampling window size.
We evaluate the proposed method with exponential increased window size and find that learning deformable convolution with a receptive field of $15\times 15$ yields the best performance.

\subsubsection{Ablation study on intermediates to be estimated}
We verify the effectiveness of the estimation of the noise $\mN$ by comparing the proposed scheme with the estimation of the noise-free low-light image $\mI-\mN$ instead.
It can be seen that, estimation of the noise reveals more noise as shown in  Fig~\ref{fig:trans} (d) \textit{v.s.} (f) and  yields better performance as shown in Table~\ref{tab:noise}.

\begin{table}[htbp!]
    \centering
    \caption{\small Ablation study on intermediates to be estimated.}
    \addtolength{\tabcolsep}{-2.5pt}
    \sisetup{detect-all=true,detect-weight=true,detect-inline-weight=math}
    \begin{tabular}{cS[table-format=2.4]S[table-format=1.4]}
        \toprule
        {Estimated intermediates} & \multicolumn{1}{c}{{PSNR(dB)}} & \multicolumn{1}{c}{{SSIM}} \\
        \midrule
        illumination $\mE$ and noise-free image $\mI-\mN$ & 21.0485  & 0.7621  \\
        illumination $\mE$ and noise $\mN$ & \BF 22.5156  & \BF 0.7864  \\
        \bottomrule
    \end{tabular}%
    \label{tab:noise}%
\end{table}%

\begin{figure*}[htbp!]\centering
\setlength{\tabcolsep}{2pt}
\begin{tabular}{ccccccc}
\vspace{3pt}
{\small $\widetilde{\mR}$}&
\adjincludegraphics[valign=m,width=0.155\linewidth,trim={{.07\width} {.16\height} {.04\width} {.04\height}},clip]{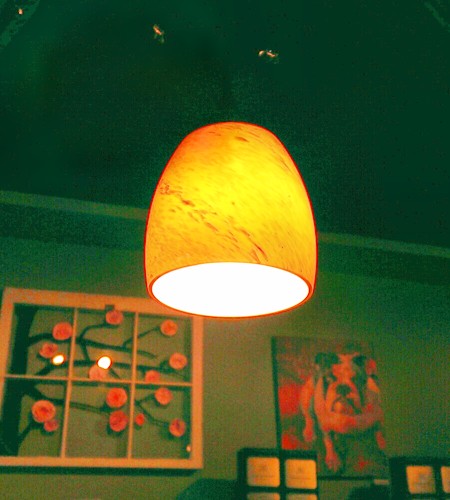}&
\adjincludegraphics[valign=m,width=0.155\linewidth,trim={{.07\width} {.16\height} {.04\width} {.04\height}},clip]{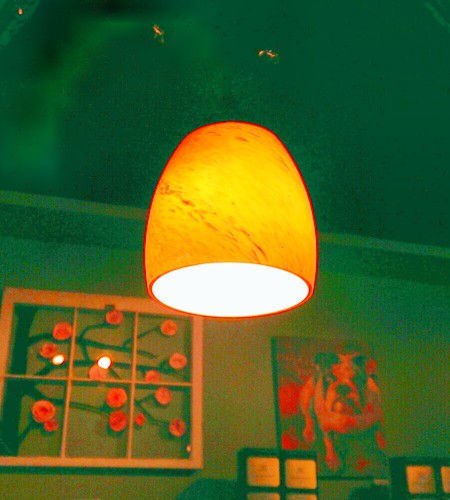}&
\adjincludegraphics[valign=m,width=0.155\linewidth,trim={{.07\width} {.16\height} {.04\width} {.04\height}},clip]{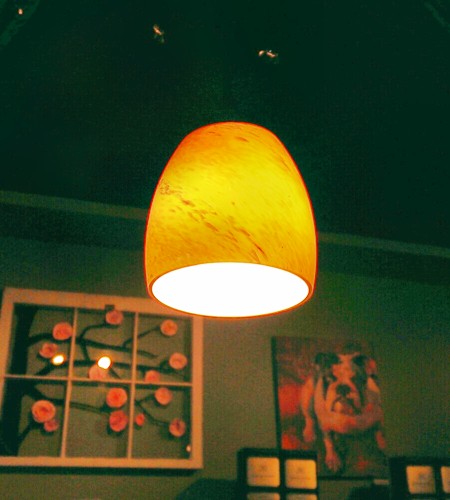}&
\adjincludegraphics[valign=m,width=0.155\linewidth,trim={{.07\width} {.16\height} {.04\width} {.04\height}},clip]{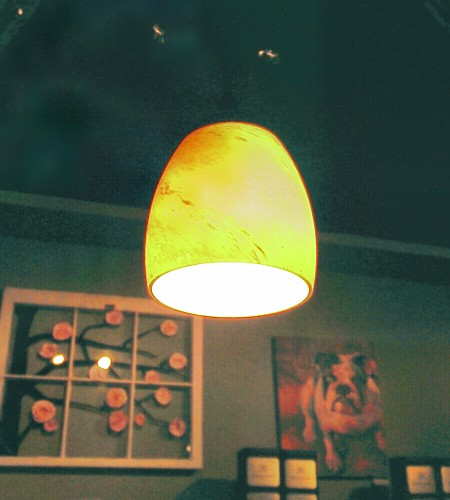}&
\adjincludegraphics[valign=m,width=0.155\linewidth,trim={{.07\width} {.16\height} {.04\width} {.04\height}},clip]{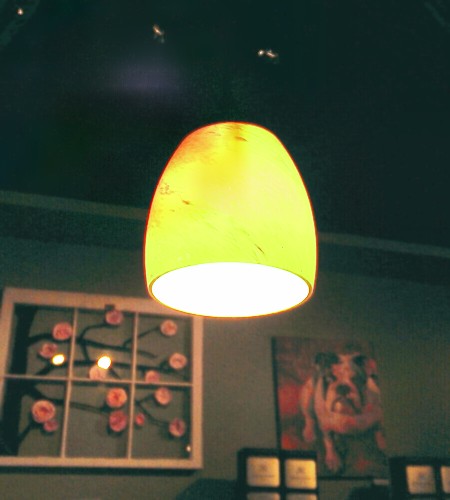}&
\adjincludegraphics[valign=m,width=0.155\linewidth,trim={{.07\width} {.16\height} {.04\width} {.04\height}},clip]{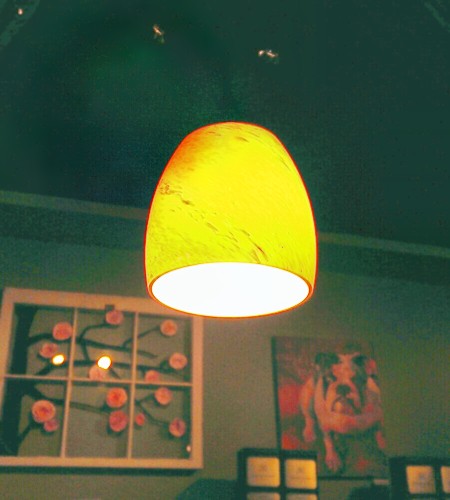}\\
\vspace{3pt}
{\small $\mE$}&
\adjincludegraphics[valign=m,width=0.155\linewidth,trim={{.07\width} {.16\height} {.04\width} {.04\height}},clip]{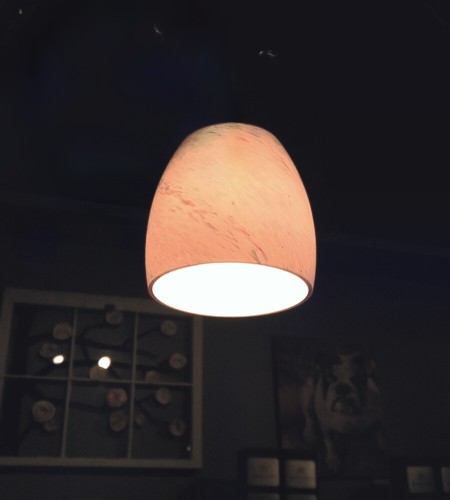}&
\adjincludegraphics[valign=m,width=0.155\linewidth,trim={{.07\width} {.16\height} {.04\width} {.04\height}},clip]{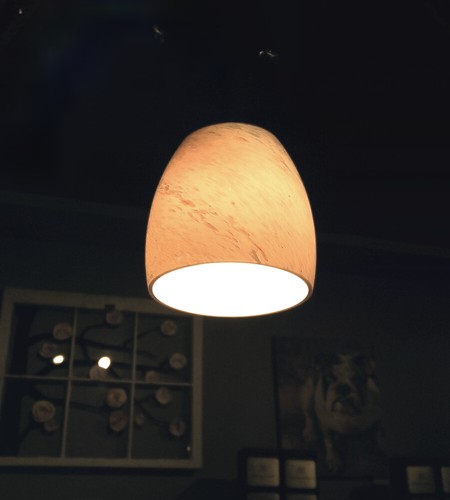}&
\adjincludegraphics[valign=m,width=0.155\linewidth,trim={{.07\width} {.16\height} {.04\width} {.04\height}},clip]{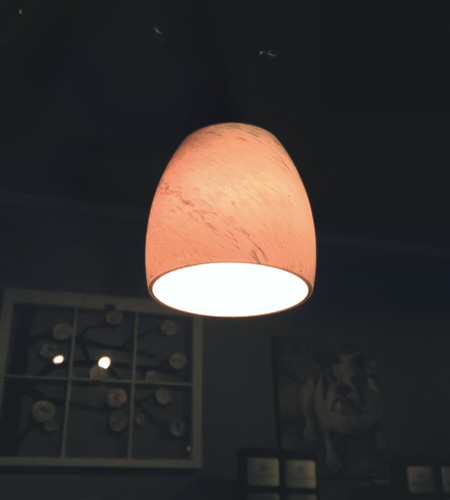}&
\adjincludegraphics[valign=m,width=0.155\linewidth,trim={{.07\width} {.16\height} {.04\width} {.04\height}},clip]{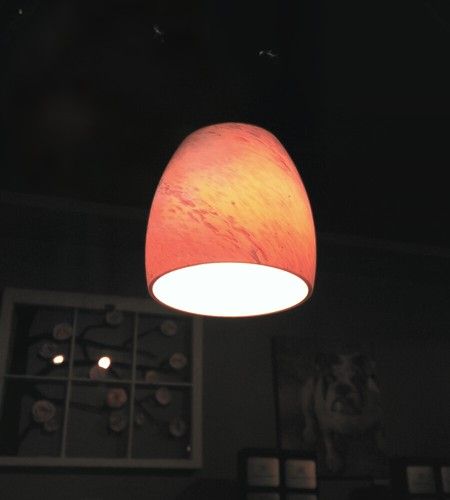}&
\adjincludegraphics[valign=m,width=0.155\linewidth,trim={{.07\width} {.16\height} {.04\width} {.04\height}},clip]{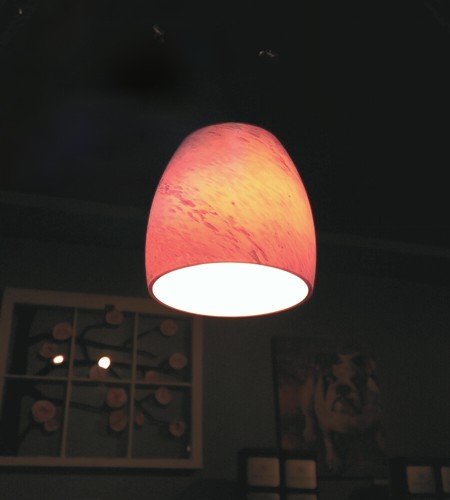}&
\adjincludegraphics[valign=m,width=0.155\linewidth,trim={{.07\width} {.16\height} {.04\width} {.04\height}},clip]{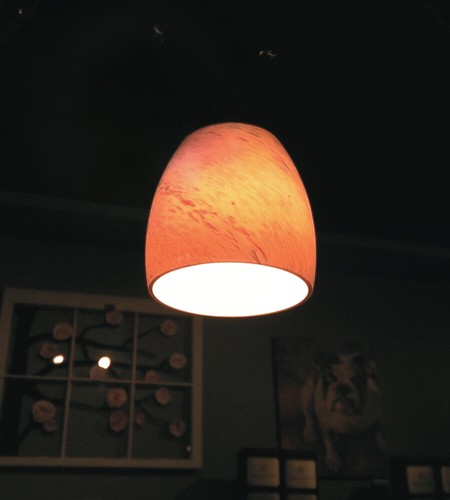}
\\
\vspace{3pt}
{\small $\mN$}&
\adjincludegraphics[valign=m,width=0.155\linewidth,trim={{.07\width} {.16\height} {.04\width} {.04\height}},clip]{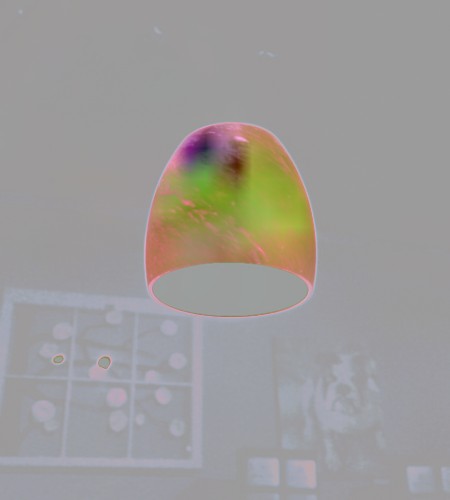}&
\adjincludegraphics[valign=m,width=0.155\linewidth,trim={{.07\width} {.16\height} {.04\width} {.04\height}},clip]{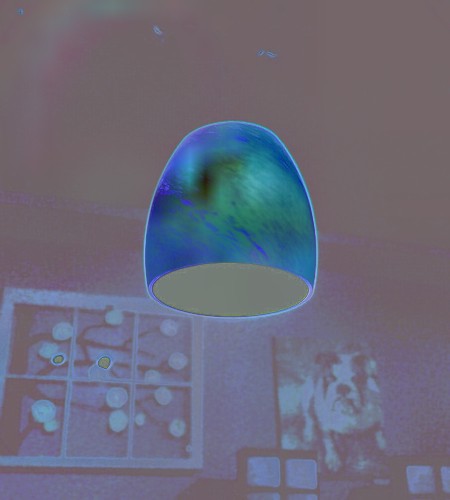}&
\adjincludegraphics[valign=m,width=0.155\linewidth,trim={{.07\width} {.16\height} {.04\width} {.04\height}},clip]{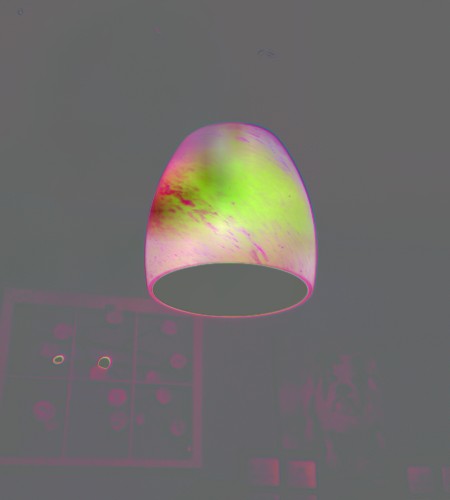}&
\adjincludegraphics[valign=m,width=0.155\linewidth,trim={{.07\width} {.16\height} {.04\width} {.04\height}},clip]{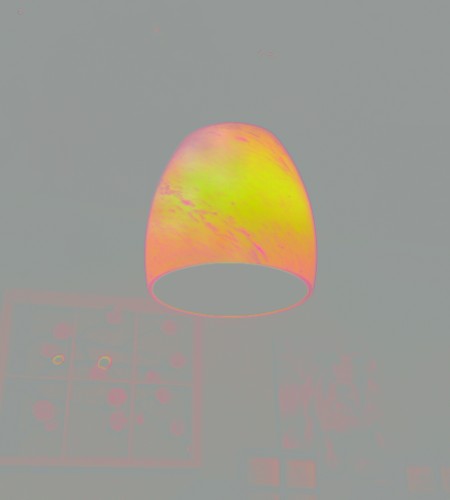}&
\adjincludegraphics[valign=m,width=0.155\linewidth,trim={{.07\width} {.16\height} {.04\width} {.04\height}},clip]{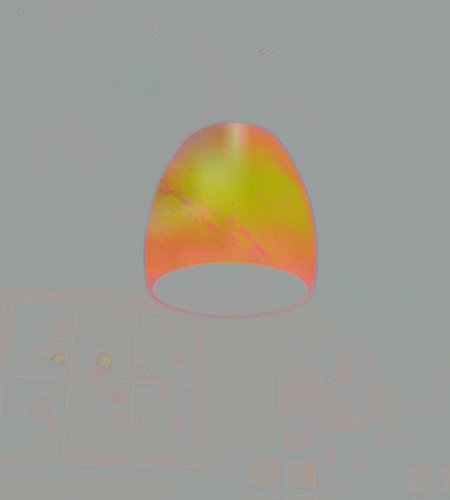}&
\adjincludegraphics[valign=m,width=0.155\linewidth,trim={{.07\width} {.16\height} {.04\width} {.04\height}},clip]{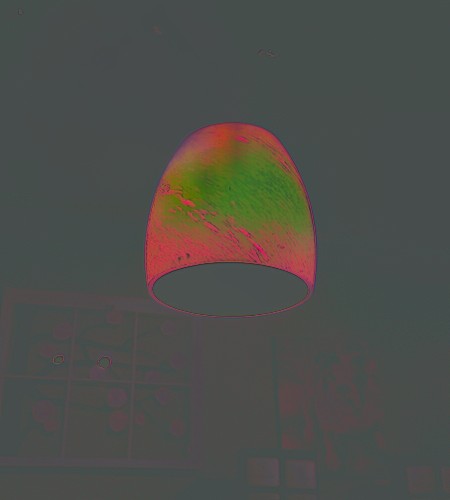}
\\
 &  {\small(a) $\opt{L}_r$ w/o $\ell_{\rm grad}$ } &  {\small(b) $\ell_{\rm grad}\to\ell_{\rm ssim}$ } &  {\small(c) default } &  {\small(d) w/o $\opt{L}_n$ } &  {\small(e) w/o $\opt{L}_e$ } &  {\small (f) $\opt{L}_e$ w/ $\ell_2$}\\
\end{tabular}
\caption{\small 
Ablation study on the loss function.
}
\label{fig:loss}
\end{figure*}

\subsubsection{Ablation study on the loss function}
We conduct ablation study on the proposed loss function.  See Table~\ref{tab:loss} and Fig.~\ref{fig:loss} for the quantitative and qualitative results respectively.
Firstly, as shown in Table~\ref{tab:loss}, in all settings, our results are better than the most recent existing work DeepUPE~\cite{wang2019underexposed} which only estimates illumination map.
It indicates the effectiveness of the proposed framework, which handles color and noise simultaneously and performs image-to-noise mapping by edge-aware deformable convolution.

It can be seen from the first three rows in Table~\ref{tab:loss} that the performance degenerates without $\ell_{\rm grad}$. 
The edge-aware loss $\opt{L}_r$ is effective in the proposed framework. 
The results also demonstrat the superiority in our setting over the perceptually motivated SSIM loss with  $\ell_{\rm ssim}=1-{\rm SSIM}$.

The last three rows demonstrate the effectiveness of the proposed priors on illumination and noise, without which the results are inferior, especially when there are artifacts in the flat regions of the images and the smoothness of the illumination is hard to preserve, as shown in Fig.~\ref{fig:loss} (d), (e) and (f).
Although with or without $\opt{L}_e$ yield comparable performance, without  $\opt{L}_e$ might produce some artifacts as shown in Fig.~\ref{fig:loss} (e).
Using $\ell_1$ norm to regularize the variation of illumination shows better performance over $\ell_2$ norm for $\opt{L}_e$.

\begin{table}[htbp!]
    \centering
    \caption{\small Ablation study on the loss function.}
    \addtolength{\tabcolsep}{-2.5pt}
    \sisetup{detect-all=true,detect-weight=true,detect-inline-weight=math}
    \begin{tabular}{cccS[table-format=2.4]S[table-format=1.4]}
        \toprule
        {$\opt{L}_r$} & {$\opt{L}_e$} & {$\opt{L}_n$} & \multicolumn{1}{c}{{PSNR(dB)}} & \multicolumn{1}{c}{{SSIM}} \\
        \midrule
        w/o $\ell_{\rm grad}$ & default & default & 20.9161  & 0.7570  \\
        $\ell_{\rm grad}\to\ell_{\rm ssim}$ & default & default & 20.7780  & 0.7614  \\
        default & default & default & \BF 22.5156  & \BF 0.7864  \\
        default & default & \textit{None} & 21.3178  & 0.7811  \\
        default & \textit{None} & default & 22.4539  & 0.7822  \\
        default & $\ell_{2}$ & default & 21.1428  & 0.7380  \\
        \bottomrule
    \end{tabular}%
    \label{tab:loss}%
\end{table}%

\section{Summary}
\label{sec:summary}
This paper develops a deep learning method for  low-light image enhancement, with the focus on the handling of measurement noise.  Motivated by the inherently coupled relationship between illumination and measurement noise, we proposed a novel deep bilateral Retinex method, which performs Retinex decomposition in the bilateral space of low-light images. The proposed method is extensively evaluated in several benchmark datasets and compared to several representative related methods. The experiments show that the proposed method outperforms the compared methods, especially in the case of very low lighting conditions. In future, we plan to investigate possible applications of the proposed method in other image processing tasks involving Retinex decomposition.


\ifCLASSOPTIONcaptionsoff
\newpage
\fi



\begin{thebibliography}{10}\itemsep=-1pt

\bibitem{arici2009histogram}
T.~Arici, S.~Dikbas, and Y.~Altunbasak.
\newblock A {{Histogram Modification Framework}} and {{Its Application}} for
  {{Image Contrast Enhancement}}.
\newblock {\em IEEE Trans. Image Process.}, 18(9):1921--1935, Sept. 2009.

\bibitem{bako2017kernelpredicting}
S.~Bako, T.~Vogels, B.~Mcwilliams, M.~Meyer, J.~Nov{\'a}K, A.~Harvill, P.~Sen,
  T.~Derose, and F.~Rousselle.
\newblock Kernel-predicting {{Convolutional Networks}} for {{Denoising Monte
  Carlo Renderings}}.
\newblock {\em ACM Trans. Graph.}, 36(4):97:1--97:14, July 2017.

\bibitem{cai2018learning}
J.~Cai, S.~Gu, and L.~Zhang.
\newblock Learning a {{Deep Single Image Contrast Enhancer}} from
  {{Multi}}-{{Exposure Images}}.
\newblock {\em IEEE Trans. Image Process.}, 27(4):2049--2062, Apr. 2018.

\bibitem{celik2011contextual}
T.~Celik and T.~Tjahjadi.
\newblock Contextual and {{Variational Contrast Enhancement}}.
\newblock {\em IEEE Trans. Image Process.}, 20(12):3431--3441, Dec. 2011.

\bibitem{chen2018learning}
C.~Chen, Q.~Chen, J.~Xu, and V.~Koltun.
\newblock Learning to {{See}} in the {{Dark}}.
\newblock In {\em Proc. IEEE Int. Conf. Comput. Vis. Pattern Recognit. (CVPR)},
  pages 3291--3300, June 2018.

\bibitem{chen2016bilateral}
J.~Chen, A.~Adams, N.~Wadhwa, and S.~W. Hasinoff.
\newblock Bilateral {{Guided Upsampling}}.
\newblock {\em ACM Trans. Graph.}, 35(6):203:1--203:8, Nov. 2016.

\bibitem{dong2010fast}
X.~Dong, Y.~A. Pang, and J.~G. Wen.
\newblock Fast {{Efficient Algorithm}} for {{Enhancement}} of {{Low Lighting
  Video}}.
\newblock In {\em {{ACM SIGGRAPH}}}, pages 69:1--69:1, 2010.

\bibitem{elad2005retinex}
M.~Elad.
\newblock Retinex by {{Two Bilateral Filters}}.
\newblock In {\em Proc. Scale Space and PDE Methods in Computer Vision
  (Scale-Space)}, pages 217--229, 2005.

\bibitem{fu2015probabilistic}
X.~Fu, Y.~Liao, D.~Zeng, Y.~Huang, X.~Zhang, and X.~Ding.
\newblock A {{Probabilistic Method}} for {{Image Enhancement With Simultaneous
  Illumination}} and {{Reflectance Estimation}}.
\newblock {\em IEEE Trans. Image Process.}, 24(12):4965--4977, Dec. 2015.

\bibitem{fu2016fusionbased}
X.~Fu, D.~Zeng, Y.~Huang, Y.~Liao, X.~Ding, and J.~Paisley.
\newblock A fusion-based enhancing method for weakly illuminated images.
\newblock {\em Signal Process.}, 129:82--96, Dec. 2016.

\bibitem{fu2016weighted}
X.~Fu, D.~Zeng, Y.~Huang, X.-P. Zhang, and X.~Ding.
\newblock A {{Weighted Variational Model}} for {{Simultaneous Reflectance}} and
  {{Illumination Estimation}}.
\newblock In {\em Proc. IEEE Int. Conf. Comput. Vis. Pattern Recognit. (CVPR)},
  pages 2782--2790, 2016.

\bibitem{gharbi2017deep}
M.~Gharbi, J.~Chen, J.~T. Barron, S.~W. Hasinoff, and F.~Durand.
\newblock Deep {{Bilateral Learning}} for {{Real}}-time {{Image Enhancement}}.
\newblock {\em ACM Trans. Graph.}, 36(4):118:1--118:12, July 2017.

\bibitem{guo2017lime}
X.~Guo, Y.~Li, and H.~Ling.
\newblock {{LIME}}: {{Low}}-{{Light Image Enhancement}} via {{Illumination Map
  Estimation}}.
\newblock {\em IEEE Trans. Image Process.}, 26(2):982--993, Feb. 2017.

\bibitem{he2015delving}
K.~He, X.~Zhang, S.~Ren, and J.~Sun.
\newblock Delving {{Deep}} into {{Rectifiers}}: {{Surpassing Human}}-{{Level
  Performance}} on {{ImageNet Classification}}.
\newblock In {\em Proc. IEEE Int. Conf. Comput. Vis. (ICCV)}, pages 1026--1034,
  Dec. 2015.

\bibitem{huang2013efficient}
S.~Huang, F.~Cheng, and Y.~Chiu.
\newblock Efficient {{Contrast Enhancement Using Adaptive Gamma Correction With
  Weighting Distribution}}.
\newblock {\em IEEE Trans. Image Process.}, 22(3):1032--1041, Mar. 2013.

\bibitem{jobson1997multiscale}
D.~J. Jobson, Z.~Rahman, and G.~A. Woodell.
\newblock A multiscale retinex for bridging the gap between color images and
  the human observation of scenes.
\newblock {\em IEEE Trans. Image Process.}, 6(7):965--976, July 1997.

\bibitem{jobson1997properties}
D.~J. Jobson, Z.~Rahman, and G.~A. Woodell.
\newblock Properties and performance of a center/surround retinex.
\newblock {\em IEEE Trans. Image Process.}, 6(3):451--462, Mar. 1997.

\bibitem{kim2019deformable}
B.~Kim, J.~Ponce, and B.~Ham.
\newblock Deformable {{Kernel Networks}} for {{Joint Image Filtering}}.
\newblock {\em arXiv:1910.08373 [cs]}, Oct. 2019.

\bibitem{kimmel2003variational}
R.~Kimmel, M.~Elad, D.~Shaked, R.~Keshet, and I.~Sobel.
\newblock A {{Variational Framework}} for {{Retinex}}.
\newblock {\em Int. J. Comput. Vision}, 52(1):7--23, Apr. 2003.

\bibitem{kingma2015adam}
D.~P. Kingma and J.~Ba.
\newblock Adam: {{A Method}} for {{Stochastic Optimization}}.
\newblock In {\em Proc. Int. Conf. Learn. Represent. (ICLR)}, 2015.

\bibitem{land1977retinex}
E.~H. Land.
\newblock The {{Retinex Theory}} of {{Color Vision}}.
\newblock {\em Sci. Amer.}, 237(6):108--129, 1977.

\bibitem{lee2013contrast}
C.~Lee, C.~Lee, and C.-S. Kim.
\newblock Contrast {{Enhancement Based}} on {{Layered Difference
  Representation}} of {{2D Histograms}}.
\newblock {\em IEEE Trans. Image Process.}, 22(12):5372--5384, Dec. 2013.

\bibitem{li2018structurerevealing}
M.~Li, J.~Liu, W.~Yang, X.~Sun, and Z.~Guo.
\newblock Structure-{{Revealing Low}}-{{Light Image Enhancement Via Robust
  Retinex Model}}.
\newblock {\em IEEE Trans. Image Process.}, 27(6):2828--2841, June 2018.

\bibitem{lore2017llnet}
K.~G. Lore, A.~Akintayo, and S.~Sarkar.
\newblock {{LLNet}}: {{A}} deep autoencoder approach to natural low-light image
  enhancement.
\newblock {\em Pattern Recognit.}, 61:650--662, Jan. 2017.

\bibitem{loza2013automatic}
A.~{\L}oza, D.~R. Bull, P.~R. Hill, and A.~M. Achim.
\newblock Automatic contrast enhancement of low-light images based on local
  statistics of wavelet coefficients.
\newblock {\em Digital Signal Process.}, 23(6):1856--1866, Dec. 2013.

\bibitem{ma2015perceptual}
K.~Ma, K.~Zeng, and Z.~Wang.
\newblock Perceptual {{Quality Assessment}} for {{Multi}}-{{Exposure Image
  Fusion}}.
\newblock {\em IEEE Trans. Image Process.}, 24(11):3345--3356, Nov. 2015.

\bibitem{ma20111based}
W.~Ma, J.-M. Morel, S.~Osher, and A.~Chien.
\newblock An {{L}} 1-based variational model for {{Retinex}} theory and its
  application to medical images.
\newblock In {\em Proc. IEEE Int. Conf. Comput. Vis. Pattern Recognit. (CVPR)},
  pages 153--160, June 2011.

\bibitem{mildenhall2018burst}
B.~Mildenhall, J.~T. Barron, J.~Chen, D.~Sharlet, R.~Ng, and R.~Carroll.
\newblock Burst {{Denoising With Kernel Prediction Networks}}.
\newblock In {\em Proc. IEEE Int. Conf. Comput. Vis. Pattern Recognit. (CVPR)},
  pages 2502--2510, 2018.

\bibitem{mittal2012noreference}
A.~Mittal, A.~K. Moorthy, and A.~C. Bovik.
\newblock No-{{Reference Image Quality Assessment}} in the {{Spatial Domain}}.
\newblock {\em IEEE Trans. Image Process.}, 21(12):4695--4708, Dec. 2012.

\bibitem{mittal2013making}
A.~Mittal, R.~Soundararajan, and A.~C. Bovik.
\newblock Making a ``{{Completely Blind}}'' {{Image Quality Analyzer}}.
\newblock {\em IEEE Signal. Proc. Let.}, 20(3):209--212, Mar. 2013.

\bibitem{venkatanathn2015blind}
V.~N, P.~D, M.~C. Bh, S.~S. Channappayya, and S.~S. Medasani.
\newblock Blind image quality evaluation using perception based features.
\newblock In {\em {{21st Nat. Conf.}} {{Commun.}} ({{NCC}})}, pages 1--6, Feb.
  2015.

\bibitem{ng2011total}
M.~Ng and W.~Wang.
\newblock A {{Total Variation Model}} for {{Retinex}}.
\newblock {\em SIAM J. Imag. Sci.}, 4(1):345--365, Jan. 2011.

\bibitem{niklaus2017videoa}
S.~Niklaus, L.~Mai, and F.~Liu.
\newblock Video {{Frame Interpolation}} via {{Adaptive Convolution}}.
\newblock In {\em Proc. IEEE Int. Conf. Comput. Vis. Pattern Recognit. (CVPR)},
  pages 670--679, 2017.

\bibitem{niklaus2017video}
S.~Niklaus, L.~Mai, and F.~Liu.
\newblock Video {{Frame Interpolation}} via {{Adaptive Separable Convolution}}.
\newblock In {\em Proc. IEEE Int. Conf. Comput. Vis. (ICCV)}, pages 261--270,
  2017.

\bibitem{paszke2019pytorch}
A.~Paszke, S.~Gross, F.~Massa, A.~Lerer, J.~Bradbury, G.~Chanan, T.~Killeen,
  Z.~Lin, N.~Gimelshein, L.~Antiga, A.~Desmaison, A.~Kopf, E.~Yang, Z.~DeVito,
  M.~Raison, A.~Tejani, S.~Chilamkurthy, B.~Steiner, L.~Fang, J.~Bai, and
  S.~Chintala.
\newblock {{PyTorch}}: {{An Imperative Style}}, {{High}}-{{Performance Deep
  Learning Library}}.
\newblock In {\em Proc. Annu. Conf. Neural Inf. Process. Syst. (NeurIPS)},
  pages 8026--8037, 2019.

\bibitem{ren2018joint}
X.~Ren, M.~Li, W.~Cheng, and J.~Liu.
\newblock Joint {{Enhancement}} and {{Denoising Method}} via {{Sequential
  Decomposition}}.
\newblock In {\em IEEE Int. Symp. Circuits Syst. (ISCAS)}, pages 1--5, May
  2018.

\bibitem{tomasi1998bilateral}
C.~Tomasi and R.~Manduchi.
\newblock Bilateral filtering for gray and color images.
\newblock In {\em Proc. IEEE Int. Conf. Comput. Vis. (ICCV)}, pages 839--846,
  Jan. 1998.

\bibitem{wang2019underexposed}
R.~Wang, Q.~Zhang, C.-W. Fu, X.~Shen, W.-S. Zheng, and J.~Jia.
\newblock Underexposed {{Photo Enhancement}} using {{Deep Illumination
  Estimation}}.
\newblock In {\em Proc. IEEE Int. Conf. Comput. Vis. Pattern Recognit. (CVPR)},
  page~9, 2019.

\bibitem{wang2018naturalness}
S.~Wang and G.~Luo.
\newblock Naturalness {{Preserved Image Enhancement Using}} a {{Priori
  Multi}}-{{Layer Lightness Statistics}}.
\newblock {\em IEEE Trans. Image Process.}, 27(2):938--948, Feb. 2018.

\bibitem{wang2013naturalness}
S.~Wang, J.~Zheng, H.~Hu, and B.~Li.
\newblock Naturalness {{Preserved Enhancement Algorithm}} for {{Non}}-{{Uniform
  Illumination Images}}.
\newblock {\em IEEE Trans. Image Process.}, 22(9):3538--3548, Sept. 2013.

\bibitem{wang2018gladnet}
W.~Wang, C.~Wei, W.~Yang, and J.~Liu.
\newblock {{GLADNet}}: {{Low}}-{{Light Enhancement Network}} with {{Global
  Awareness}}.
\newblock In {\em Proc. IEEE Int. Conf. {{Automat. Face Gesture Recognit.}}
  ({{FG}})}, pages 751--755, May 2018.

\bibitem{wang2004image}
Z.~Wang, A.~C. Bovik, H.~R. Sheikh, and E.~P. Simoncelli.
\newblock Image quality assessment: From error visibility to structural
  similarity.
\newblock {\em IEEE Trans. Image Process.}, 13(4):600--612, Apr. 2004.

\bibitem{wei2018deep}
C.~Wei, W.~Wang, W.~Yang, and J.~Liu.
\newblock Deep {{Retinex Decomposition}} for {{Low}}-{{Light Enhancement}}.
\newblock In {\em Br. Mac. Vis. Conf. (BMVC)}, Aug. 2018.

\bibitem{wei2020physicsbased}
K.~Wei, Y.~Fu, J.~Yang, and H.~Huang.
\newblock A {Physics}-based {Noise} {Formation} {Model} for {Extreme}
  {Low}-light {Raw} {Denoising}.
\newblock In {\em Proc. IEEE Int. Conf. Comput. Vis. Pattern Recognit. (CVPR)},
  Apr. 2020.

\bibitem{xu2012structure}
L.~Xu, Q.~Yan, Y.~Xia, and J.~Jia.
\newblock Structure extraction from texture via relative total variation.
\newblock {\em ACM Trans. Graph.}, 31(6):139:1--139:10, Nov. 2012.

\bibitem{xu2019learning}
X.~Xu, M.~Li, and W.~Sun.
\newblock Learning {{Deformable Kernels}} for {{Image}} and {{Video
  Denoising}}.
\newblock {\em arXiv:1904.06903 [cs]}, Apr. 2019.

\bibitem{ying2017bioinspired}
Z.~Ying, G.~Li, and W.~Gao.
\newblock A {{Bio}}-{{Inspired Multi}}-{{Exposure Fusion Framework}} for
  {{Low}}-light {{Image Enhancement}}.
\newblock {\em arXiv:1711.00591 [cs]}, Nov. 2017.

\bibitem{yuan2012automatic}
L.~Yuan and J.~Sun.
\newblock Automatic {{Exposure Correction}} of {{Consumer Photographs}}.
\newblock In {\em Proc. IEEE Eur. Conf. Comput. Vis. (ECCV)}, pages 771--785,
  2012.

\bibitem{zhang2019kindling}
Y.~Zhang, J.~Zhang, and X.~Guo.
\newblock Kindling the {{Darkness}}: {{A Practical Low}}-light {{Image
  Enhancer}}.
\newblock In {\em Proc. ACM Int. Conf. Multimed. (ACM MM)}, May 2019.

\bibitem{zhao2012closedform}
Q.~Zhao, P.~Tan, Q.~Dai, L.~Shen, E.~Wu, and S.~Lin.
\newblock A {{Closed}}-{{Form Solution}} to {{Retinex}} with {{Nonlocal Texture
  Constraints}}.
\newblock {\em IEEE Trans. Pattern Anal. Mach. Intell.}, 34(7):1437--1444, July
  2012.

\end{thebibliography}


\end{document}